\documentclass{aa}  
\usepackage{graphicx}
\usepackage{txfonts}
\usepackage{natbib}

\defcitealias{DiFolco07}{Paper~I}
\defcitealias{Absil08}{Paper~II}

\begin{document}
   \title{A near-infrared interferometric survey of debris disc stars}

   \subtitle{III. First statistics based on 42 stars observed with CHARA/FLUOR}

   \author{O. Absil\inst{1}
          \and
          D. Defr\`ere\inst{2,3}
          \and
          V. Coud\'e du Foresto\inst{4}
          \and
          E. Di Folco\inst{5}
          \and
          A. M\'erand\inst{6}
          \and
          J.-C. Augereau\inst{7}
          \and
          S. Ertel\inst{7}
          \and
          C. Hanot\inst{1}
          \and
          P.~Kervella\inst{4}
          \and
          B.~Mollier\inst{4}
          \and
          N.~Scott\inst{8,4}
          \and
          X. Che\inst{9}
          \and
          J.~D. Monnier\inst{9}
          \and
          N. Thureau\inst{10}
          \and
          P.~G. Tuthill\inst{11}
          \and
          T.~A.~ten~Brummelaar\inst{8}
          \and
          H.~A.~McAlister\inst{8}
          \and
          J. Sturmann\inst{8}
          \and
          L. Sturmann\inst{8}
          \and
          N. Turner\inst{8}
          }


\institute{D\'epartement d'Astrophysique, G\'eophysique et Oc\'eanographie, Universit\'e de Li\`ege, 17 All\'ee du Six Ao\^ut, B-4000 Li\`ege, Belgium \\ \email{absil(at)astro.ulg.ac.be}
\and Steward Observatory, Department of Astronomy, University of Arizona, 933 N.\ Cherry Avenue, Tucson, AZ 85721, USA
\and Max Planck Institut f\"ur Radioastronomie, Auf dem H\"ugel 69, D-53121 Bonn, Germany
\and LESIA, Observatoire de Paris, CNRS/UMR 8109, UPMC, Universit\'e Paris Diderot, 5 place J.\ Janssen, F-92195 Meudon, France
\and Observatoire Aquitain des Sciences de l'Univers, Universit\'e de Bordeaux, 2 rue de l'Observatoire BP 89, F-33271 Floirac, France
\and European Southern Observatory, Alonso de Cordova 3107, Casilla 19001, Vitacura, Santiago 19, Chile
\and UJF-Grenoble 1 / CNRS, Institut de Plan\'etologie et d'Astrophysique de Grenoble (IPAG) UMR 5274, F-38041 Grenoble, France
\and Center for High Angular Resolution Astronomy, Georgia State University, PO Box 3969, Atlanta, Georgia 30302-3965, USA
\and Astronomy Department, University of Michigan (Astronomy), 500 Church St, Ann Arbor, MI 48109, USA
\and Department of Physics and Astronomy, University of St.\ Andrews, North Haugh, St.\ Andrews, Fife KY16 9SS, UK
\and Sydney Institute for Astronomy (SIfA), School of Physics, University of Sydney, NSW 2006, Australia}

\date{Received April 9, 2013; accepted May 24, 2013}

\abstract
   {Dust is expected to be ubiquitous in extrasolar planetary systems owing to the dynamical activity of minor bodies. Inner dust populations are, however, still poorly known because of the high contrast and small angular separation with respect to their host star, and yet, a proper characterisation of \emph{exozodiacal} dust is mandatory for the design of future Earth-like planet imaging missions.}
   {We aim to determine the level of near-infrared exozodiacal dust emission around a sample of 42 nearby main sequence stars with spectral types ranging from A to K and to investigate its correlation with various stellar parameters and with the presence of cold dust belts.}
   {We use high-precision K-band visibilities obtained with the FLUOR interferometer on the shortest baseline of the CHARA array. The calibrated visibilities are compared with the expected visibility of the stellar photosphere to assess whether there is an additional, fully resolved circumstellar emission source.}
   {Near-infrared circumstellar emission amounting to about $1\%$ of the stellar flux is detected around 13 of our 42 target stars. Follow-up observations showed that one of them (eps Cep) is associated with a stellar companion, while another one was detected around what turned out to be a giant star (kap CrB). The remaining 11 excesses found around single main sequence stars are most probably associated with hot circumstellar dust, yielding an overall occurrence rate of $28^{+8}_{-6}\%$ for our (biased) sample. We show that the occurrence rate of bright exozodiacal discs correlates with spectral type, K-band excesses being more frequent around A-type stars. It also correlates with the presence of detectable far-infrared excess emission in the case of solar-type stars.}
   {This study provides new insight into the phenomenon of bright exozodiacal discs, showing that hot dust populations are probably linked to outer dust reservoirs in the case of solar-type stars. For A-type stars, no clear conclusion can be made regarding the origin of the detected near-infrared excesses.}


   \maketitle

\section{Introduction}

Debris discs, which are produced by the collisional grinding down of large rocky bodies, represent one of the most prominent signposts for planetary systems around main sequence stars. Nowadays, debris discs have been detected around hundreds of stars. These detections are mostly based on the reprocessing of stellar light by the disc, which produces excess emission in the system's spectral energy distribution (SED) at infrared and (sub-)millimetric wavelengths. Cold discs are more readily detectable than warm discs, because the far-infrared excess they produce generally overwhelms the stellar emission. Space-based missions, such as \textsc{Spitzer} and \textsc{Herschel}, have been particularly successful in this context, providing a robust statistical view on cold circumstellar dust around nearby main sequence stars \citep[e.g.,][]{Trilling08,Carpenter09,Eiroa13}. These surveys have shown that about 20\% of solar-type stars are surrounded by cold dust discs with fractional luminosities similar to or higher than the expected integrated luminosity of our own Edgeworth-Kuiper belt. In contrast, warm, inner dust populations (``exozodiacal dust''), with grain temperatures of about 300\,K or higher, are still poorly known. This only reflects the challenge that their detection generally represents.

While mid-infrared spectro-photometry can be used to detect and characterise particularly bright exozodiacal discs \citep[e.g.,][]{Beichman05,Morales11,Lisse12}, resolved imaging is the tool of choice for studying these dust populations, since it allows the disc emission to be disentangled from the large stellar flux at near- to mid-infrared wavelengths. A few warm discs have been resolved by mid-infrared single-aperture imaging instruments, especially around A- and F-type stars \citep[e.g.,][]{Moerchen07,Moerchen10}. The statistical aspect of resolved exozodiacal dust populations has, however, only been tackled using interferometry so far, in particular in the mid-infrared domain using the Keck Interferometer Nuller \citep[KIN,][]{MillanGabet11}. Although the statistical value of the sample remains modest, with only 25 stars observed, the mean exozodi level in the KIN study could be constrained to be $<150$~zodis\footnote{The unit ``zodi'' refers here to the equivalent luminosity of a disc identical to the solar system's zodiacal dust around the target star, i.e., a ``zodiacal-twin disc'' in \citet{Roberge12}.} at $3\sigma$. Despite this pioneering study, the occurrence of exozodiacal discs as a function of spectral type, age, and other stellar parameters mostly remains unknown.

The present study aims to significantly increase our grasp of the statistical aspects of exozodiacal discs around nearby main sequence stars, in an attempt to improve our understanding of their origin and evolution. In particular, we aim to study the occurrence of exozodiacal discs as a function of spectral type and age, and as a function of the presence of detectable dust reservoirs in the outer parts of the planetary system. Our study is based on the use of high-precision near-infrared interferometry, which provides a way to identify extended emission within the first astronomical units (AU) around nearby main sequence stars (Sect.~\ref{sec:method}). A magnitude-limited sample of 42 stars with spectral types ranging from A to K was built for this specific purpose (Sect.~\ref{sec:method}) and was observed at the CHARA array using the FLUOR beam-combiner (Sect.~\ref{sec:obs}). Some of the early results of this survey have already been presented in previous papers, including the first two of this series: \citet[][hereafter \citetalias{DiFolco07}]{DiFolco07} and \citet[][hereafter \citetalias{Absil08}]{Absil08}. The comparison of our observations with simple stellar models provides a direct estimation of the flux ratio between the circumstellar environment and the photosphere of the star at K band (Sect.~\ref{sec:results}). Several detections are identified, and the results are discussed both on an individual basis (Sect.~\ref{sec:results}) and in a statistical way (Sect.~\ref{sec:stat}), providing the first statistical study of the circumstellar environment of main sequence stars in the near-infrared. Based on the results of the statistical analysis, possible origins of the detected near-infrared excesses are finally discussed in Sect.~\ref{sec:discussion}.

\section{Methodology and stellar sample} \label{sec:method}

The principle of debris disc detection by stellar interferometry is based on the fact that stellar photospheres and circumstellar dust discs have different spatial scales. As discussed previously \citepalias{DiFolco07,Absil08}, a circumstellar emission with uniform surface brightness filling the interferometric field-of-view creates a drop in the measured squared visibility compared to the expected visibility of the star:
\begin{equation}
{\cal V}^2(b) \simeq (1-2f_{\rm CSE}) {\cal V}_{\star}^2(b) \, , \label{eq:visdisk}
\end{equation}
where ${\cal V}^2$ and ${\cal V}_{\star}^2$ are respectively the squared visibility of the star-disc system and of the stellar photosphere, $b$ is the interferometer baseline length, and $f_{\rm CSE} \ll 1$ is the flux ratio between the integrated circumstellar emission within the field-of-view and the stellar photospheric emission. Due to the expected faintness of the circumstellar emission ($f_{\rm CSE} \lesssim 1$\%), detection needs high accuracy both on the measured ${\cal V}^2$ and on the estimation of the squared visibility ${\cal V}_{\star}^2$ of the photosphere. The former condition is met by state-of-the-art near-infrared single-mode interferometers, while the latter can be ensured to a large extent by performing the observations at short baselines, where the star is mostly unresolved. The stellar visibility is then close to one and depends weakly on the photospheric model (including the diameter).

Two main sources of errors are encountered in stellar interferometry: statistical errors, which are related to the detection process, and systematic errors, associated with the calibration of the squared visibilities. The contribution of the former to the overall signal-to-noise ratio decreases as $N^{-0.5}$, with $N$ the number of statistically independent observations of the same star, while the contribution of the latter can be minimised by using several different calibrator stars in the calibration process. Using this strategy, we have already demonstrated in previous studies that the disc/star flux ratio can be measured with sub-percent accuracy.

	\subsection{Stellar sample}

To carry out an exozodiacal disc survey with such an observing method requires the interferometric instrument to be used in a regime where it delivers its highest accuracy. In the case of the FLUOR interferometer at the CHARA array \citep{Coude03}, the highest accuracy can be reached for near-infrared magnitudes up to $K\simeq 4$. Besides this sensitivity limit and the observability from Mount Wilson (${\rm dec} >-16\degr$ for transit at least $40\degr$ above horizon), the other target selection criteria are mostly related to the main goals of the survey. First, we limited our sample to luminosity classes IV and V, and tried to balance our stars between the three spectral categories of interest: A, F, and G-K (taken together). Second, because we are interested in the correlation between hot and cold dust populations, we chose a similar quantity of stars known to harbour outer dust reservoirs and stars without any detected dust within the sensitivity of the \textsc{Spitzer} and \textsc{Herschel} far-infrared space missions. (We also relied on earlier results from the IRAS and/or ISO missions for three stars in our sample.) Here, we denote ``outer reservoir'' to be any dust located beyond a few AU from its host star, detected at mid- to far-infrared wavelengths. In practice, some of the outer dust reservoirs are actually rather warm, with temperatures up to a few hundred Kelvin. These populations are, however, too cold to produce any significant thermal emission at K band. Third, we removed from our sample tight binary stars with separations $<5\arcsec$, which may not only affect our observations by creating excess emission within the field-of-view, but also bias our survey by cleaning up the inner system of dust due to gravitational interactions. For this reason, we removed from our sample the eclipsing binary star alf~CrB, which was observed during an eclipse in \citetalias{Absil08}. Last, we tried to include stars of various ages, ranging between about 100\,Myr for the youngest ones (in order not to include the late phases of planet formation) and several Gyr for the oldest ones. Most of the stars younger than 1\,Gyr in our sample are, however, of spectral type A; only four FGK stars have ages $<1$\,Gyr.

\begin{table*}[p]
\caption{Main parameters of the selected targets: name, HD number, spectral type, distance, V and K magnitudes, rotational velocity, estimated limb-darkened diameter, apparent oblateness, measured limb-darkened diameter (if available), and age. The 1$\sigma$ errors are given in superscript.} \label{tab:sample}
\centering
\begin{tabular}{cccccccccccc}
\hline \hline Name & HD & Type & Dist. & $V$ & $K$ &$v \sin i$ &  Mean\tablefootmark{a} $\theta_{\rm LD}$ & $\rho$ & Meas.\ $\theta_{\rm LD}$ & Age & Refs.\tablefootmark{b}
\\ & & & (pc) & (mag) & (mag) & (km\,s$^{-1}$) & (mas) &  & (mas) & (Gyr) &
\\ \hline
\object{bet Cas}	&	432		&	F2IV	&	16.8	&	2.269	&	$1.41^{0.02}$	&	71	&	$1.996^{0.025}$	&	1.02	&	$2.071^{0.055}$	&	0.9		&	S88,C11,H09 \\
\object{54 Psc}		&	3651	&	K0V		&	11.1	&	5.875	&	$4.00^{0.20}$	&	1.1	&	$0.723^{0.068}$	&	1.00	&	$0.790^{0.027}$	&	6.5		&	K09,B08,G13 \\
\object{eta Cas A}	&	4614	&	G3V		&	6.0 	&	3.450	&	$2.05^{0.02}$	&	2.8	&	$1.633^{0.022}$	&	1.00	&	$1.623^{0.004}$	&	5.4		&	S88,B12,G13 \\
\object{ups And}	&	9826	&	F9V		&	13.5	&	4.093	&	$2.84^{0.01}$	&	9.6	&	$1.106^{0.012}$	&	1.00	&	$1.114^{0.009}$	&	4.0		&	S88,B08,G13 \\
\object{107 Psc}	&	10476	&	K1V		&	7.5 	&	5.235	&	$3.29^{0.03}$	&	1.7	&	$1.014^{0.018}$	&	1.00	&	---				&	5.0		&	A89,---,G13 \\
\object{tau Cet}	&	10700	&	G8V		&	3.7 	&	3.489	&	$1.68^{0.01}$	&	1.3	&	$2.078^{0.025}$	&	1.00	&	$2.015^{0.004}$	&   10		&	S88,D07,D04 \\
\object{tet Per}	&	16895	&	F7V		&	11.1	&	4.099	&	$2.78^{0.09}$	&	8.9	&	$1.150^{0.049}$	&	1.00	&	$1.103^{0.009}$	&	6.0		&	V09,B12,G13 \\
\object{eps Eri}	&	22049	&	K2V		&	3.2 	&	3.721	&	$1.67^{0.01}$	&	2.4	&	$2.177^{0.024}$	&	1.00	&	$2.126^{0.009}$	&	0.9		&	D02,D07,W04 \\
\object{10 Tau}		&	22484	&	F9V		&	14.0	&	4.290	&	$2.90^{0.01}$	&	4.4	&	$1.102^{0.012}$	&	1.00	&	$1.081^{0.014}$	&	6.7		&	S88,B12,G13 \\
\object{1 Ori}		&	30652	&	F6V		&	8.1 	&	3.183	&	$2.08^{0.02}$	&	17	&	$1.530^{0.021}$	&	1.00	&	$1.526^{0.004}$	&	1.2		&	D02,B12,G13 \\
\object{zet Lep}	&	38678	&	A2V 	&	21.6	&	3.536	&	$3.31^{0.02}$	&	229	&	$0.746^{0.010}$	&	1.14	&	$0.670^{0.140}$	&	0.2		&	A91,A09,G13 \\
\object{eta Lep}	&	40136	&	F1V		&	14.9	&	3.707	&	$2.91^{0.02}$	&	17	&	$0.990^{0.013}$	&	1.00	&	$1.022^{0.011}$	&	2.3		&	A91,A13,G13 \\
\object{ksi Gem}	&	48737	&	F5IV	&	18.0	&	3.336	&	$2.13^{0.06}$	&	66	&	$1.539^{0.045}$	&	1.03	&	$1.401^{0.009}$	&	1.7 	&	N69,B12,B12 \\
\object{lam Gem}	&	56537	&	A3V		&	31.0	&	3.572	&	$3.38^{0.20}$	&	154	&	$0.651^{0.081}$	&	1.07	&	$0.835^{0.013}$	&	0.5		&	R05,B12,G13 \\
\object{HD 69830}	&	69830	&	K0V		&	12.5	&	5.945	&	$4.16^{0.04}$	&	0.3	&	$0.659^{0.012}$	&	1.00	&	---				&	6.0		&	K09,---,G13 \\
\object{30 Mon}		&	71155	&	A0V		&	37.6	&	3.881	&	$3.93^{0.01}$	&	134	&	$0.534^{0.006}$	&	1.04	&	---				&	0.2		&	M94,---,R05 \\
\object{bet UMa}	&	95418	&	A1V		&	24.5	&	2.341	&	$2.38^{0.06}$	&	46	&	$1.093^{0.032}$	&	1.01	&	$1.149^{0.014}$	&	0.3		&	N69,B12,G13 \\
\object{del Leo}	&	97603	&	A4V		&	17.9	&	2.549	&	$2.24^{0.06}$	&	180	&	$1.238^{0.036}$	&	1.11	&	$1.328^{0.009}$	&	0.7		&	N69,B12,G13 \\
\object{bet Leo}	&	102647	&	A3V		&	11.0	&	2.121	&	$1.88^{0.19}$	&	128	&	$1.442^{0.130}$	&	1.04	&	$1.342^{0.013}$	&	0.1		&	K09,A09,G13 \\
\object{bet Vir}	&	102870	&	F9V		&	10.9	&	3.589	&	$2.31^{0.02}$	&	4.0	&	$1.418^{0.020}$	&	1.00	&	$1.431^{0.006}$	&	4.4		&	A91,B12,G13 \\
\object{del UMa}	&	106591	&	A3V		&	24.7	&	3.295	&	$3.10^{0.02}$	&	233	&	$0.817^{0.011}$	&	1.16	&	---				&	0.5		&	A91,---,G13 \\
\object{eta Crv}	&	109085	&	F2V		&	18.3	&	4.302	&	$3.37^{0.30}$	&	92	&	$0.819^{0.119}$	&	1.03	&	---				&	2.4		&	K09,---,G13 \\
\object{70 Vir}		&	117176	&	G5V		&	18.0	&	4.966	&	$3.25^{0.01}$	&	2.7	&	$0.992^{0.011}$	&	1.00	&	$1.009^{0.024}$	&	5.4		&	S88,B08,W04 \\
\object{iot Vir}	&	124850	&	F7V		&	22.3	&	4.068	&	$2.79^{0.04}$	&	4.4	&	$1.137^{0.024}$	&	1.00	&	---				&	1.8		&	S88,---,B96 \\
\object{sig Boo}	&	128167	&	F3V		&	15.9	&	4.467	&	$3.49^{0.06}$	&	9.3	&	$0.782^{0.023}$	&	1.00	&	$0.841^{0.013}$	&	1.7		&	B79,B12,G13 \\
\object{ksi Boo}	&	131156	&	G8V		&	6.7 	&	4.663	&	$2.96^{0.03}$	&	4.6	&	$1.132^{0.019}$	&	1.00	&	$1.196^{0.014}$	&	0.3		&	A89,B12,G13 \\
\object{lam Ser}	&	141004	&	G0V		&	12.1	&	4.413	&	$3.02^{0.02}$	&	3.1	&	$1.057^{0.018}$	&	1.00	&	$0.838^{0.120}$	&	5.3		&	A89,V09,G13 \\
\object{kap CrB}	&	142091	&	K1IV	&	30.6	&	4.802	&	$2.49^{0.02}$	&	1.5	&	$1.562^{0.021}$	&	1.00	&	---				&	2.5		&	S88,---,J08 \\
\object{chi Her}	&	142373	&	F9V		&	15.9	&	4.605	&	$3.12^{0.01}$	&	2.8	&	$1.012^{0.011}$	&	1.00	&	---				&	6.2		&	S88,---,G13 \\
\object{gam Ser}	&	142860	&	F6V		&	11.3	&	3.828	&	$2.62^{0.02}$	&	11	&	$1.215^{0.018}$	&	1.00	&	$1.217^{0.005}$	&	4.6		&	A91,B12,G13 \\
\object{mu Her}		&	161797	&	G5IV	&	8.3 	&	3.413	&	$1.74^{0.016}$	&	2.7	&	$1.975^{0.025}$	&	1.00	&	$1.953^{0.039}$	&	8.3		&	S88,M03,W04 \\
\object{gam Oph}	&	161868	&	A0V		&	31.6	&	3.740	&	$3.67^{0.04}$	&	210	&	$0.615^{0.013}$	&	1.12	&	---				&	0.3		&	S88,---,R05\\
\object{70 Oph A}	&	165341	&	K0V		&	5.1 	&	4.085	&	$2.29^{0.04}$	&	4.6	&	$1.566^{0.018}$	&	1.00	&	---				&	1.1		&	C06,---,M09 \\
\object{alf Lyr}	&	172167	&	A0V		&	7.7 	&	0.030	&	$0.00^{0.01}$	&	22	&	$3.309^{0.037}$	&	1.00	&	$3.305^{0.010}$	&	0.7		&	D02,A08,M12 \\
\object{110 Her}	&	173667	&	F6V		&	19.2	&	4.202	&	$3.06^{0.01}$	&	17	&	$0.981^{0.011}$	&	1.00	&	$1.000^{0.006}$	&	3.4		&	S88,B12,W04 \\
\object{zet Aql}	&	177724	&	A0V 	&	25.5	&	2.980	&	$2.94^{0.08}$	&	317	&	$0.856^{0.033}$	&	1.30	&	$0.895^{0.017}$	&	0.8 	&	N69,B12,B12 \\
\object{sig Dra}	&	185144	&	K0V		&	5.8 	&	4.664	&	$2.83^{0.08}$	&	1.4	&	$1.229^{0.047}$	&	1.00	&	$1.254^{0.012}$	&	3.2		&	N69,B12,W04 \\
\object{alf Aql}	&	187642	&	A7V		&	5.1 	&	0.770	&	$0.24^{0.03}$	&	240	&	$3.232^{0.057}$	&	1.14	&	$3.305^{0.015}$	&	1.3		&	D02,M07,D05 \\
\object{61 Cyg A}	&	201091	&	K5V		&	3.5 	&	5.195	&	$2.25^{0.32}$	&	4.7	&	$1.950^{0.299}$	&	1.00	&	$1.775^{0.013}$	&	6.0		&	K09,K08,K08 \\
\object{61 Cyg B}	&	201092	&	K7V		&	3.5		&	6.013	&	$2.54^{0.33}$	&	0	&	$1.863^{0.296}$	&	1.00	&	$1.581^{0.022}$	&	6.0		&	K09,K08,K08 \\
\object{alf Cep}	&	203280	&	A7IV	&	15.0	&	2.456	&	$1.85^{0.06}$	&	225	&	$1.560^{0.045}$	&	1.19	&	$1.561^{0.027}$	&	0.8		&	N69,Z09,R05\\
\object{eps Cep}	&	211336	&	F0IV	&	26.2	&	4.177	&	$3.54^{0.31}$	&	91	&	$0.721^{0.108}$	&	1.03	&	---				&	0.6		&	K09,---,R07 \\
\hline
\end{tabular}
\tablefoot{The V magnitudes are taken from Simbad or \citet{Kharchenko09}, and the $v \sin i$ from the catalogues of \citet{Royer07}, \citet{Valenti05}, \citet{Reiners06} and \citet{MartinezArnaiz10}, except in the cases of kap CrB \citep[from][]{Johnson08}, alf Lyr \citep[from][]{Monnier12}, alf Aql \citep[from][]{Monnier07} and alf Cep \citep[from][]{Zhao09}.
\tablefoottext{a}{Geometric mean of the minor and major axes of the projected elliptical photosphere.}
\tablefoottext{b}{References are given for the K-band magnitude, the measured diameter and the age, using the notations here below.}}
\tablebib{(A89) \citet{Arribas89}; (A91) \citet{Aumann91};  (A08) \citet{Absil08}; (A09) \citet{Akeson09}; (A13) this work; (B79) \citet{Blackwell79}; (B96) \citet{Baliunas96}; (B08) \citet{Baines08}; (B12) \citet{Boyajian12a}; (C06) \citet{Christou06}; (C11) \citet{Che11}; (D02) \citet{Ducati02}; (D05) \citet{Domiciano05}; (D04) \citet{DiFolco04};  (D07) \citet{DiFolco07}; (G13) \citet{Gaspar13}; (H09) \citet{Holmberg09}; (J08) \citet{Johnson08}; (K08) \citet{Kervella08}; (K09) \citet{Kharchenko09}; (M94) \citet{McGregor94}; (M03) \citet{Mozurkewich03}; (M07) \citet{Monnier07}; (M12) \citet{Monnier12}; (N69) \citet{Neugebauer69}; (R07) \citet{Rhee07}; (R05) \citet{Rieke05}; (S88) \citet{Selby88}; (V05) \citet{Valenti05}; (V09) \citet{vanBelle09}; (W04) \citet{Wright04}; (Z09) \citet{Zhao09}.}
\end{table*}

\begin{table}
\caption{Distribution of our star sample in terms of spectral type and presence or not of outer dust reservoirs.}
\label{tab:split}
\centering
\begin{tabular}{c c c c c}
\hline\hline
 & A & F & G-K & Total\\
\hline
Outer reservoir & 7 & 7 & 5 & 19 \\
No outer reservoir & 5 & 8 & 10 & 23 \\
\hline
Total & 12 & 15 & 15 & 42 \\
\hline
\end{tabular}
\end{table}

Based on the data available to us at this time (April 2013), including early information from the \textsc{Herschel}/DUNES and DEBRIS surveys \citep{Eiroa10,Eiroa13,Matthews10}, a total of 28 main sequence stars with outer dust reservoirs match our selection criteria, among which 11 A-type, 11 F-type, and 6 GK-type stars. Our final sample could therefore contain up to 56 stars, by adding a similar number of stars without outer reservoir (which are much more abundant). Due to the limited available observing time and to the fact that some of the 28 stars with outer reservoirs were not classified as such when our observations were carried out, our final sample, described in Table~\ref{tab:sample}, comprises 42 stars. Their distribution between different statistical categories (spectral type, presence of outer dust reservoir) is given in Table~\ref{tab:split}. The belonging to the ``outer reservoir'' category (also referred to as the ``dusty'' category hereafter for the sake of conciseness) is based on the literature, as described in Table~\ref{tab:excess}, where mid- to far-infrared excess (non-)detections are listed for all target stars. Some stars were originally observed in our survey as part of the dusty category, but finally turned out to have no detectable outer reservoir based on new far-infrared observations (due e.g.\ to source confusion in IRAS or ISO data), explaining in part the imbalance between the dusty and non-dusty samples.

As far as dusty targets are concerned, we consider our final sample to be representative of the magnitude-limited 28-star sample defined above, since more than half of the targets with outer dust reservoir were observed in each of the A, F, and GK spectral type categories. However, we cannot make certain that this is also the case for non-dusty targets: if we consider in our non-dusty master sample all those stars shown by the \textsc{Spitzer} and \textsc{Herschel} missions not to have any detectable amount of outer dust, only about 25\% (A and F stars) to 40\% (GK stars) of the available stars were observed within this survey. For instance, it could turn out that, by (lack of) chance, the five non-dusty A-type stars that we have observed have specific characteristics that are not representative of the $\sim 20$ non-dusty A-type stars matching our selection criteria. Although we tried to avoid such a bias as much as possible, this is a limitation to the present survey directly related to the small number of targets in the observed sample, which can only be addressed by extending the sample. This can be done in two different ways, which are both being pursued: (i) fainter stars in the northern hemisphere will be made accessible thanks to an ongoing major refurbishing of the FLUOR instrument, and (ii) a hundred more stars are being observed in the southern hemisphere with the PIONIER instrument at VLTI \citep{LeBouquin11}. In the present paper, we only consider targets observed in a homogeneous way with one single instrument (CHARA/FLUOR) whose characteristics have not changed during the whole survey, and purposely discard other recent studies conducted with other near-infrared interferometers, such as for Fomalhaut \citep{Absil09} and bet~Pic \citep{Defrere12}.

\begin{table}[!t]
\caption{Presence of a mid- to far-infrared excess around the target stars.}
\label{tab:excess}
\centering
\begin{tabular}{c c c c}
\hline\hline
Name & $10^6 \times L_{\rm d}/L_{\star}$\tablefootmark{a} & Excess\tablefootmark{b} & Refs. \\
\hline
bet	Cas	& $25$ & far  & R07 \\ 
54	Psc	& $<1.7$ & ---  & E13 \\
eta	Cas	A	& $<1.9$ & --- & E13 \\
ups	And	& $<0.53$ & --- & B06,E13 \\
107	Psc	& $<5.5$ & --- & C06,T08,M13 \\ 
tau	Cet	& $25$ & far & H01 \\
tet	Per	& $<0.61$ & --- & L09,E13 \\
eps	Eri	& $110$ & mid,far & B09 \\
10	Tau	& $12$ & far & T08,M13 \\ 
1	Ori	& $<0.81$ & --- & T08,M13 \\ 
zet	Lep	& $97$ & mid,far & C06,S06,T13 \\ 
eta	Lep	& $63$ & mid,far & L09,E13 \\
ksi	Gem	& --- & --- & MIPS24\tablefootmark{c} \\
lam	Gem	& $<0.47$ & --- & M09,T13 \\
HD	69830	& $200$ & mid & B05,E13 \\
30	Mon	& $89$ & mid,far & C06,S06,T13 \\ 
bet	UMa	& $14$ &  mid,far & C06,M10 \\
del	Leo	& $<0.20$ & --- & T13 \\
bet	Leo	& $30$ & mid,far & C11 \\
bet	Vir	& $0.36$ & far & T08,M13 \\ 
del	UMa	& $4.9$ & mid & S06,T13 \\ 
eta	Crv	& $360$ & mid,far & C06,M10 \\
70	Vir	& $10$ & mid,far & D11,E13 \\
iot	Vir	& $<2.3$ & --- & MIPS24-70\tablefootmark{c} \\ 
sig	Boo	& $4.9$ & far & C06,R07,M13 \\ 
ksi	Boo	& $<1.2$ & --- & M13 \\ 
lam	Ser	& $<0.95$ & --- & M13 \\ 
kap	CrB	& $50$ & far & B13 \\
chi	Her	& $<2.8$ & --- & B06,T08,M13 \\ 
gam	Ser	& $<4.8$ & --- & L09,T08,M13 \\ 
mu	Her	& $<25$ & --- & H01,R07 \\
gam	Oph	& $90$ & mid,far & S08 \\
70	Oph	A	& $<0.45$ & --- & E13 \\
alf	Lyr & $23$ & mid,far & S06,S13 \\
110	Her	& $1.8$ & far & B06,E13 \\ 
zet	Aql	& $<1.6$ & --- & P09 \\
sig	Dra	& $<0.89$ & --- & B06,E13 \\
alf	Aql & --- & --- & T13 \\
61	Cyg	A	& $<1.2$ & --- & E13 \\
61	Cyg	B	& $<1.8$ & --- & E13 \\
alf	Cep	& $<2.1$ & --- & C05 \\
eps	Cep	& --- & --- & MIPS70\tablefootmark{c,d} \\ 
\hline
\end{tabular}
\tablefoot{\tablefoottext{a}{Fractional luminosity of the cold debris discs, or $3\sigma$ (far-infrared) upper limit when no dust is detected.} 
\tablefoottext{b}{Wavelength range at which the excess is detected (mid: $10-35\,\mu$m, far: $>60\,\mu$m).}
\tablefoottext{c}{No excess in \textsc{Spitzer}/MIPS 24 and/or 70\,$\mu$m photometry (G.~Bryden, pers.\ comm.).}
\tablefoottext{d}{The IRAS excess reported by \citet{Rhee07} is due to a nearby background galaxy (G.~Bryden, pers.\ comm.).}}
\tablebib{(B05) \citet{Beichman05}; (B06)~\citet{Beichman06}; (B09)~\citet{Backman09}; (B13)~\citet{Bonsor13a}; (C06)~\citet{Chen06}; (C11)~\citet{Churcher11}; (D11)~\citet{Dodson11}; (E13)~\citet{Eiroa13}; (H01)~\citet{Habing01}; (L09)~\citet{Lawler09}; (M09)~\citet{Morales09}; (M10)~\citet{Matthews10}; (M13)~B.~Matthews et al.\ (in prep); (P09)~\citet{Plavchan09}; (R07)~\citet{Rhee07}; (S06)~\citet{Su06}; (S08)~\citet{Su08}; (S13)~\citet{Su13}; (T08) \citet{Trilling08}; (T13)~N.~Thureau et al.\ (in prep.).}
\end{table}

	\subsection{Stellar diameters and photospheric model}

Our photospheric model, consisting in an oblate limb-darkened disc, was thoroughly described in \citetalias{Absil08}. For the present study, limb-darkened diameters were estimated for all our target stars using surface-brightness relationships \citep[SBR,][]{Kervella04}, which provide a sufficient accuracy for our purpose \citepalias[see][]{Absil08}. Robust diameter estimation with these relationships requires using accurate visible and near-infrared magnitudes. Because all our stars are saturated in the 2MASS catalogue, we had to rely on other published near-infrared photometric studies to collect accurate K-band magnitudes (see Table~\ref{tab:sample}). Our SBR diameter estimations were checked against diameter measurements from long-baseline interferometry, where available, generally showing consistency within $3\sigma$ at most. Only del Leo shows a significant discrepancy ($>3\sigma$) between the estimated and measured diameters. We note, however, that for this particular star, two different interferometric studies do not give consistent results either: a limb-darkened diameter $\theta_{\rm LD}=1.165 \pm 0.022$~mas is found by \citet{Akeson09} based on CHARA/FLUOR data, while \citet{Boyajian12a} derived a diameter of $\theta_{\rm LD}=1.328 \pm 0.009$~mas using CHARA/Classic. Our SBR diameter falls between these two measurements and will be used in the following analysis. In the other cases where a measured diameter is available, it is preferred to the SBR diameter for predicting the stellar visibilities. We note that using SBR diameters instead of measured ones would not significantly change the estimated disc/star flux ratios in our analysis.

We also derived the apparent oblateness $\rho=\theta_{\rm max}/\theta_{\rm min}$ of all the photospheres based on the rotational velocities $v \sin i$ of the target stars and on the geometrical model presented in \citetalias{Absil08}. The diameters given in Table~\ref{tab:sample} are the geometric mean of the maximum and minimum projected limb-darkened diameters, that is, the diameter of the equivalent spherical photosphere. In the following analysis, the actual projected shape of the photosphere is taken into account. In particular, in the (most frequent) cases where the position angle of the photosphere is not known, we take the full range of possible projected diameters along the observing baselines into account and fold it into the final error bars. The additional error is typically a few percent of the mean stellar diameter, and reaches 14\% for the fastest rotator. Even in that case, the additional error remains modest compared to the statistical error on our measurements, since the star itself is mostly unresolved.

	\subsection{Exozodiacal disc model}

The model that we have used to derive Eq.~\ref{eq:visdisk} consists in a star surrounded by a circumstellar emission with uniform surface brightness filling the whole field-of-view. (This model will be referred to as the ``star-disc'' model in the rest of the paper.) This model is probably not representative of exozodiacal discs in general, because it is unlikely that the circumstellar emission remains uniform on the FLUOR field-of-view, about $0\farcs8$ in diameter (see Sect.~\ref{sub:FLUOR}). However, Eq.~\ref{eq:visdisk} remains valid as long as the visibility associated to the circumstellar emission is equal (or very close) to 0; that is, the circumstellar emission is fully resolved at the considered baseline and observing wavelength. If the circumstellar emission was not fully resolved, it would reduce the visibility drop associated to a given excess emission level, thereby reducing the sensitivity of our observations to circumstellar emission. Furthermore, it would create modulations in the visibility curve as a function of spatial frequency, as illustrated in Fig.~9 of \citet{Absil09} and in Fig.~4 of \citet{Defrere11}.

\begin{figure}[t]
\centering
\includegraphics[width=\linewidth]{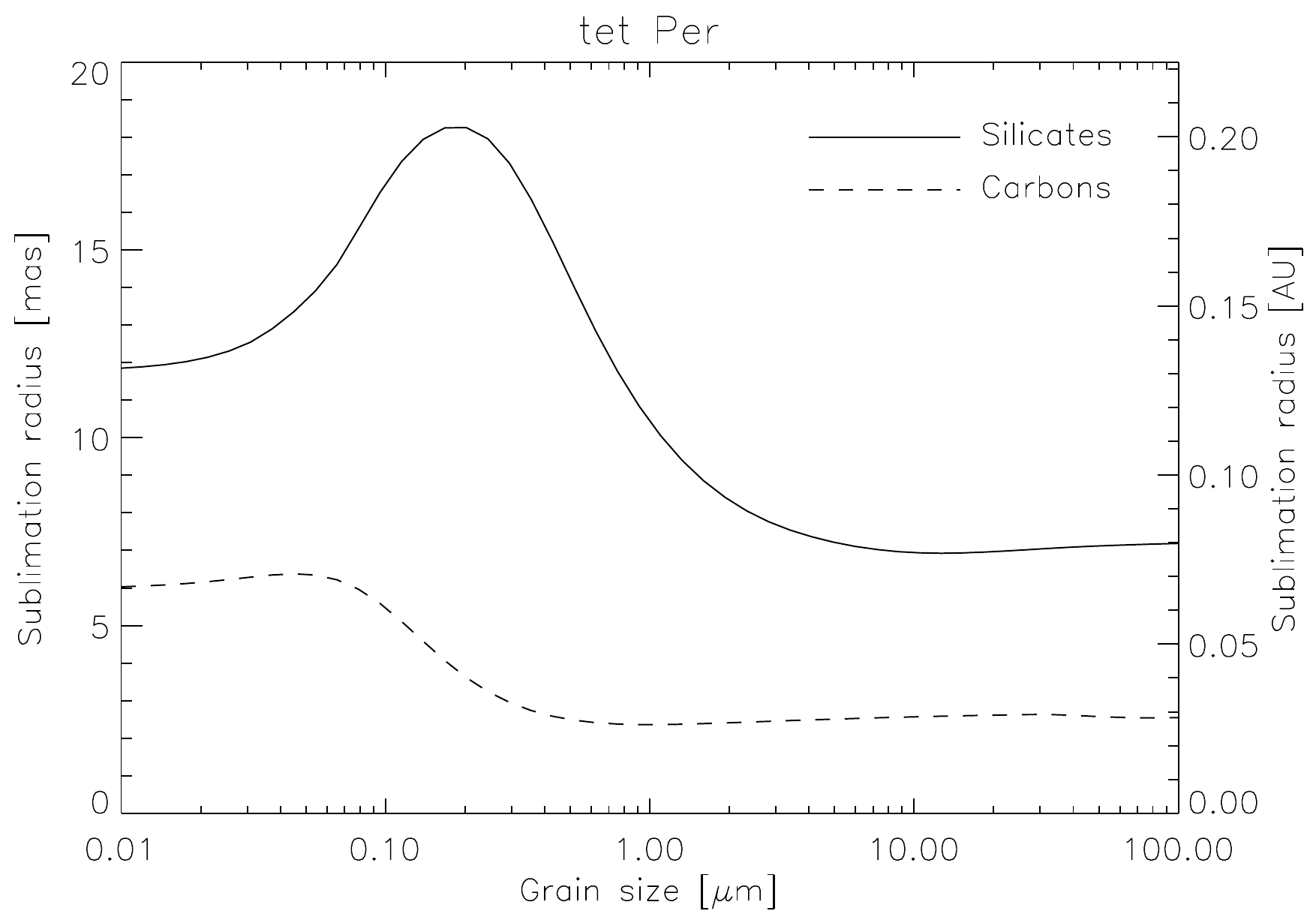}
\caption{Sublimation distance of silicate and carbonaceous dust grains as a function of grain size in the tet Per environment.} \label{fig:subradius}
\end{figure}

To check whether exozodiacal discs are indeed expected to be fully resolved around our target stars, we computed the sublimation distance associated with silicate and carbonaceous material for various grain sizes using the GRaTeR radiative transfer code \citep{Augereau99} under the assumption of thermal equilibrium of the grains with the star and constant values of the sublimation temperature (1200\,K for silicate and 2000\,K for carbonaceous grains). This is illustrated in Fig.~\ref{fig:subradius} for a representative target in our sample in terms of magnitude and spectral type: tet Per (F7V, $K=2.8$). The sublimation diameter, i.e., twice the sublimation radius given in milliarcsec in Fig.~\ref{fig:subradius}, must be compared to the angular resolution of our instrument: $\lambda/2B \simeq 7$\,mas for a 34-m baseline and an effective observing wavelength of 2.13\,$\mu$m. Figure~\ref{fig:subradius} shows that if all the circumstellar emission was located at the sublimation radius of micron-size carbonaceous grains, its diameter would only be about 5\,mas and would therefore not be fully resolved. However, all the models that we have developed so far to reproduce resolved near-infrared detections of exozodiacal dust \citep{Defrere11,Mennesson13,Lebreton13} have shown that most of the disc's thermal emission is produced at the sublimation distance of sub-micron grains, i.e., a region about 12\,mas in diameter in the case of tet Per. We therefore expect that any dusty disc around tet Per would be fully resolved with FLUOR on the shortest baseline of the CHARA array.

The sublimation distance of silicate and carbonaceous grains was computed for all the stars in our sample. We checked that the assumption of a fully resolved exozodiacal disc is valid for most of our targets. In practice, only five stars could have a dominant emission zone significantly smaller than the angular resolution of our instrument: 54 Psc, 107 Psc, HD 69830, 61 Cyg A, and 61 Cyg B. These are five of the eight K-type stars in our sample. For these five stars, we estimate that, in the worst cases, we could lose up to a factor of 4 in sensitivity to exozodiacal dust. Because the sensitivity loss is strongly model-dependent, it cannot be properly taken into account when fitting the data with our star-disc model. Since it only affects five stars in our sample, we do not consider this effect as a major limitation in our analysis, but we note that we might slightly underestimate the occurrence rate of percent-level K-band excesses around late-type stars.

\section{Observations and data reduction} \label{sec:obs}

	\subsection{Observing campaigns and strategy} \label{sub:FLUOR}

Interferometric observations were obtained in the infrared K band ($1.94 - 2.34$~$\mu$m) with FLUOR, the Fiber Linked Unit for Optical Recombination \citep{Coude03}, using the shortest baseline of the CHARA Array formed by the S1 and S2 telescopes \citep[34\,m long,][]{tenBrummelaar05}. A total of 98 observing nights were allocated to this survey between 2005 and 2011, with observing campaigns typically of eight nights carried out in May and November 2005, April and October 2006, December 2007, May and October 2008, May and November 2009, May and November 2010, and June 2011. Each observing block (OB) consists in the acquisition of typically 150 scans of the interference fringes using a scanning mirror in one arm of the interferometer. The amplitude of the optical path difference modulation within a scan was  128~$\mu$m, with 300 frames recorded on the PICNIC detector during each scan, resulting in a sampling of five frames per fringe. The detector was read in a destructive mode, using a reset-read-read sequence. While less sensitive than the non-destructive read-out mode, this mode has a higher dynamic range and has been heavily tested in the past for high-accuracy applications.

Over the seven years of the survey, a total of about 500 OBs were obtained on our scientific targets, i.e., about 12 OBs per target in average. Each scientific OB was bracketed with calibration OBs, using calibrator stars from the catalogues of \citet{Borde02} and \citet{Merand05}. These calibrators are generally K-type giants with well-known diameters and no known companion. To reduce calibration-related biases as much as possible, we used at least three different calibrators for each scientific target. In practice, our observing strategy generally consisted in a sequence of three to four bracketed OBs (e.g., CAL1-SCI-CAL2-SCI-CAL3-SCI-CAL4) on a given scientific target before moving to another one. Obtaining a dozen bracketed OBs on a given scientific target generally requires the equivalent of one full observing night. The ratio between the number of targets observed (43 including alf CrB) and the total number of observing nights (98) is representative of our observing efficiency ($\sim 44\%$), related to weather and technical down time.



The FLUOR field-of-view, limited by the use of single-mode fibres, has a Gaussian shape resulting from the overlap integral of the incoming turbulent wave fronts with the fundamental mode of the fibre. Its size depends on the atmospheric turbulence conditions. The estimation of the mean Fried parameter ($r_0$) provided by the CHARA Array during our observations ($\sim$7~cm in the visible) allows us to derive a mean value of $0\farcs8$ for the full width at half maximum (FWHM) of the field-of-view. This parameter is sensitive to the actual atmospheric conditions and typically ranges between $0\farcs7$ and $1\farcs2$ for our observations. Because the large majority of our targets were observed during different nights under different seeing conditions, we will assume a field-of-view of $0\farcs8$ (FWHM) for all of them.

	\subsection{Data reduction}

The FLUOR Data Reduction Software \citep[DRS,][]{Coude97,Kervella04,Merand06} was used to extract the raw squared modulus of the coherence factor between the two independent apertures. The extraction of the squared visibilities from the fringe packets recorded in the time domain is based on integrating of the squared fringe peak obtained by a Fourier transform of the fringe packet \citep{Coude97}, or equivalently by a wavelet analysis \citep{Kervella04}. The interferometric transfer function of the instrument was estimated by observing calibrator stars before and after each observation of a scientific target. Calibrators chosen in this study are late G or K giants, whereas our target stars have spectral types between A0 and K7. Since the visibility estimator implemented in the FLUOR DRS depends on the actual spectrum of the target star, an appropriate correction must be applied to our data, otherwise our squared visibilities would be biased at a level of about 0.3\% \citep{Coude97}. This correction can be based either on the shape factors discussed by \citet{Coude97} or on a wide band model for estimating the calibrator's visibilities and interpreting the data \citep[see e.g.][]{Kervella03,Aufdenberg06}. The latter method was chosen for this work, and all the calculations presented here take a full model of the FLUOR instrument into account, including the spectral bandwidth effects.

A significant fraction of the collected OBs had to be discarded during the data reduction process, for various reasons related to weather or technical issues. In particular, we discarded OBs with too low signal-to-noise ratio in the fringes (generally due to poor seeing conditions producing a low coupling efficiency into the fibres), OBs affected by strong mechanical vibrations or unusual electronic noise, and OBs with large unbalance in the flux levels between the two arms of the interferometer. In a few cases, we also needed to discard sequences of OBs showing strong variability in the interferometric transfer function. This process resulted in selecting about 300 high-quality OBs out of the 500 original OBs for the 42 survey targets, i.e., an average of seven OBs per scientific target. The final number of OBs per star varies significantly from one star to the next, though, not only because of the number of times each star could be observed, but also because of the data selection process (some stars losing up to 75\% of their OBs in the process, while other ones were not affected). The same selection criteria were applied to calibration OBs, resulting in a similar discard rate.

\section{Results} \label{sec:results}

\begin{table}[!t]
\caption{Estimated K-band disc/star flux ratio ($f_{\rm CSE}$) with its error bar $\sigma_f$, reduced chi square of the fit ($\chi^2_r$), and significance of the measured flux ratio ($\chi_f=f_{\rm CSE}/\sigma_f$). The last column indicates whether a significant K-band circumstellar emission is detected, based on these data.}
\label{tab:result}
\centering
\begin{tabular}{c c c c c c}
\hline\hline
Name & $f_{\rm CSE}$ (\%) & $\sigma_f$ (\%) & $\chi^2_r$ & $\chi_f$ & Excess? \\
\hline
bet Cas & 0.07 & 0.30 & 0.90 & 0.2 & NO \\
54 Psc & 0.63 & 0.42 & 3.13 & 1.5 & NO \\
eta Cas A&	0.33 & 0.13 & 0.11 & 2.6 & NO \\
ups And & 0.53 & 0.17 & 2.62 & 3.0 & NO\tablefootmark{c} \\
107 Psc & 0.75 & 0.57 & 2.07 & 1.3 & NO \\
tau Cet\tablefootmark{a} & 0.98 & 0.18 & 0.83 & 5.4 & YES \\
tet Per & 0.44 & 0.27 & 1.82 & 1.6 & NO \\
eps Eri\tablefootmark{a} & -0.10 & 0.20 & 2.44  & -0.5&	NO \\
10 Tau & 1.21 & 0.11 & 1.76  & 11.0&	YES \\
1 Ori & 0.44 & 0.23 & 2.92 & 1.9 & NO \\
zet Lep & 0.55 & 0.26 & 1.33 & 2.1 & NO \\
eta Lep & 0.89 & 0.21 & 2.20 & 4.3 & YES \\
ksi Gem & -1.36 & 0.69 & 1.41  & -2.0&	NO \\
lam Gem & 0.74 & 0.17 & 2.35 & 4.3 & YES \\
HD 69830&	-0.23 & 0.45 & 0.13  & -0.5&	NO \\
30 Mon & 0.04 & 0.45 & 0.69 & 0.1 & NO \\
bet UMa\tablefootmark{b} & -0.05 & 0.16 & 0.40 & -0.3&	NO \\
del Leo & -1.14 & 0.77 & 0.53 & -1.5&	NO \\
bet Leo & 0.94 & 0.26 & 5.50 & 3.6 & YES \\
bet Vir & 0.06 & 0.33 & 1.84 & 0.2 & NO \\
del UMa & 0.37 & 0.37 & 2.11 & 1.0&	NO \\
eta Crv\tablefootmark{b} & 0.37 & 0.54 & 1.19 & 0.7 & NO \\
70 Vir & 0.12 & 0.27 & 0.59 & 0.5 & NO \\
iot Vir & -0.75 & 0.25 & 1.20 & -3.0&	NO \\
sig Boo\tablefootmark{b} & 0.40 & 0.45 & 1.80 & 0.9 & NO \\
ksi Boo & 0.74 & 0.20 & 0.21 & 3.7 & YES \\
lam Ser & 0.55 & 0.35 & 2.25 & 1.6 & NO \\
kap CrB & 1.18 & 0.20 & 1.16 & 5.9 & YES \\
chi Her & 0.58 & 0.65 & 1.58 & 0.9 & NO \\
gam Ser & -0.06 & 0.27 & 0.11 & -0.2&	NO \\
mu Her & 1.02 & 0.33 & 2.58 & 3.1 & NO\tablefootmark{c} \\
gam Oph\tablefootmark{b} & 0.25 & 0.48 & 1.23 & 0.5 & NO \\
70 Oph A&	0.31 & 0.36 & 2.40 & 0.9 & NO \\
alf Lyr\tablefootmark{b} & 1.26 & 0.27 & 2.11 & 4.7 & YES \\
110 Her & 0.94 & 0.25 & 0.35 & 3.8 & YES \\
zet Aql\tablefootmark{b} & 1.69 & 0.27 & 0.97 & 6.3 & YES \\
sig Dra & 0.15 & 0.17 & 1.55 & 0.9 & NO \\
alf Aql & 3.07 & 0.24 & 1.75 & 12.9&	YES \\
61 Cyg A&	0.13 & 0.55 & 0.49 & 0.2 & NO \\
61 Cyg B&	-0.36 & 0.36 & 0.94 & -1.0&	NO \\
alf Cep & 0.87 & 0.18 & 1.79 & 4.7 & YES \\
eps Cep & 3.25 & 0.69 & 13.77 & 4.7 & YES \\
\hline
\end{tabular}
\tablefoot{\tablefoottext{a}{From \citetalias{DiFolco07}.}
\tablefoottext{b}{From \citetalias{Absil08}.} 
\tablefoottext{c}{See discussion in Sect.~\ref{sec:individuals}.}}
\end{table}

\begin{figure*}[t]
\centering
\begin{tabular}{cc}
\includegraphics[width=8.3cm]{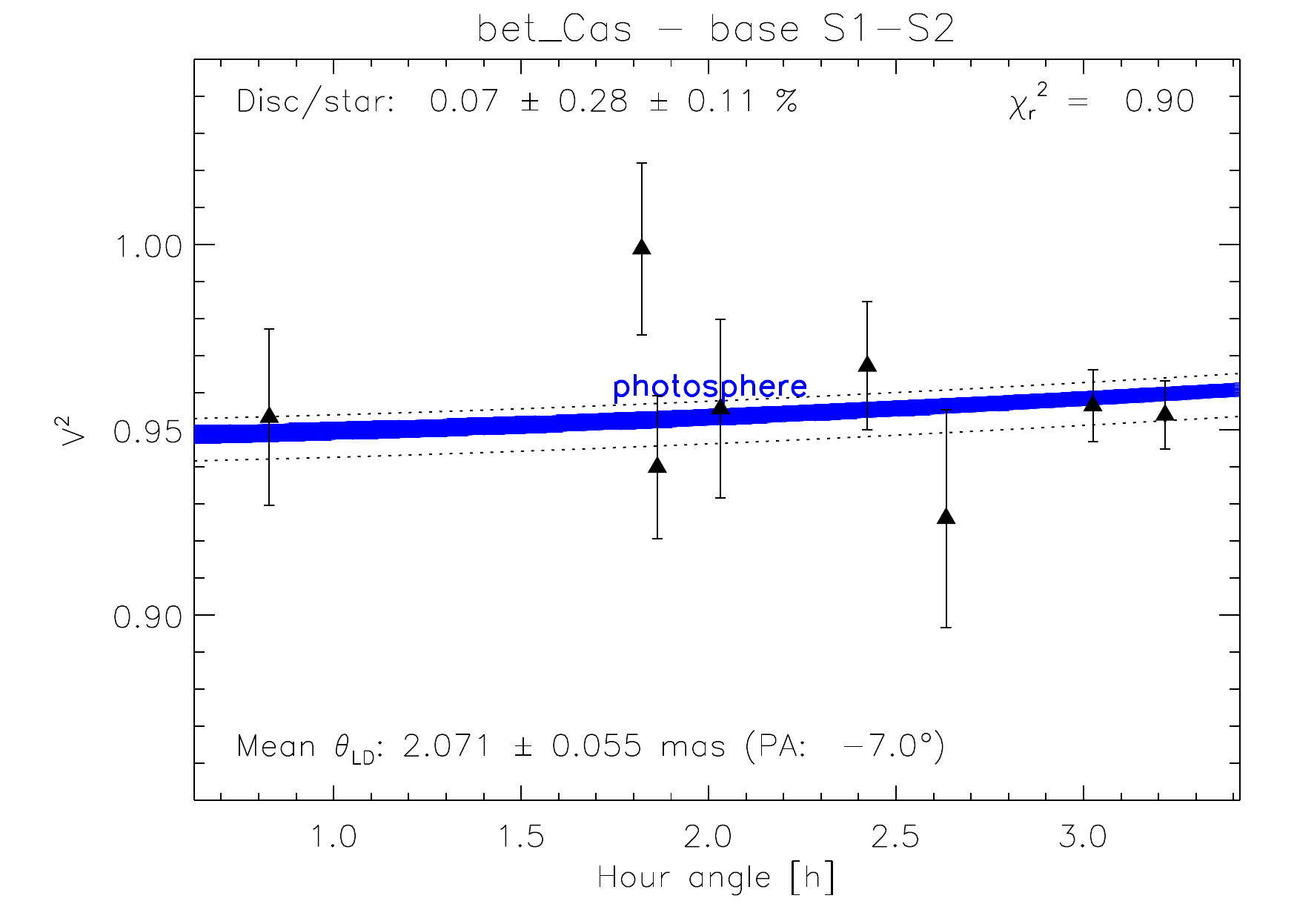} & \includegraphics[width=8.3cm]{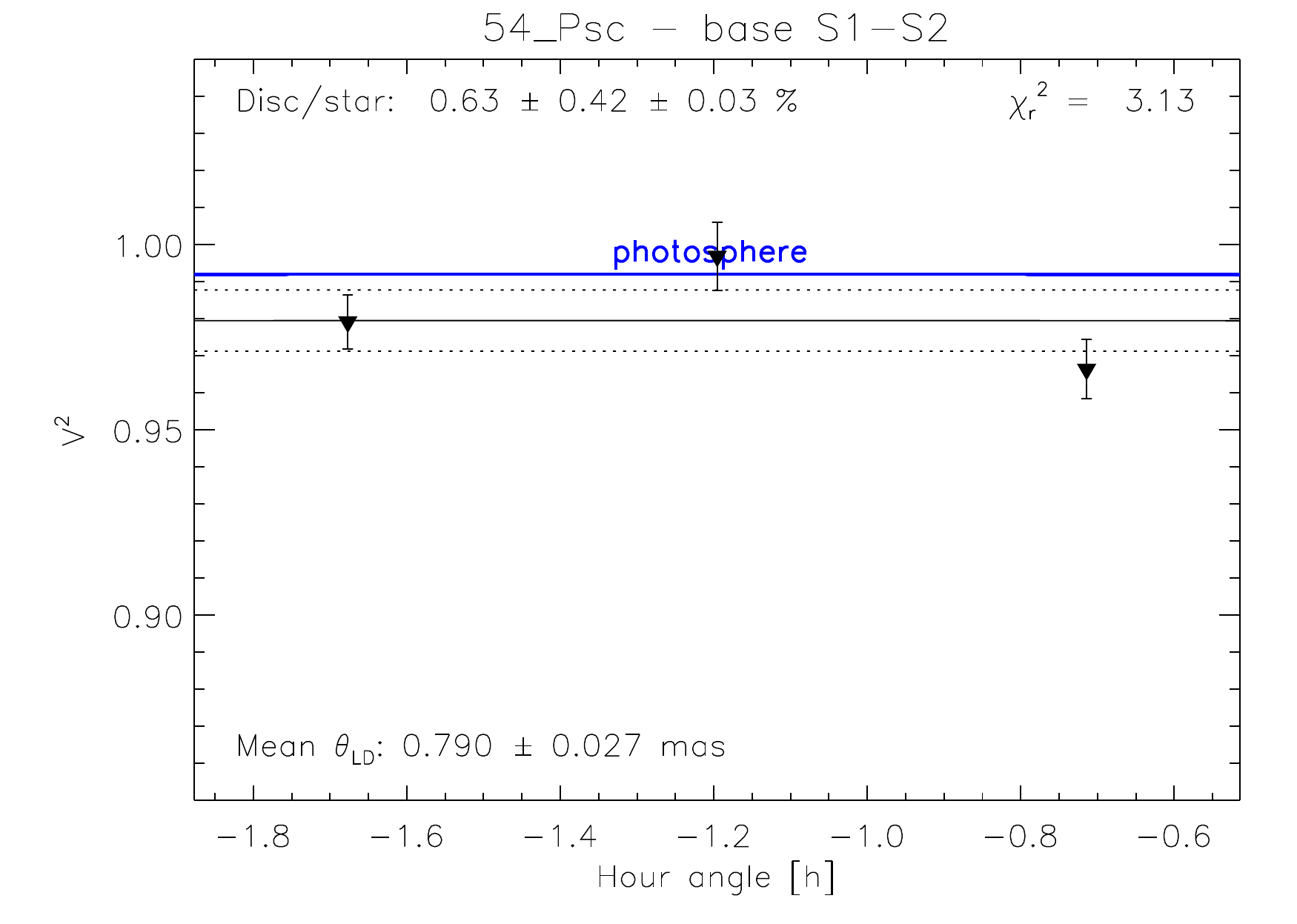} \\
\includegraphics[width=8.3cm]{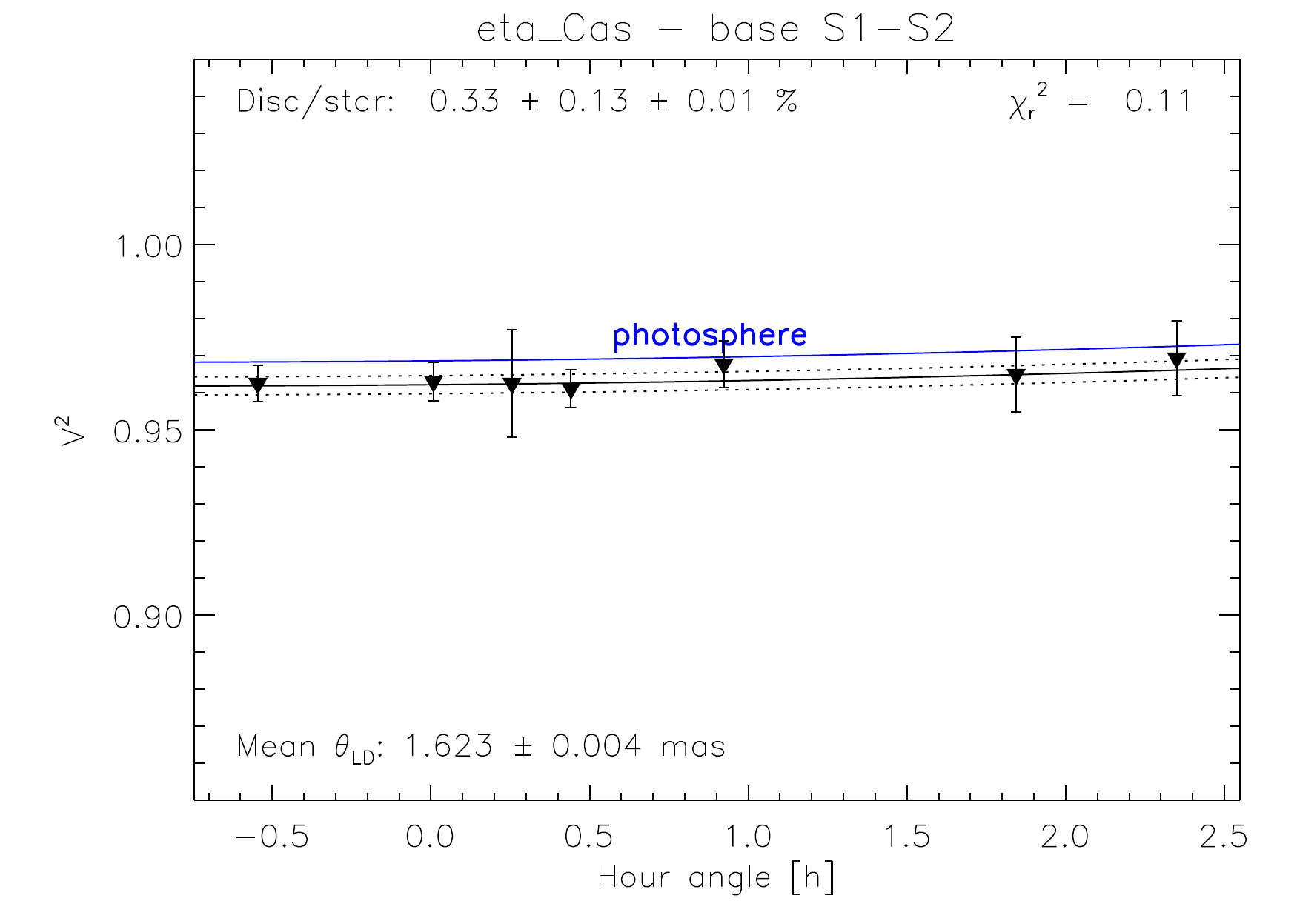} & \includegraphics[width=8.3cm]{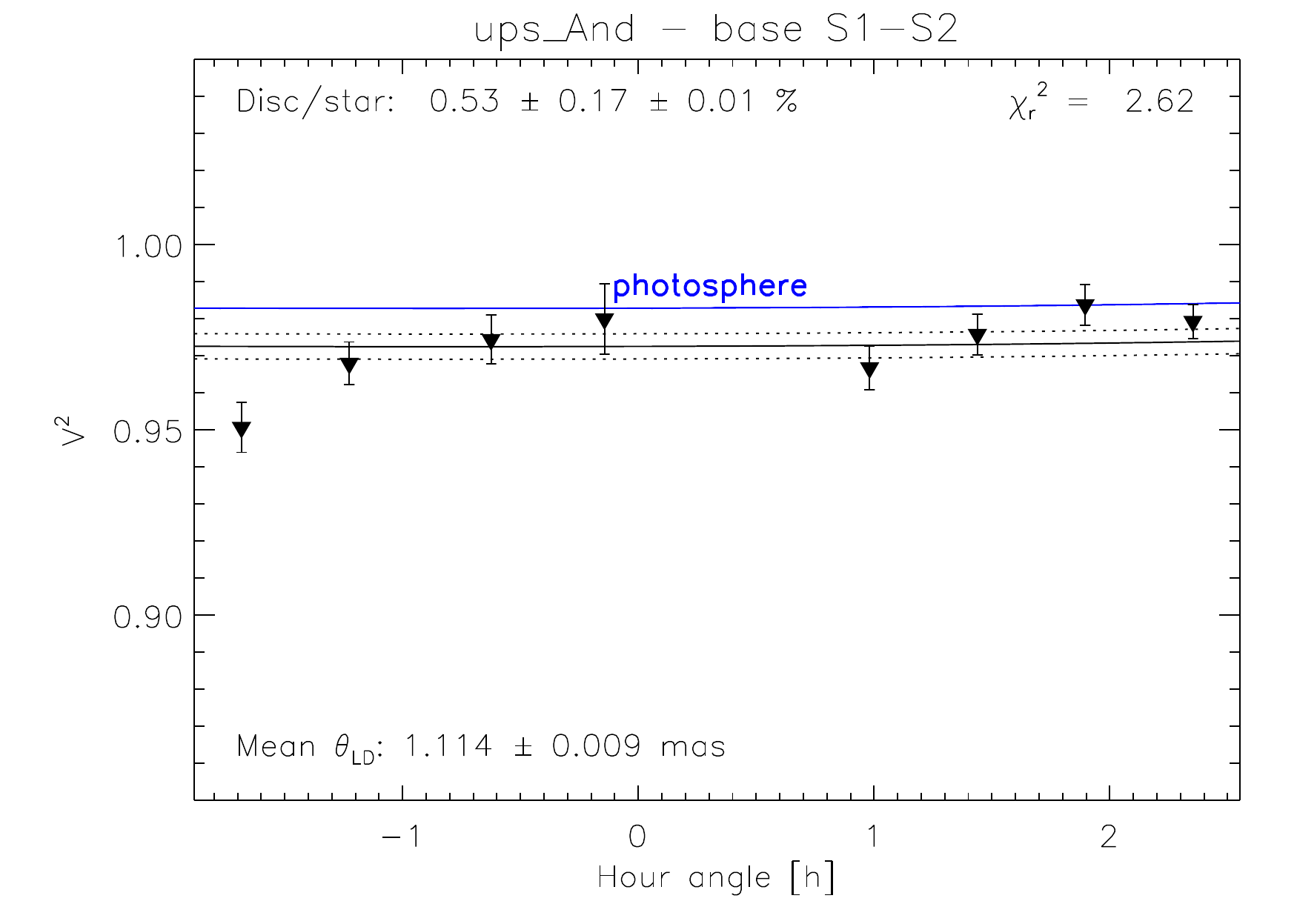} \\
\includegraphics[width=8.3cm]{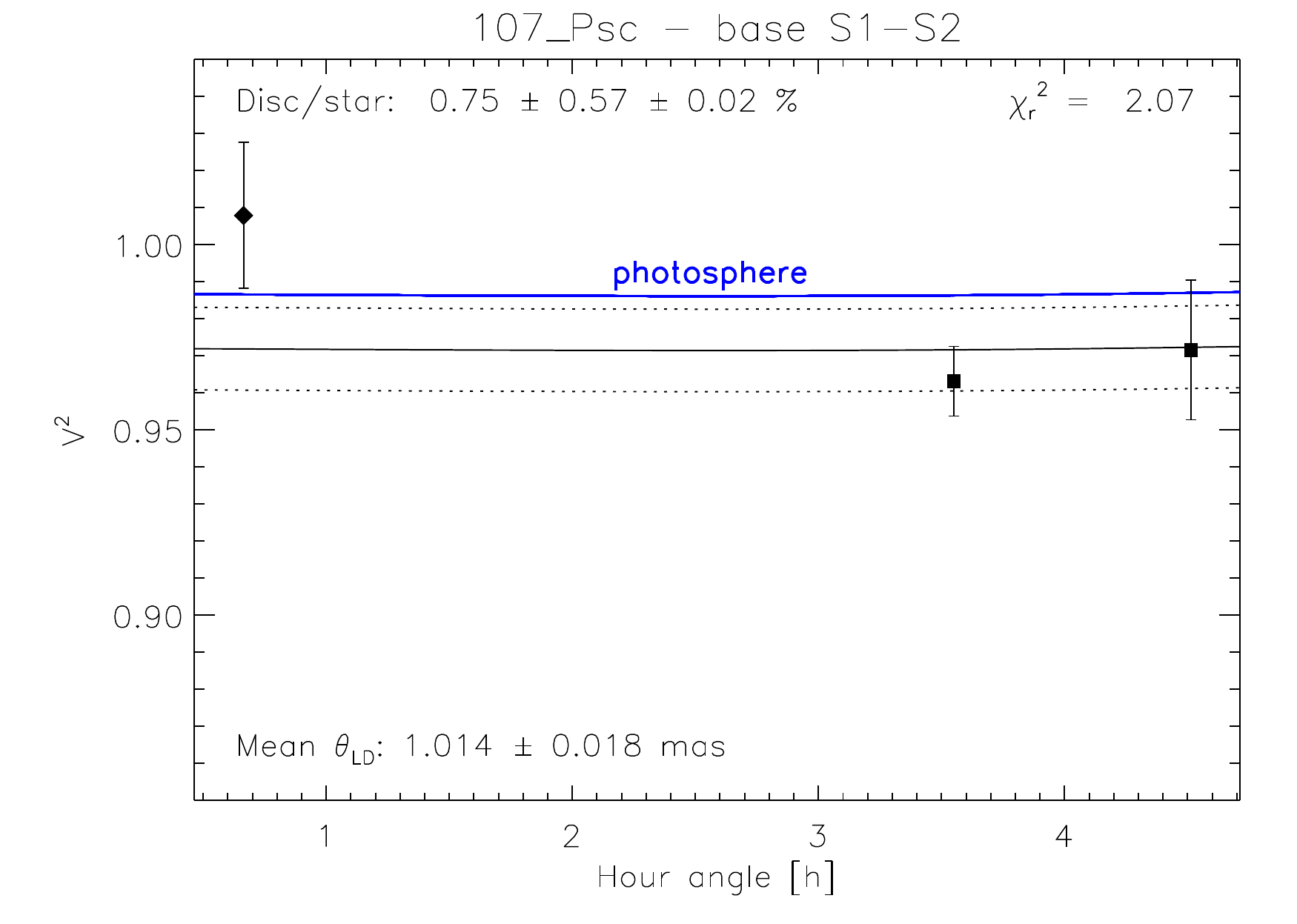} & \includegraphics[width=8.3cm]{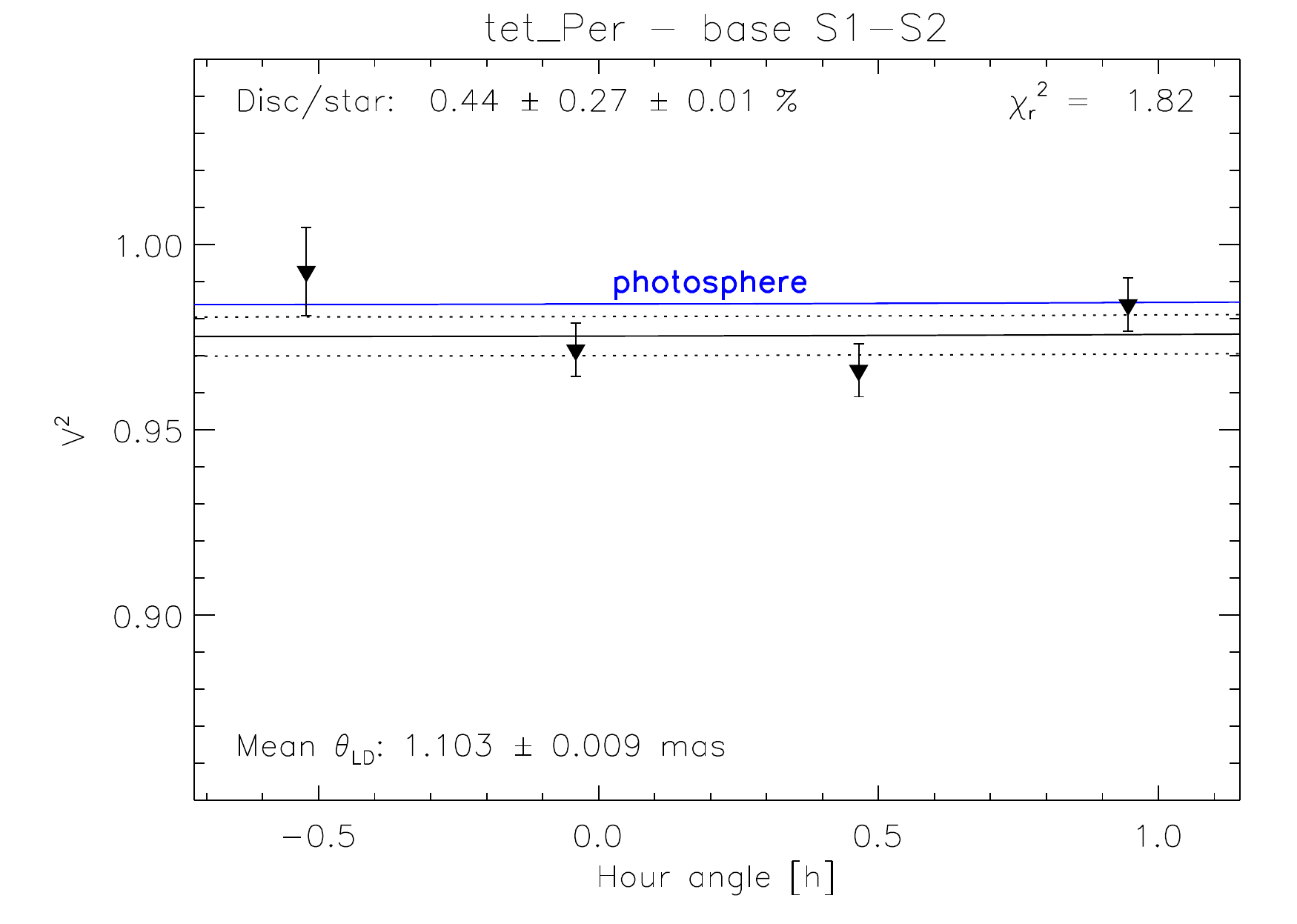}
\end{tabular}
\caption{Squared visibilities with error bars, expected photospheric squared visibility (blue shaded region) and best-fit star-disc model (solid line) with its $3\sigma$ confidence interval (dotted lines), plotted as a function of hour angle for all the targets not already presented in \citetalias{DiFolco07} or \citetalias{Absil08}. The best-fit disc/star contrast is given as an inset, together with the reduced $\chi^2$ of the fit. The two error bars given on the disc/star contrast correspond to the statistical and systematic errors (see Sect.~\ref{sec:errors}). Also given is the mean limb-darkened angular diameter used in the oblate photospheric model. The position angle (PA) of the oblate apparent photosphere is generally not known. This uncertainty is taken into account in the width of the blue line, except in the cases where the PA is known from previous studies (it is then specified in the inset). Different symbols are used for data taken in different years: circles (2006), upward triangles (2007), downward triangles (2008), squares (2009), diamonds (2010), and crosses (2011).}\label{fig:result}
\end{figure*}

\addtocounter{figure}{-1}
\begin{figure*}[p]
\centering
\begin{tabular}{cc}
\includegraphics[width=8.3cm]{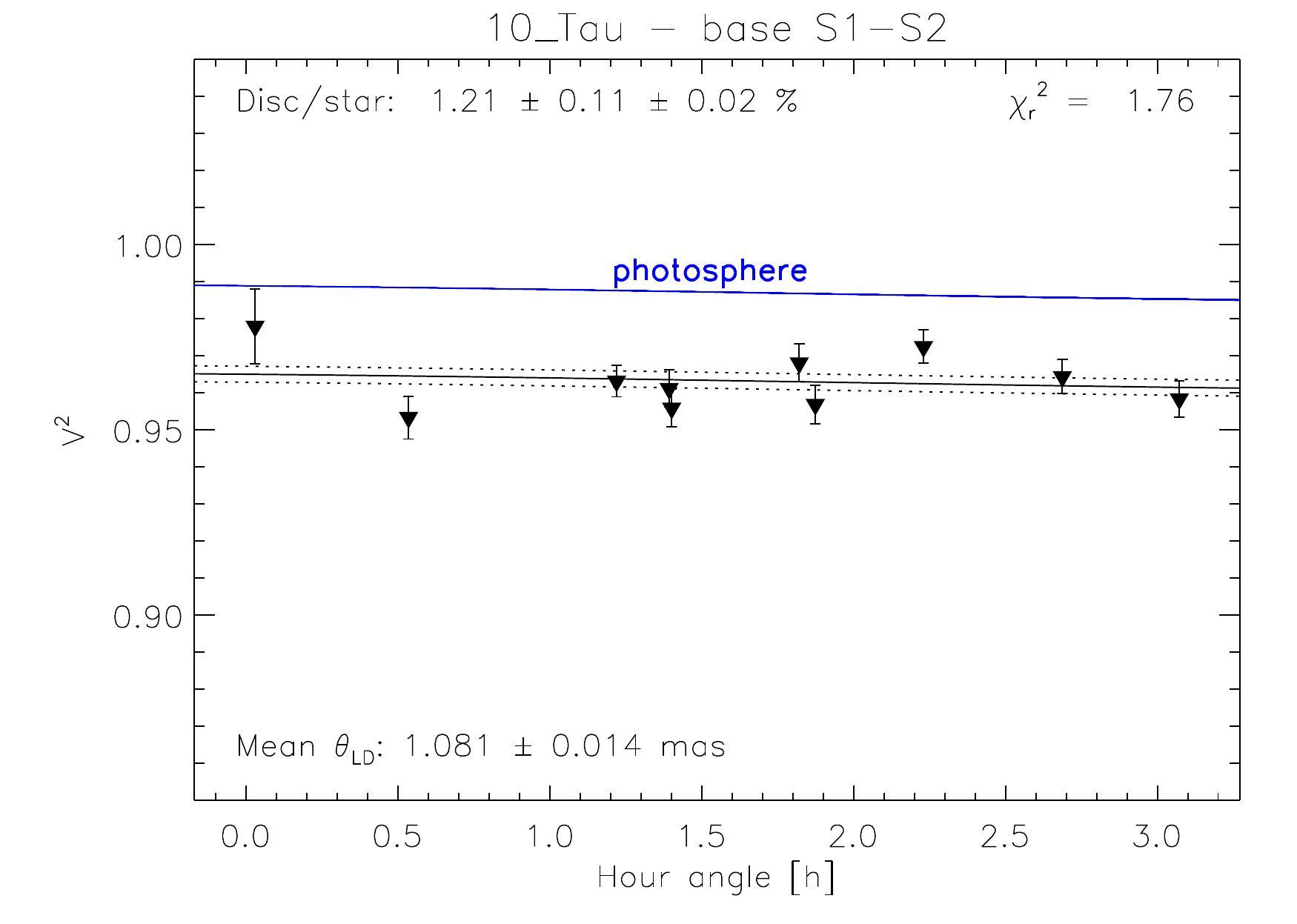} & \includegraphics[width=8.3cm]{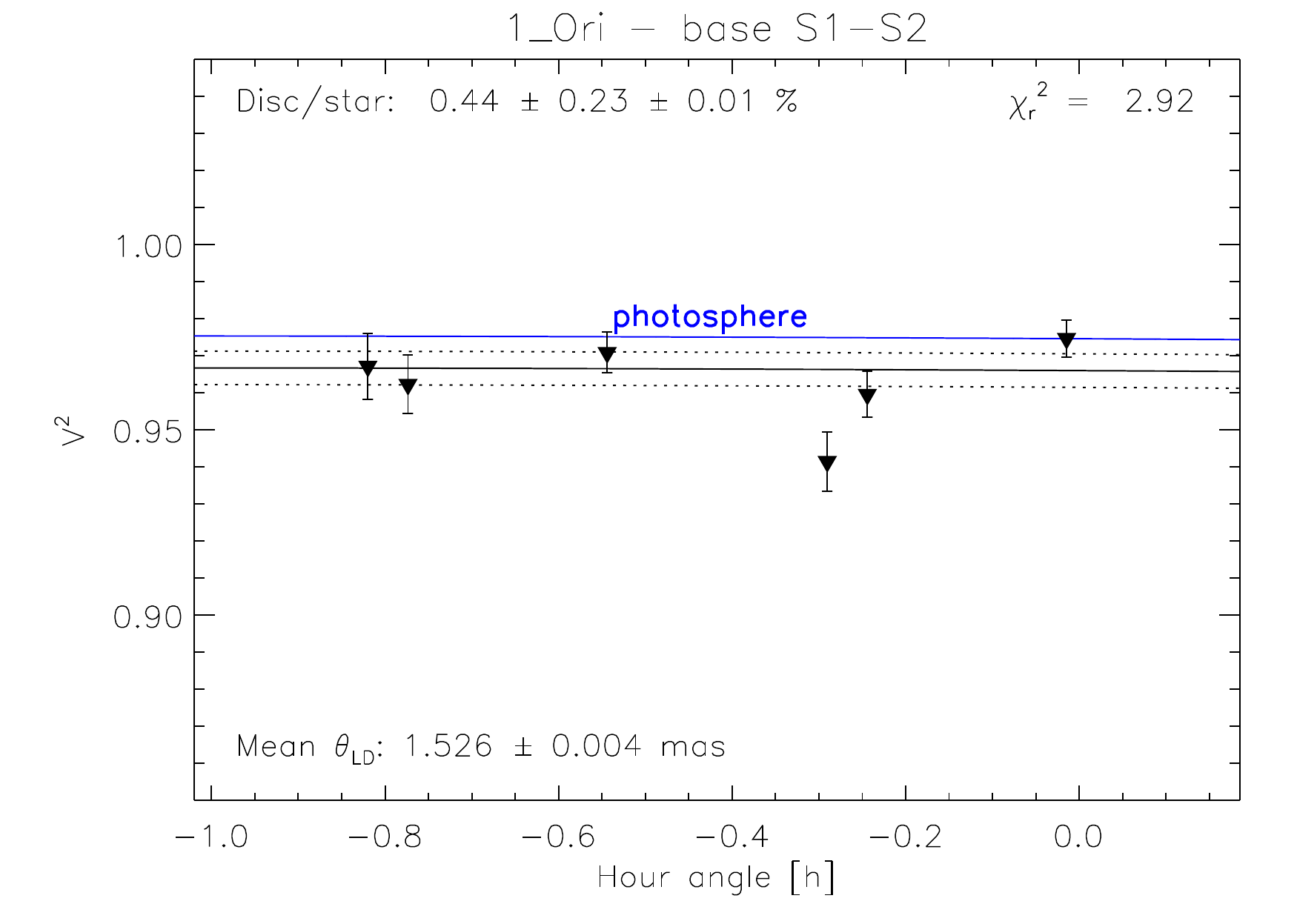} \\
\includegraphics[width=8.3cm]{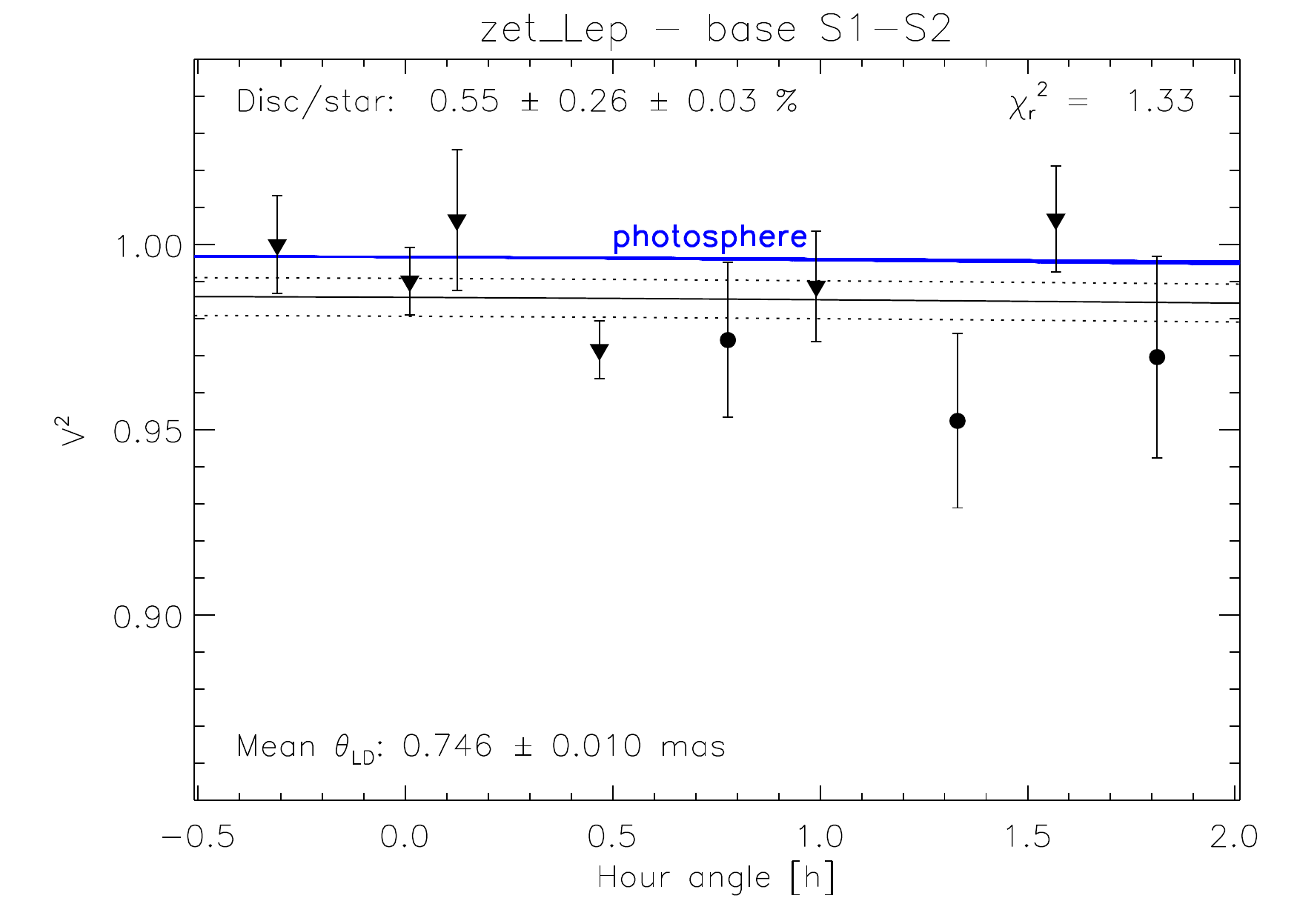} & \includegraphics[width=8.3cm]{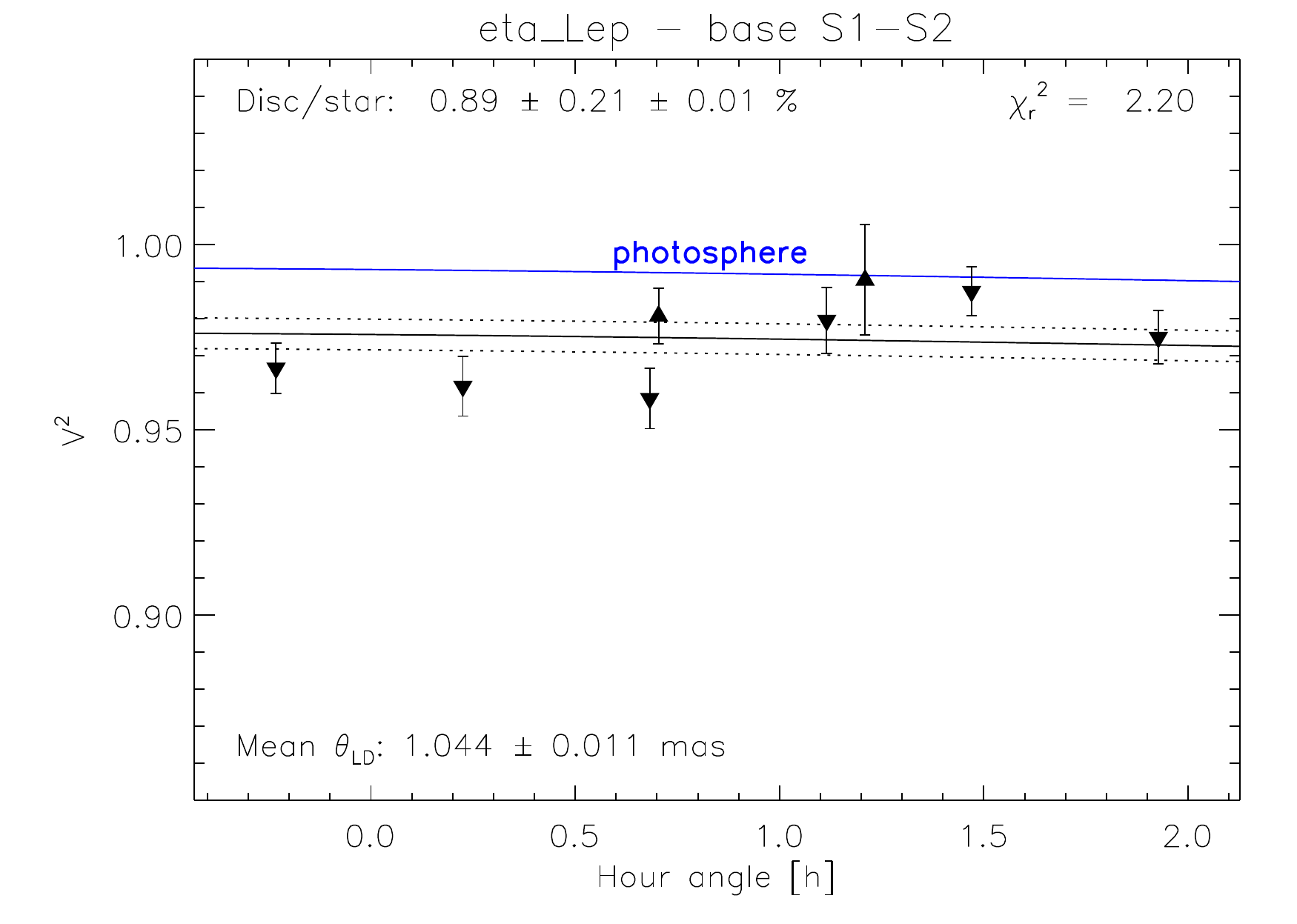} \\
\includegraphics[width=8.3cm]{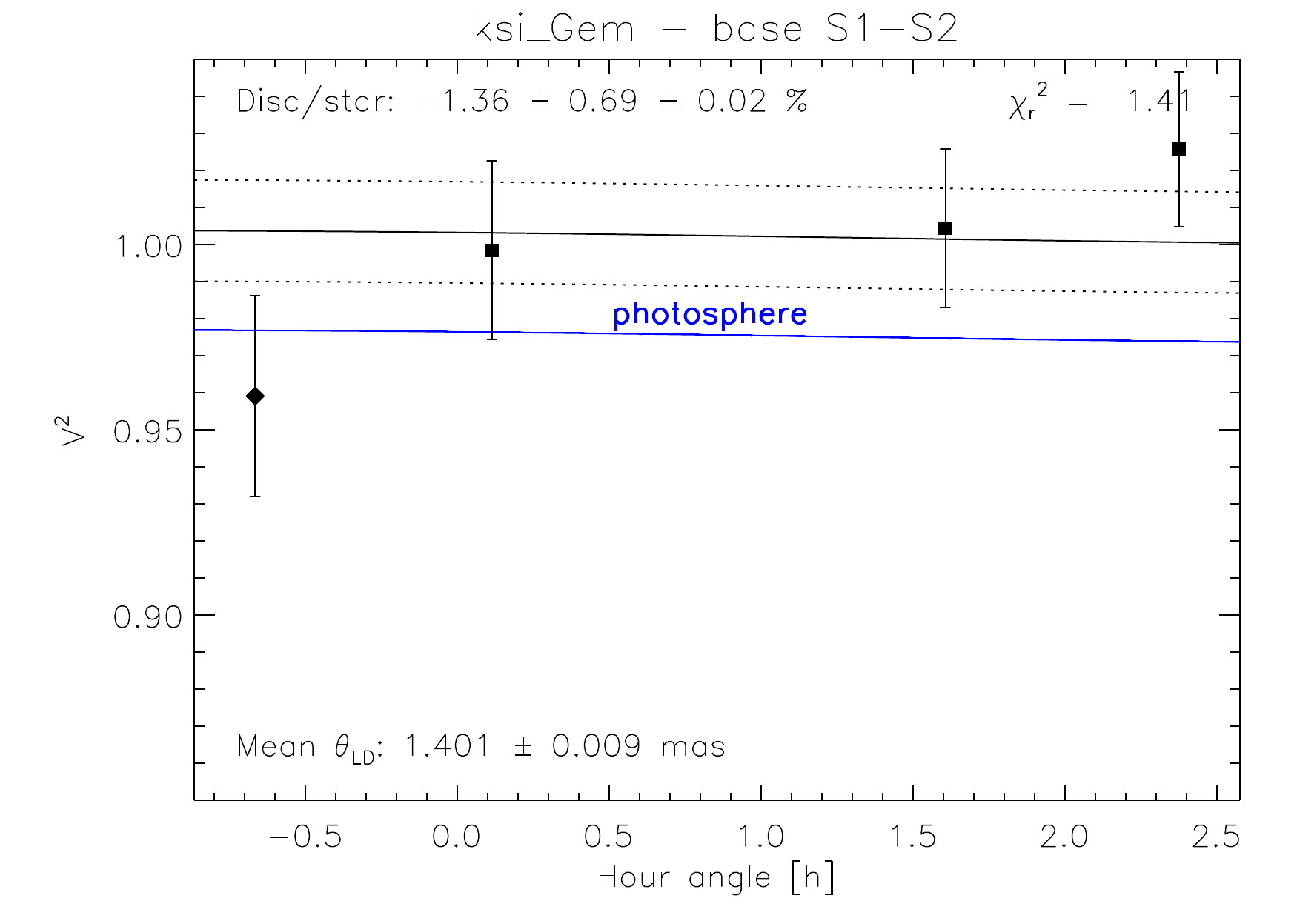} & \includegraphics[width=8.3cm]{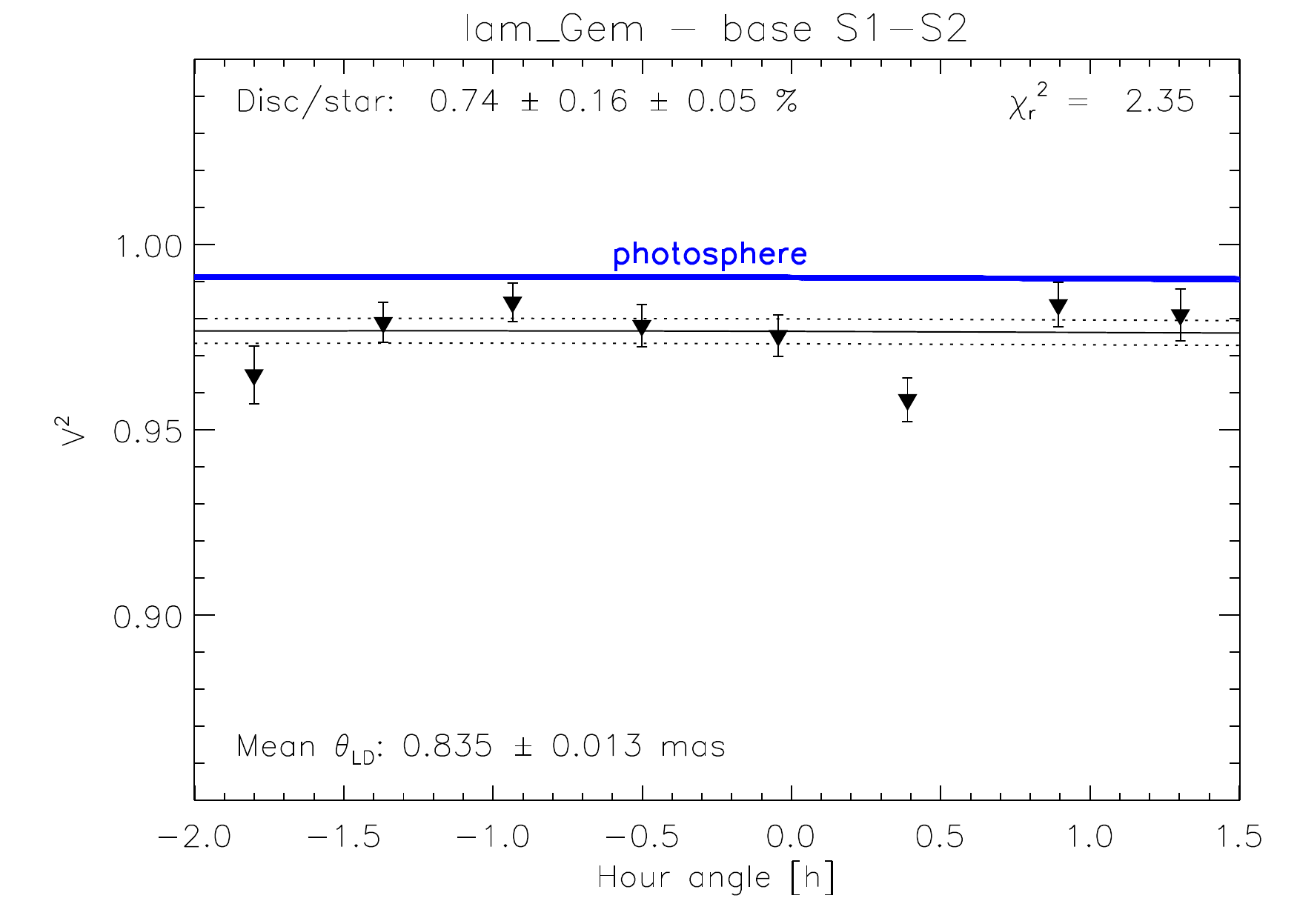} \\
\includegraphics[width=8.3cm]{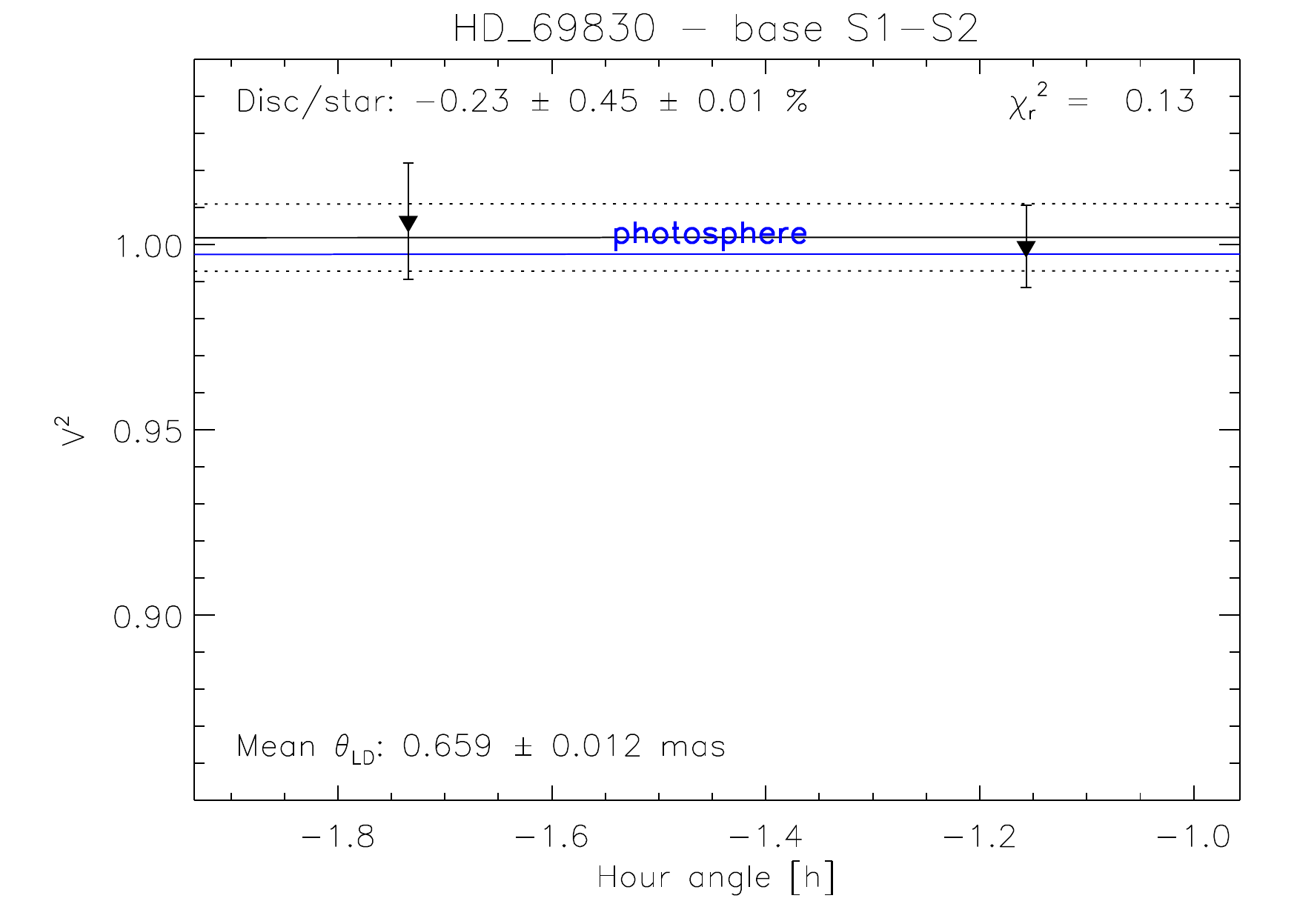} & \includegraphics[width=8.3cm]{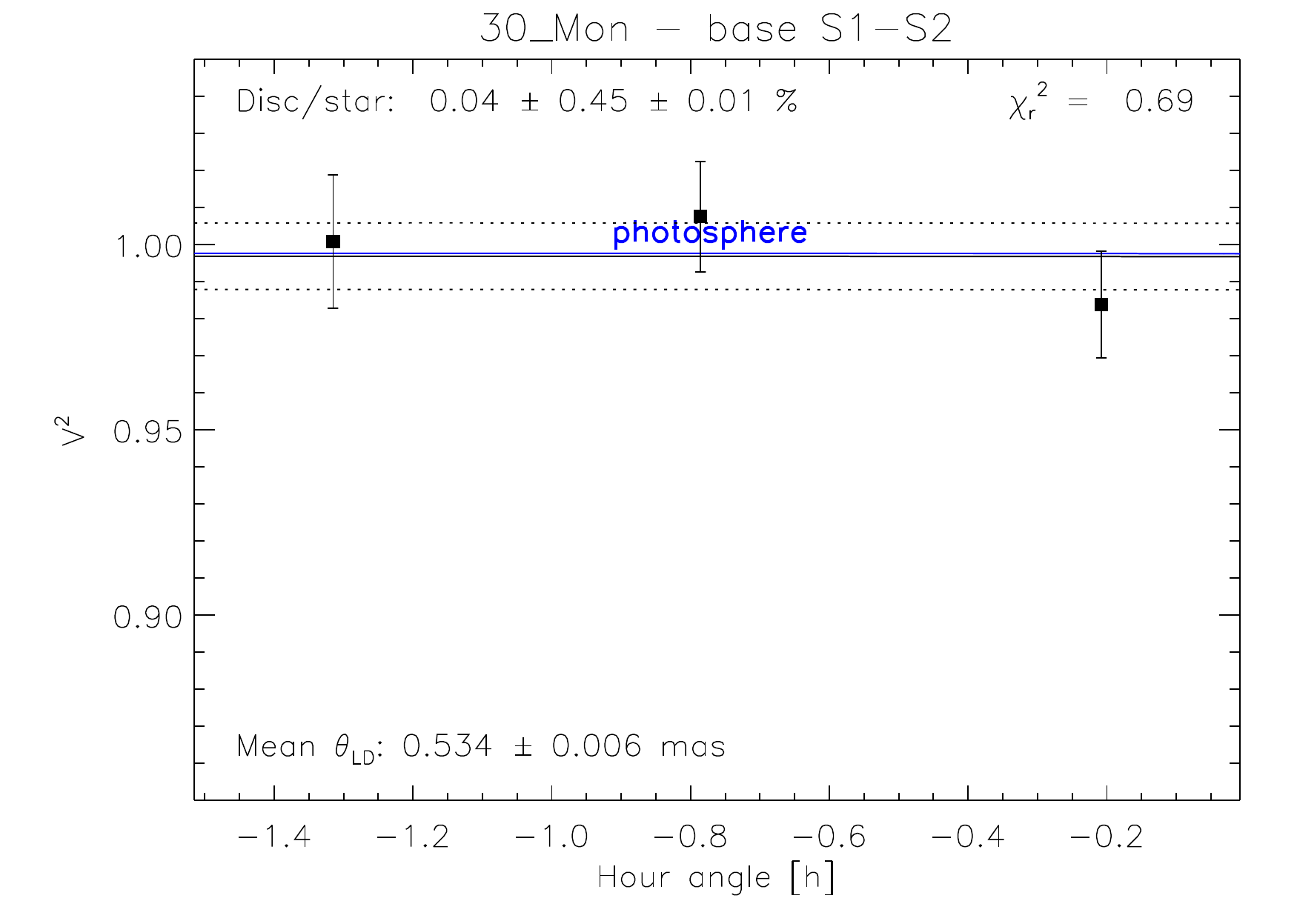}
\end{tabular}
\caption{(Continued)}
\end{figure*}

\addtocounter{figure}{-1}
\begin{figure*}[p]
\centering
\begin{tabular}{cc}
\includegraphics[width=8.3cm]{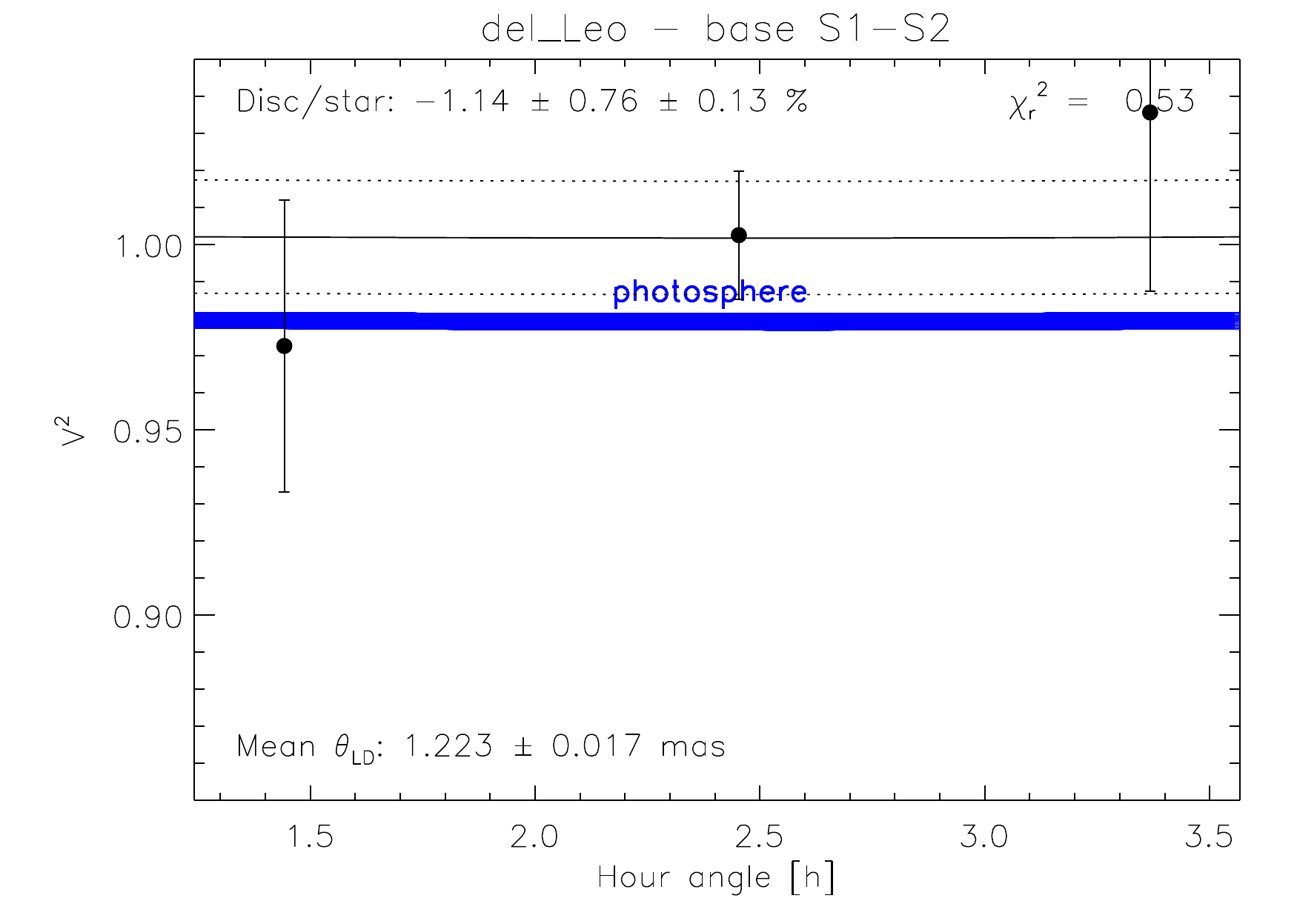} & \includegraphics[width=8.3cm]{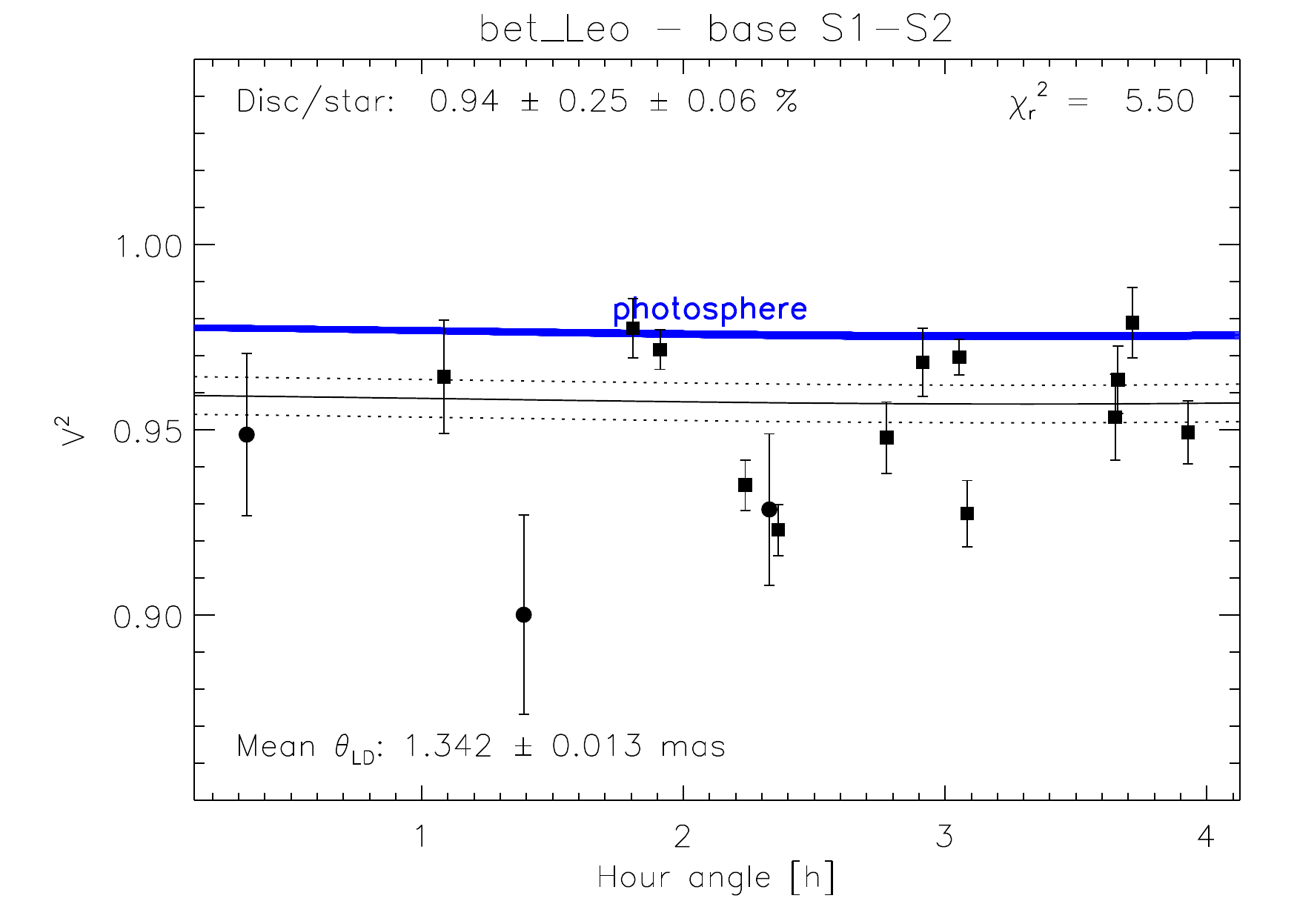} \\
\includegraphics[width=8.3cm]{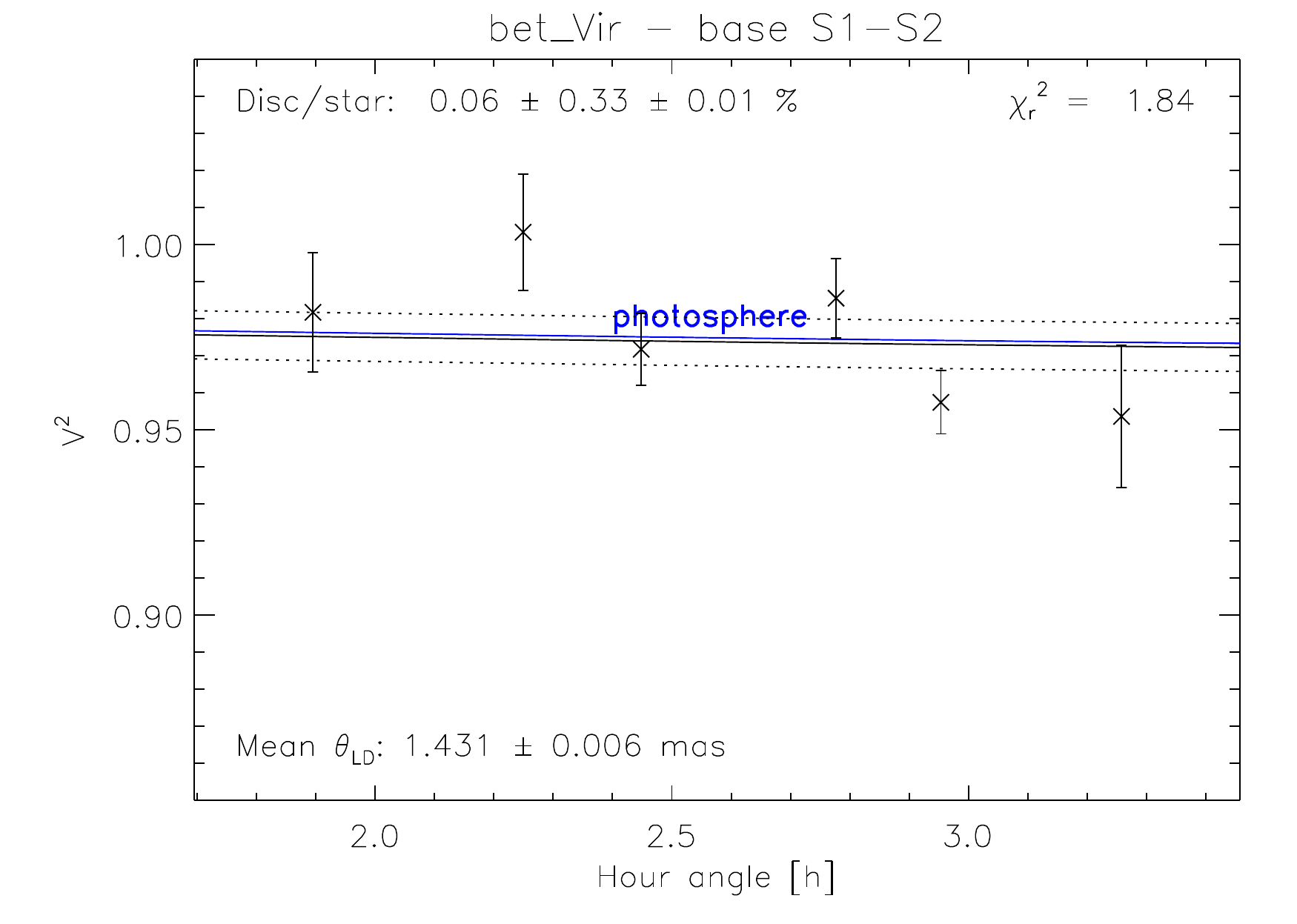} & \includegraphics[width=8.3cm]{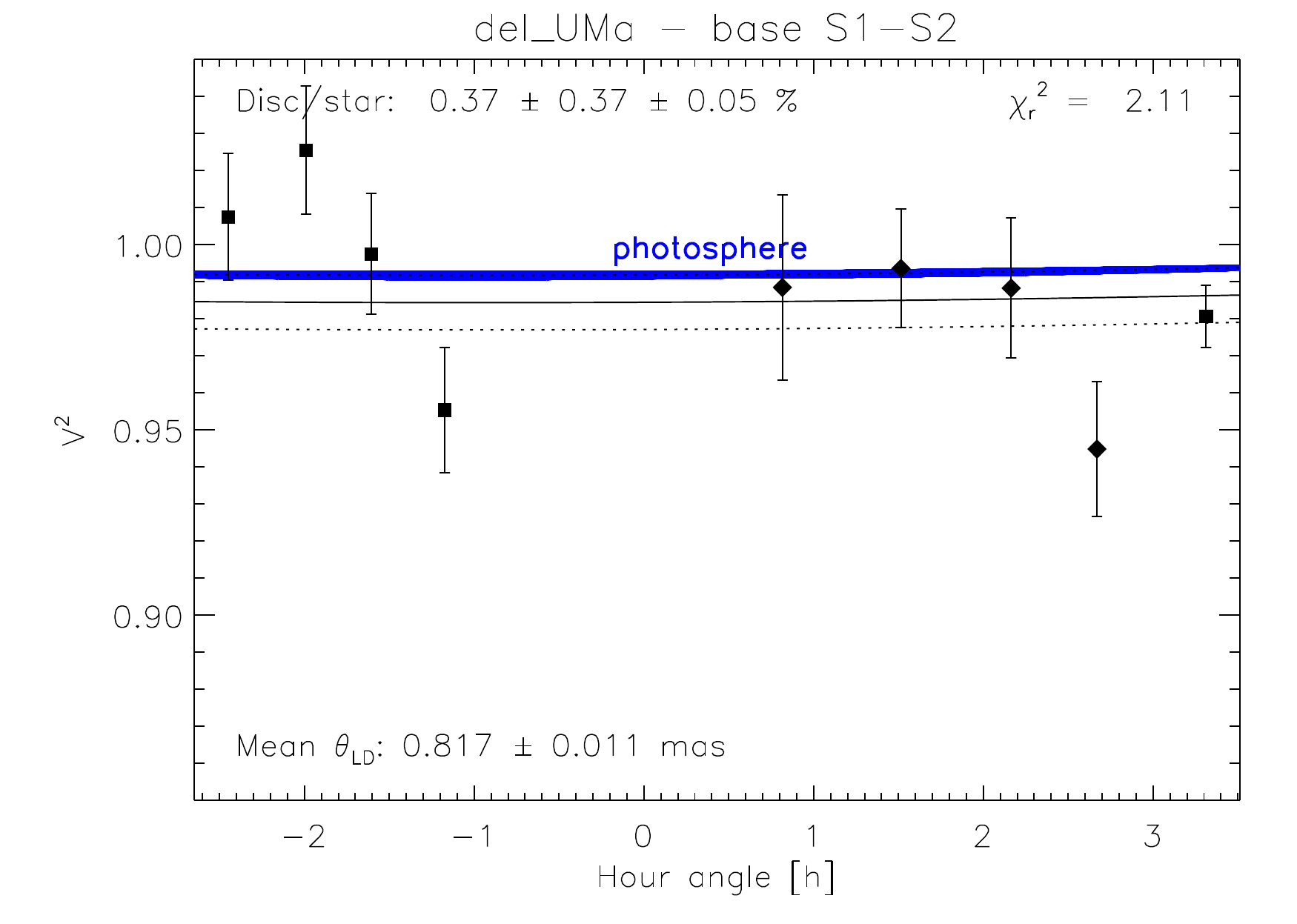} \\
\includegraphics[width=8.3cm]{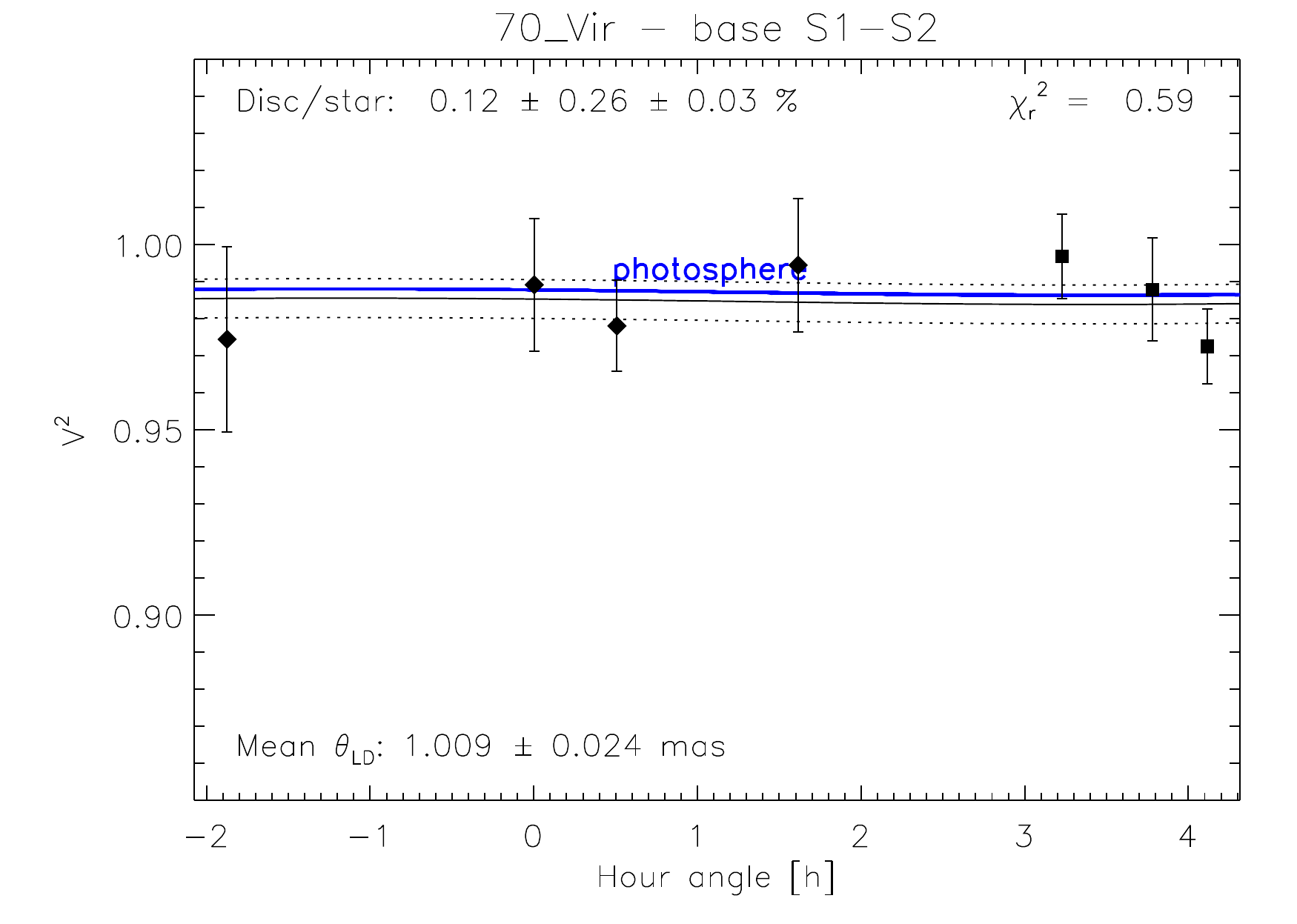} & \includegraphics[width=8.3cm]{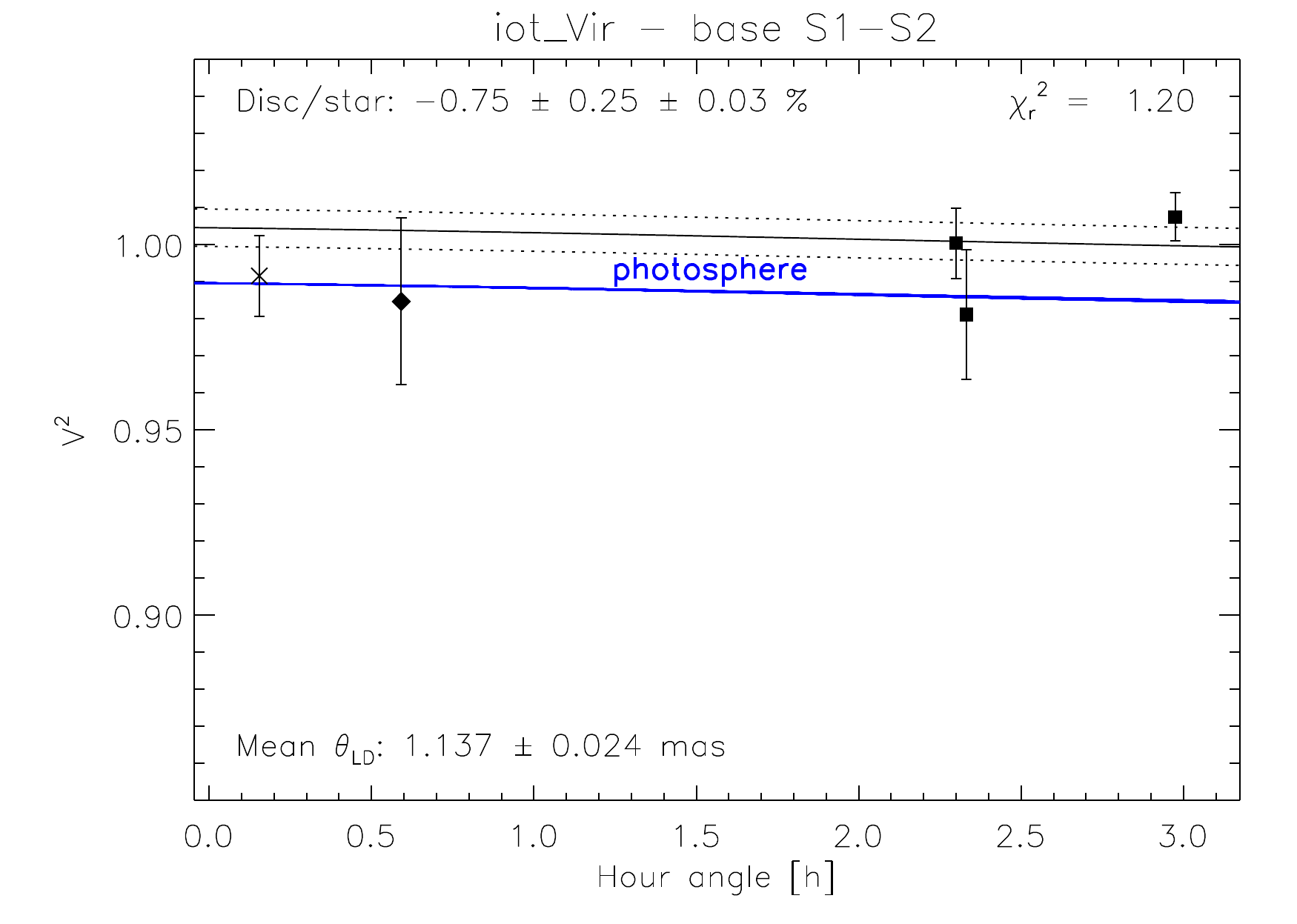} \\
\includegraphics[width=8.3cm]{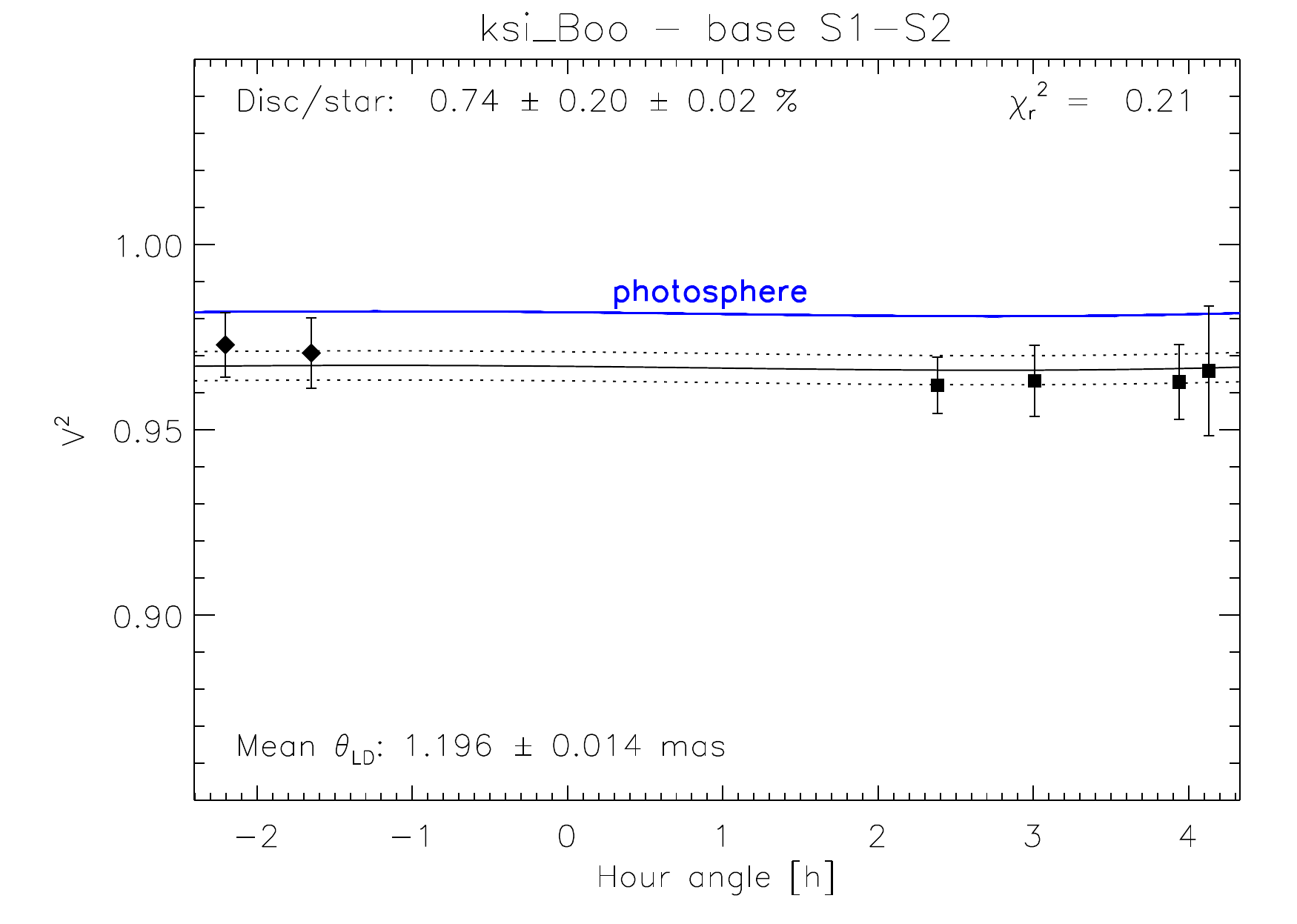} & \includegraphics[width=8.3cm]{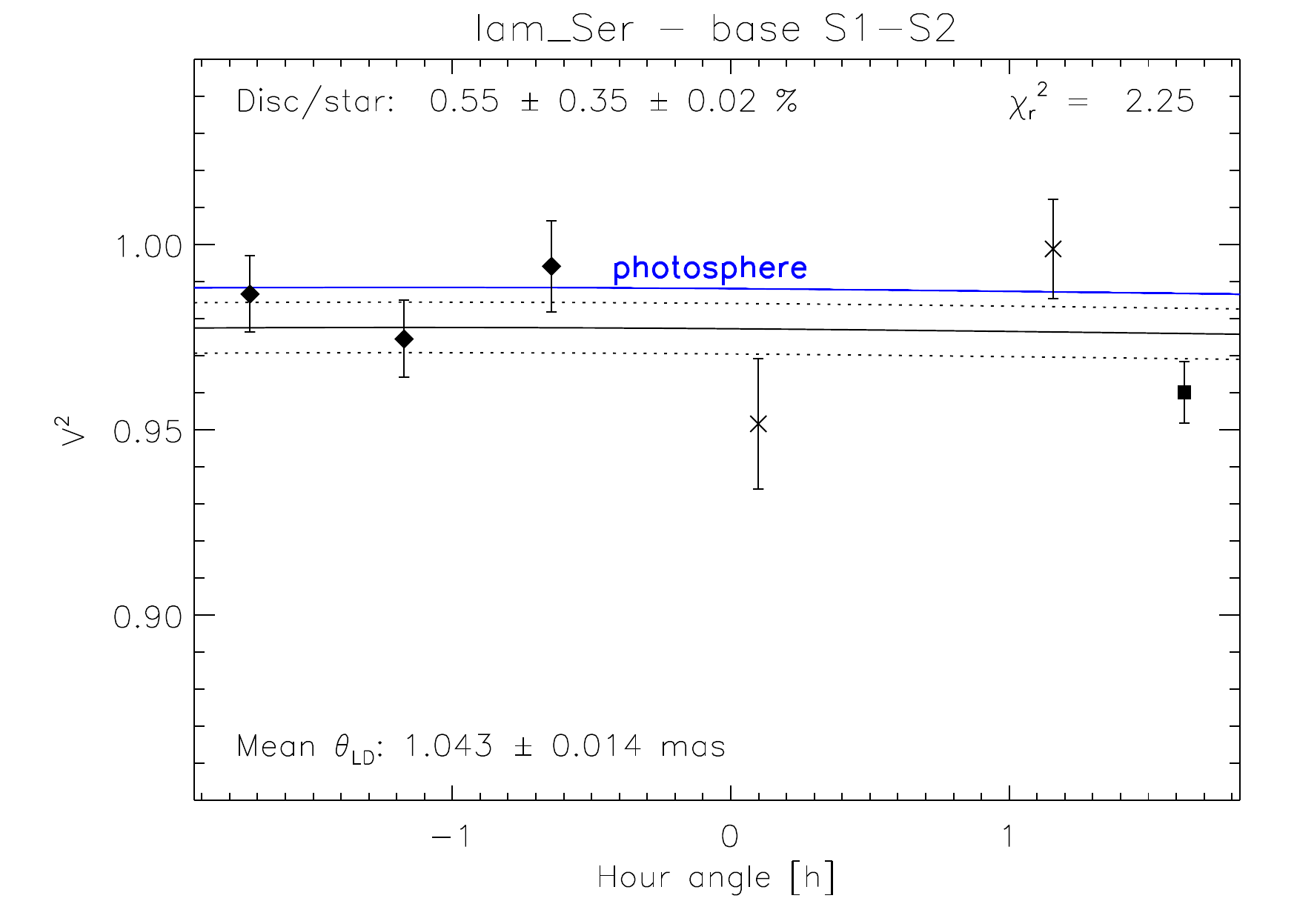}
\end{tabular}
\caption{(Continued)}
\end{figure*}

\addtocounter{figure}{-1}
\begin{figure*}[p]
\centering
\begin{tabular}{cc}
\includegraphics[width=8.3cm]{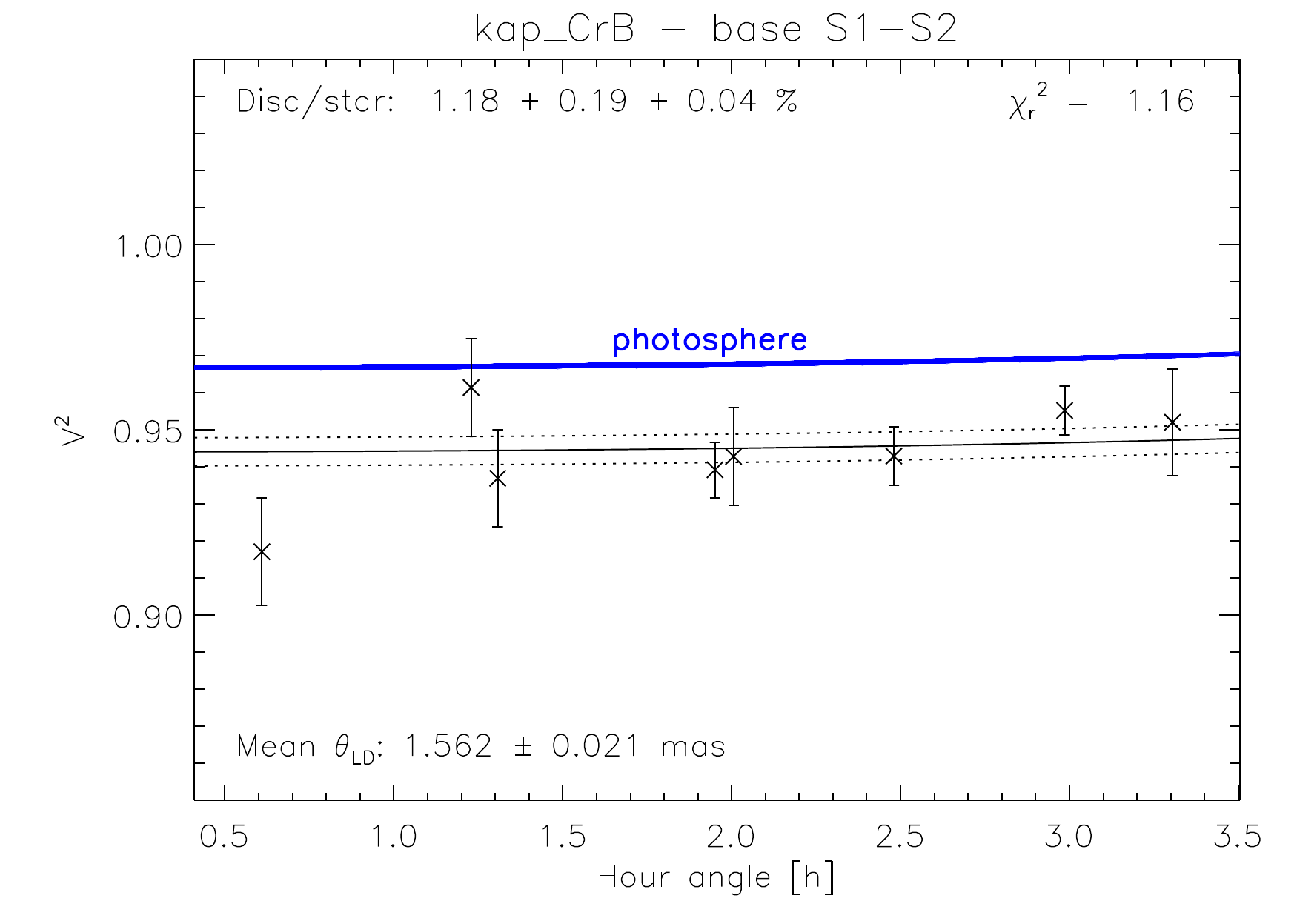} & \includegraphics[width=8.3cm]{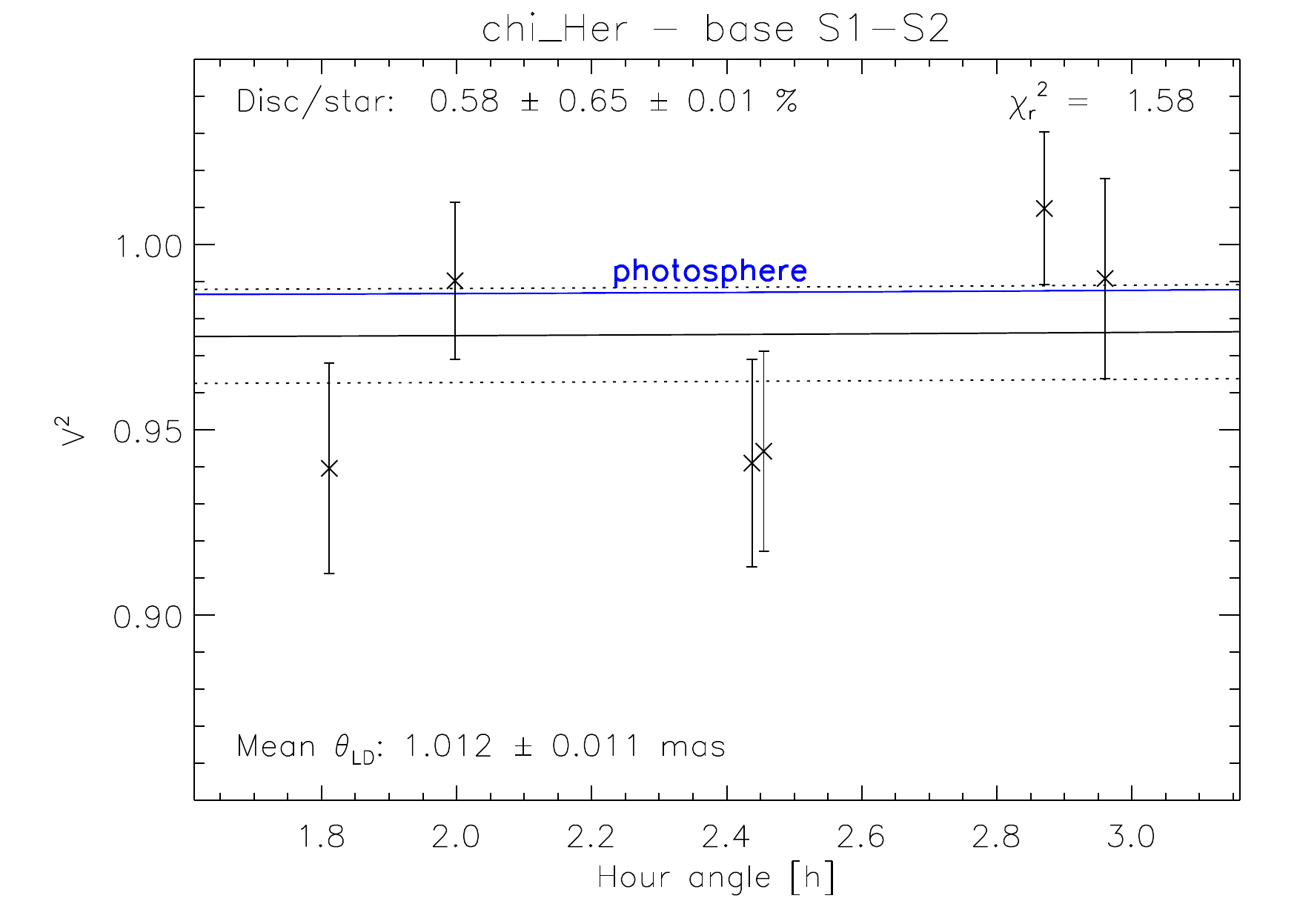} \\
\includegraphics[width=8.3cm]{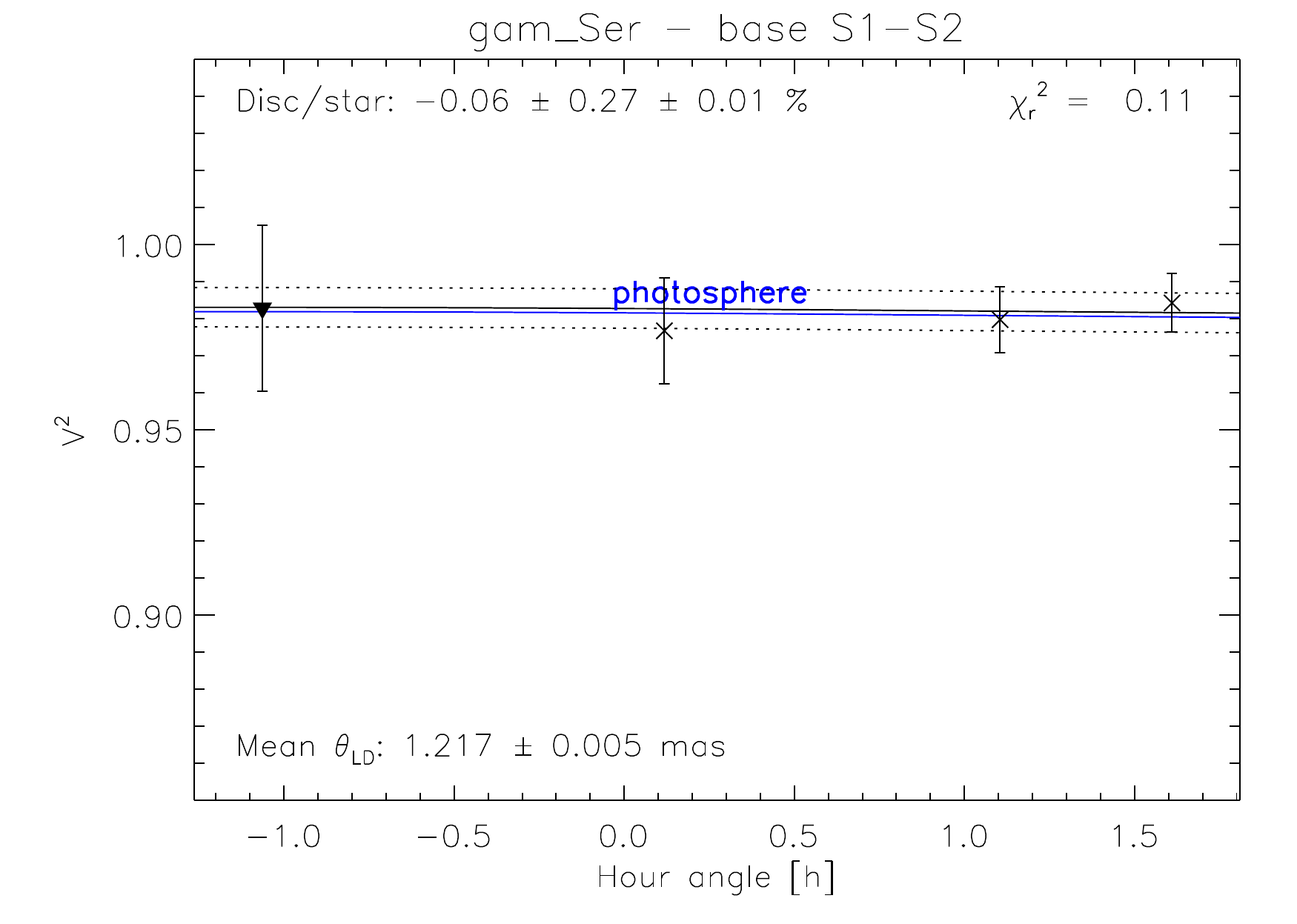} & \includegraphics[width=8.3cm]{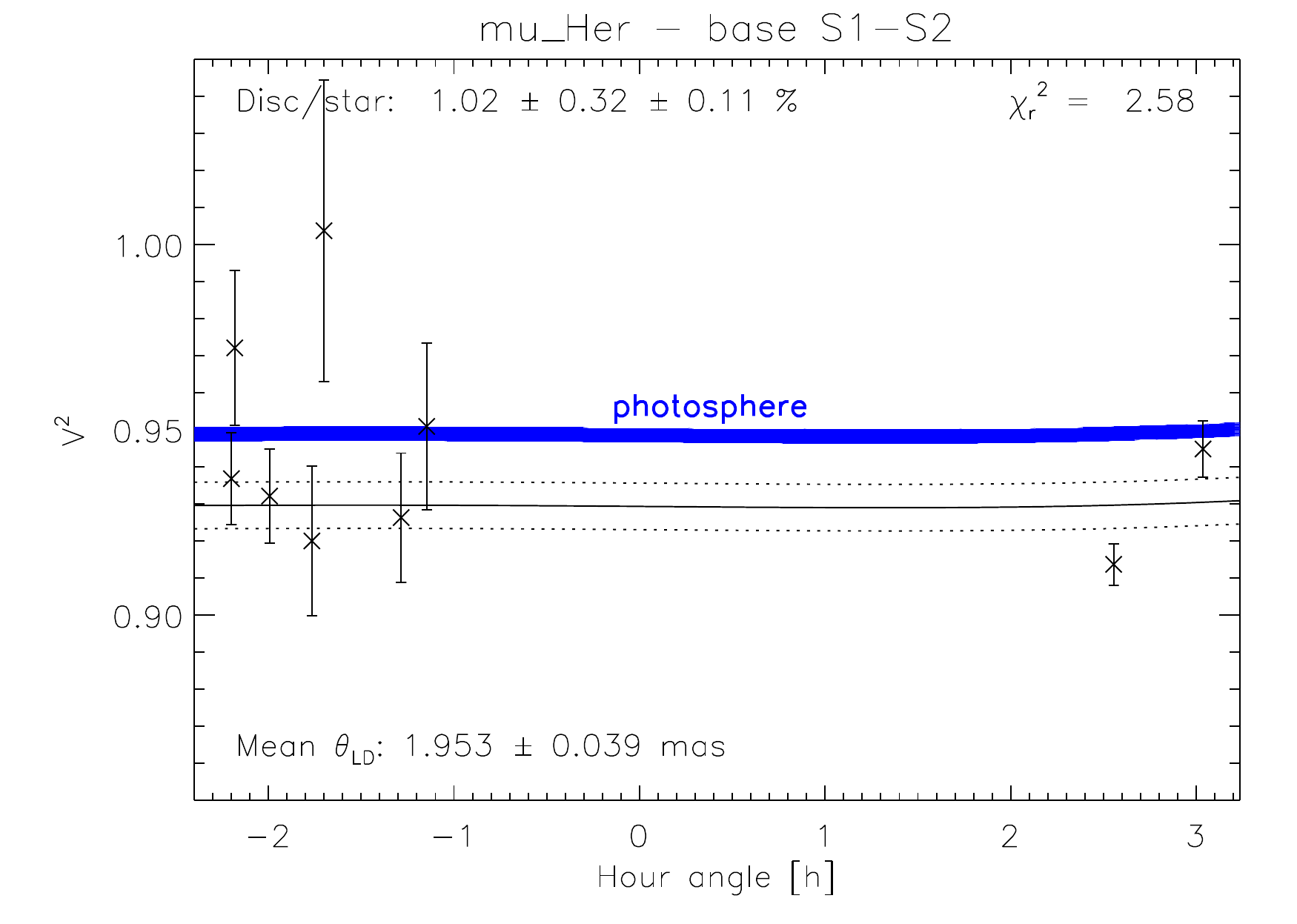} \\
\includegraphics[width=8.3cm]{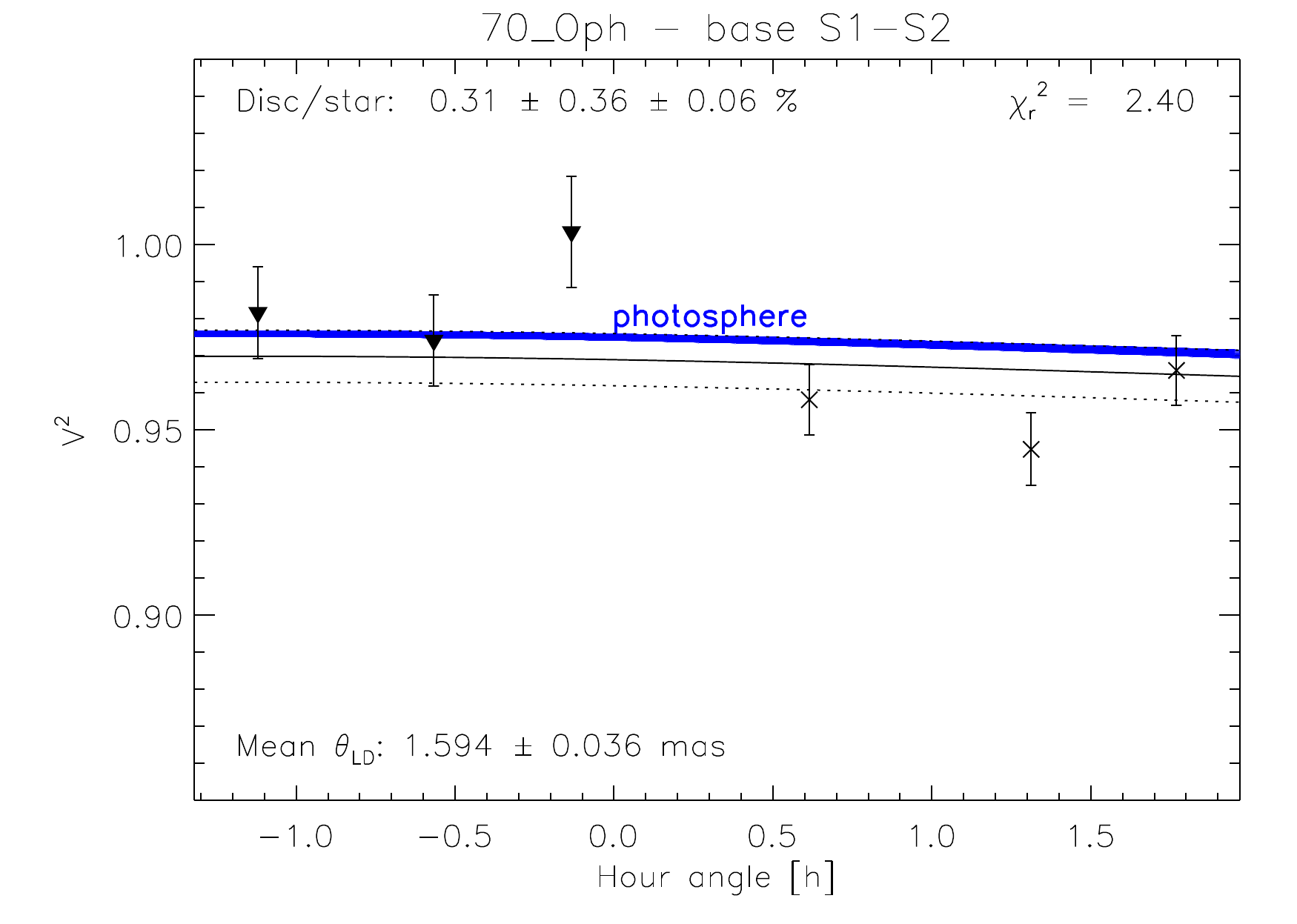} & \includegraphics[width=8.3cm]{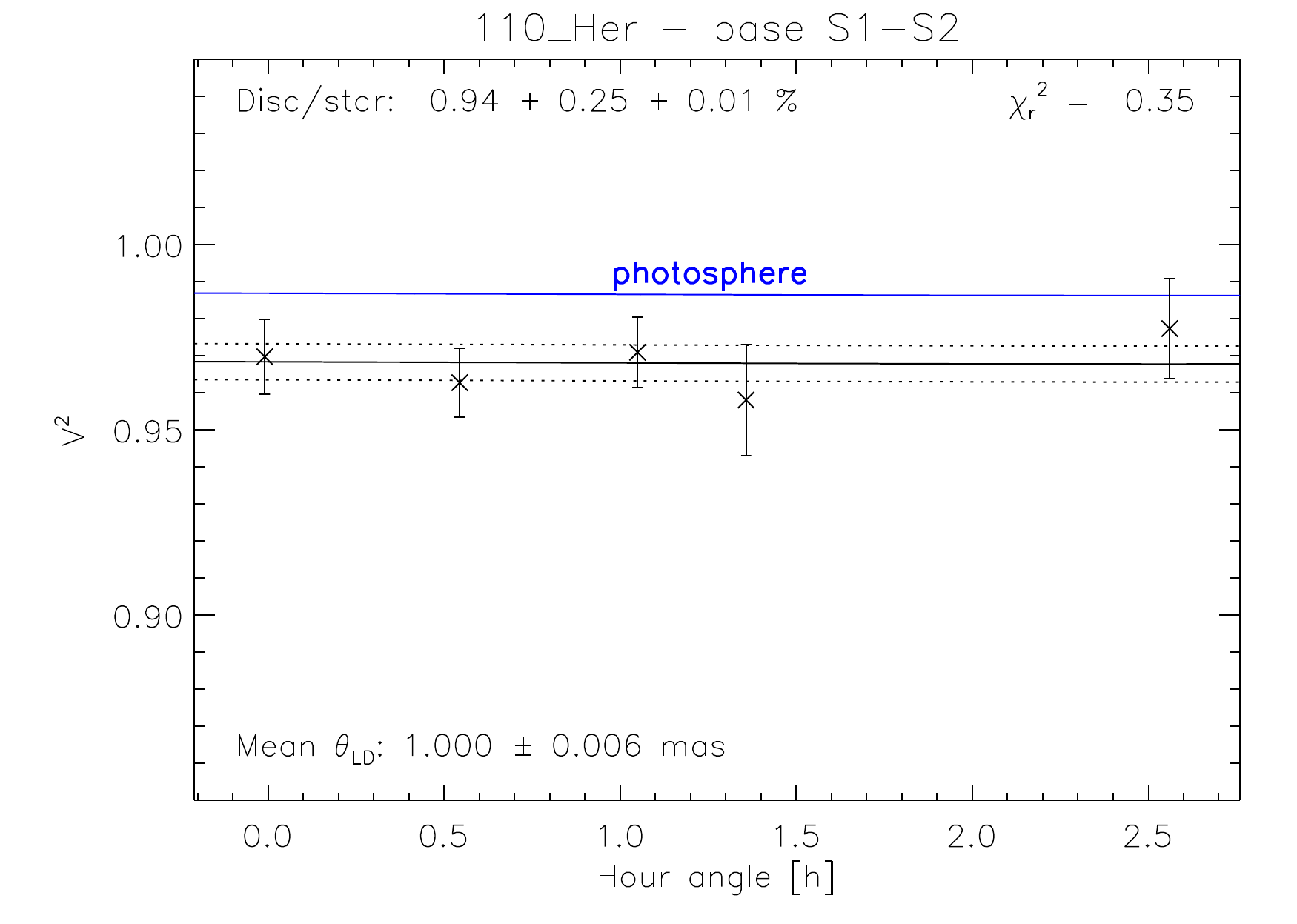} \\
\includegraphics[width=8.3cm]{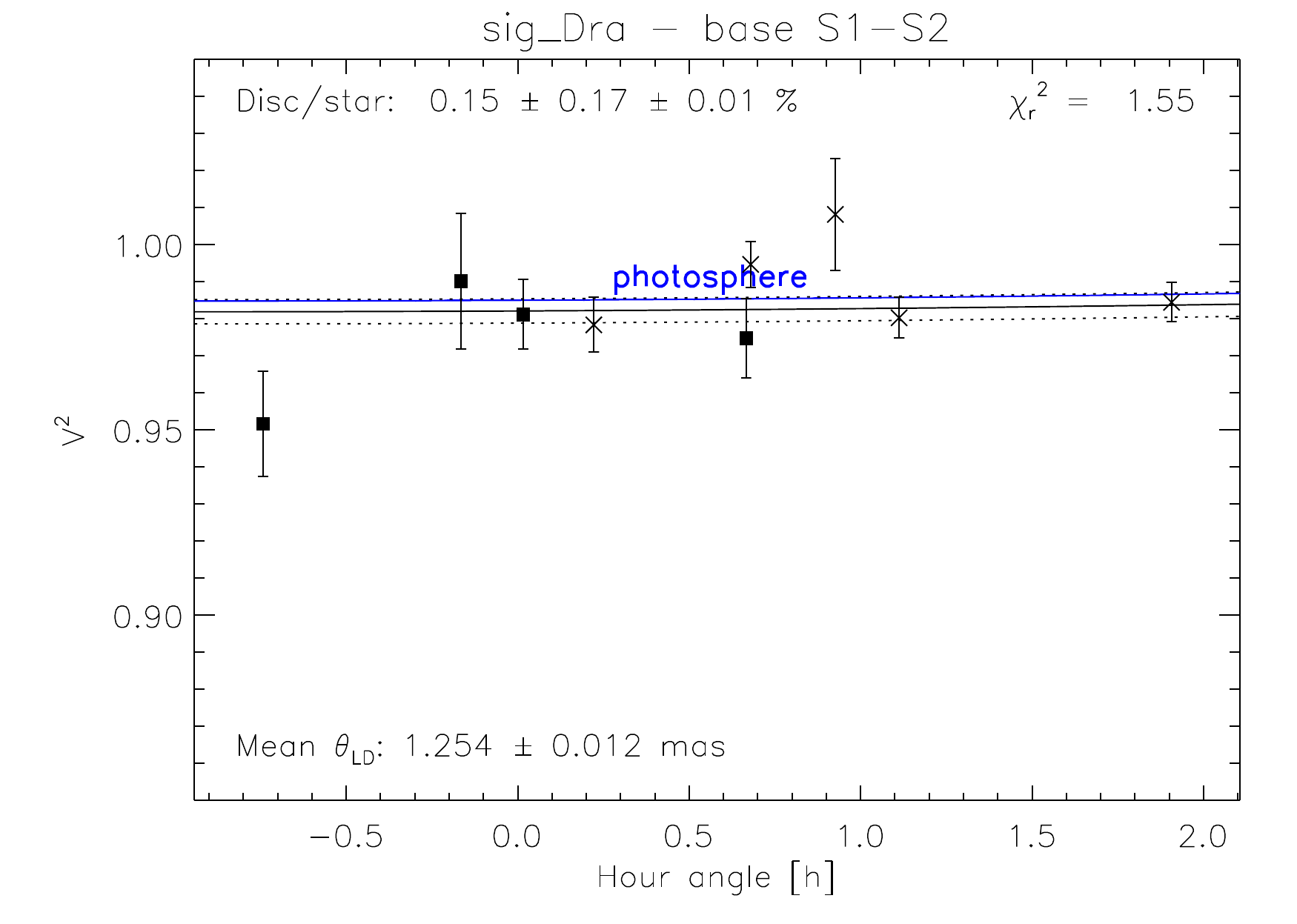} & \includegraphics[width=8.3cm]{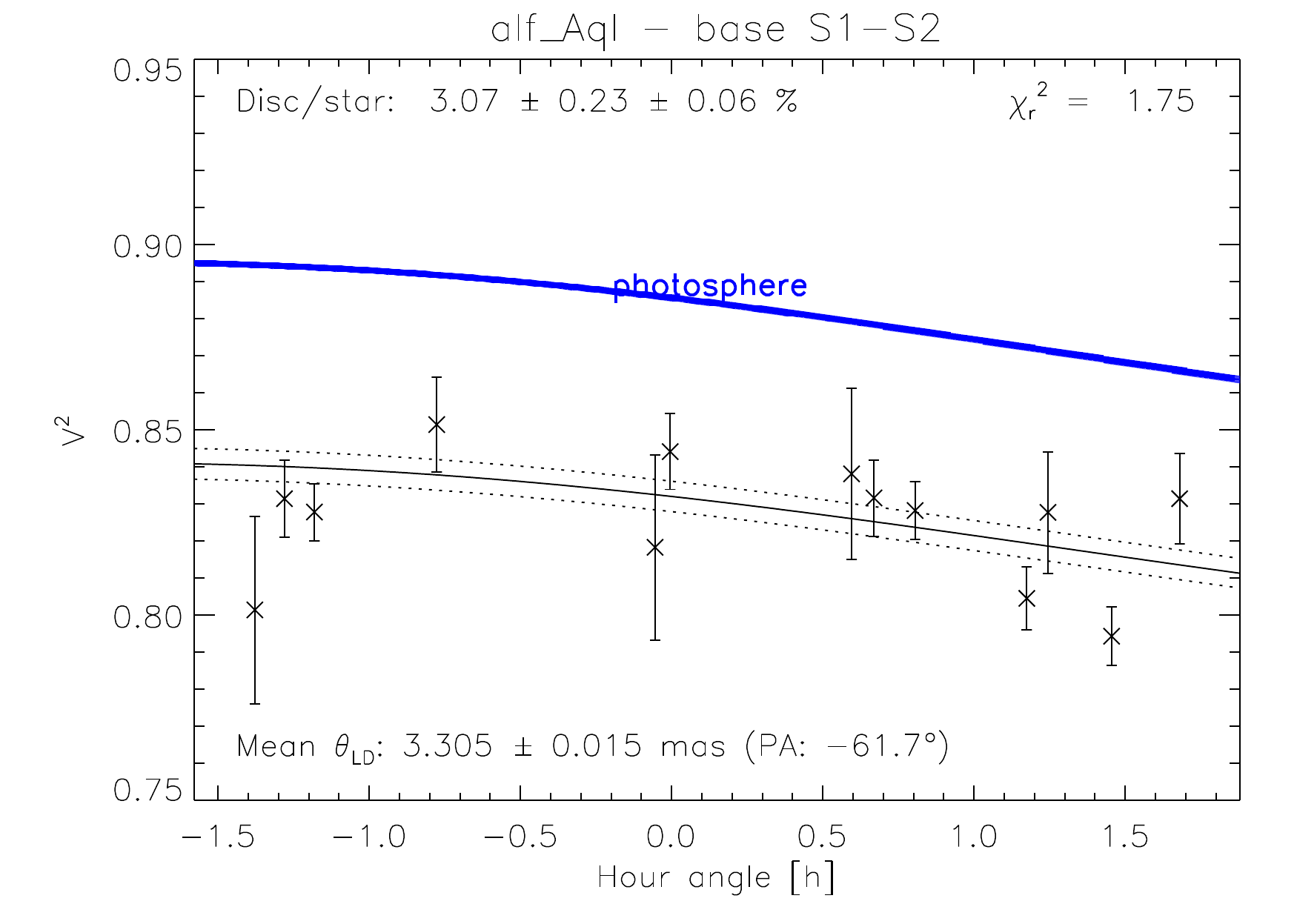}
\end{tabular}
\caption{(Continued)}
\end{figure*}

\addtocounter{figure}{-1}
\begin{figure*}[t]
\centering
\begin{tabular}{cc}
\includegraphics[width=8.3cm]{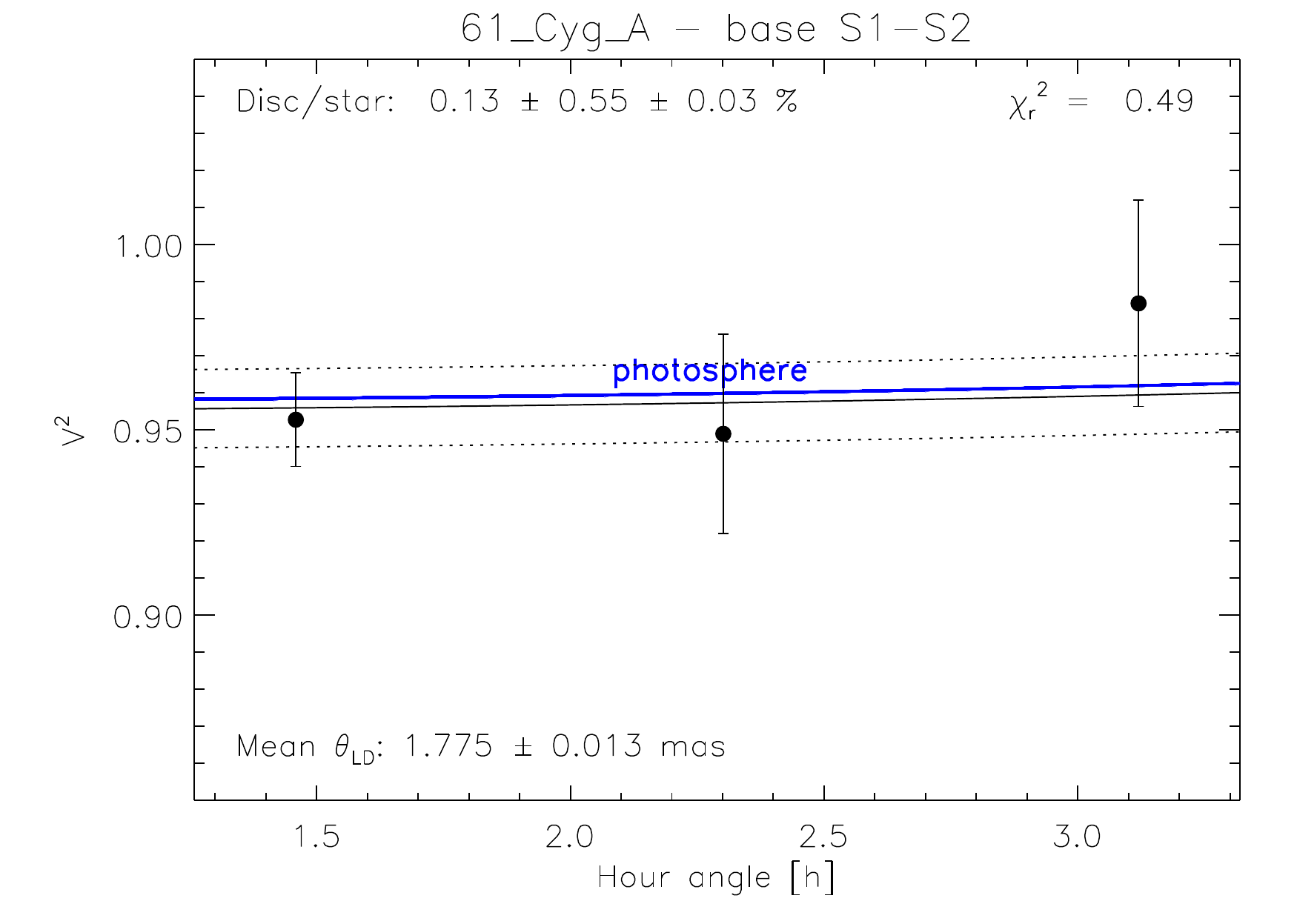} & \includegraphics[width=8.3cm]{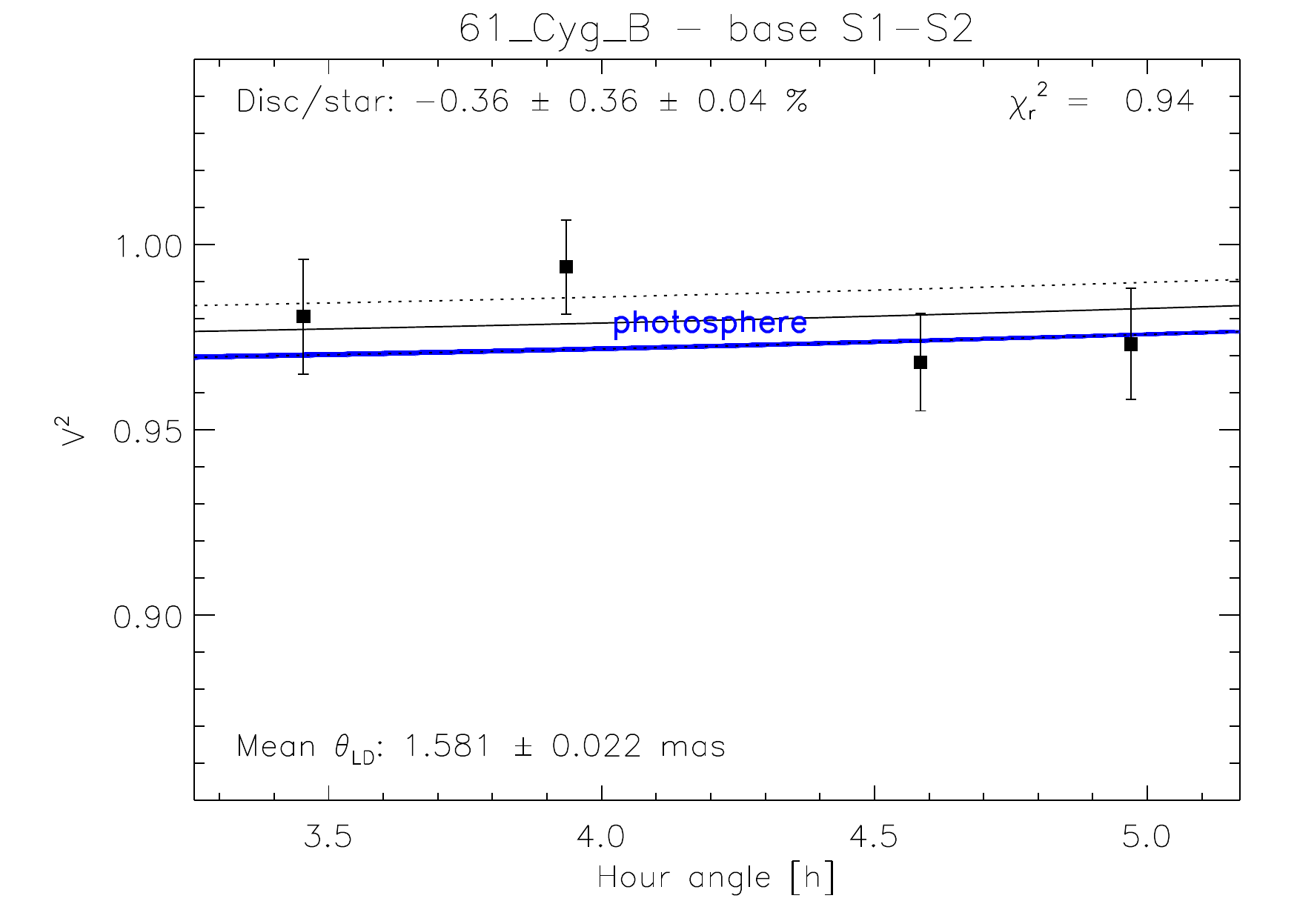} \\
\includegraphics[width=8.3cm]{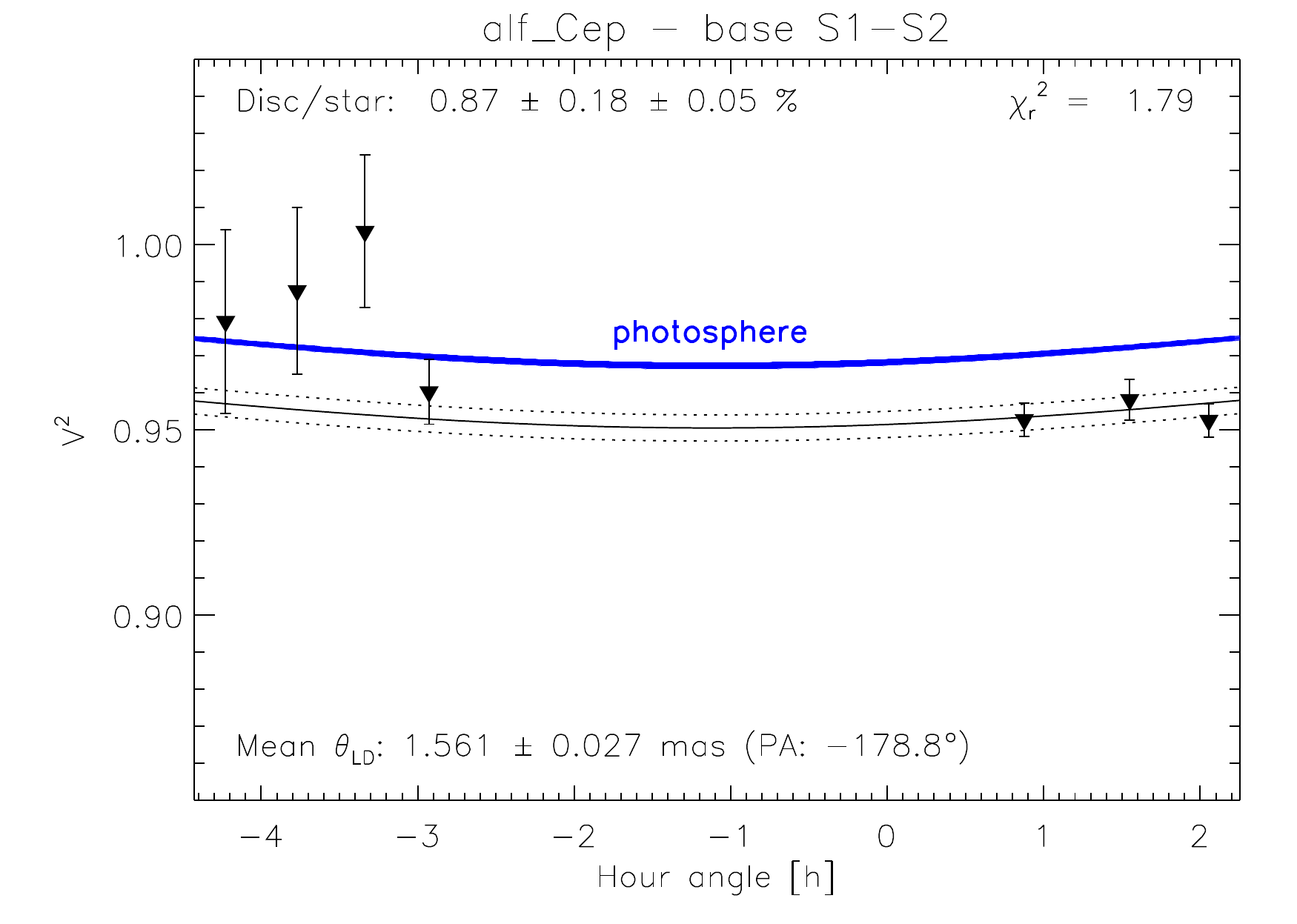} & \includegraphics[width=8.3cm]{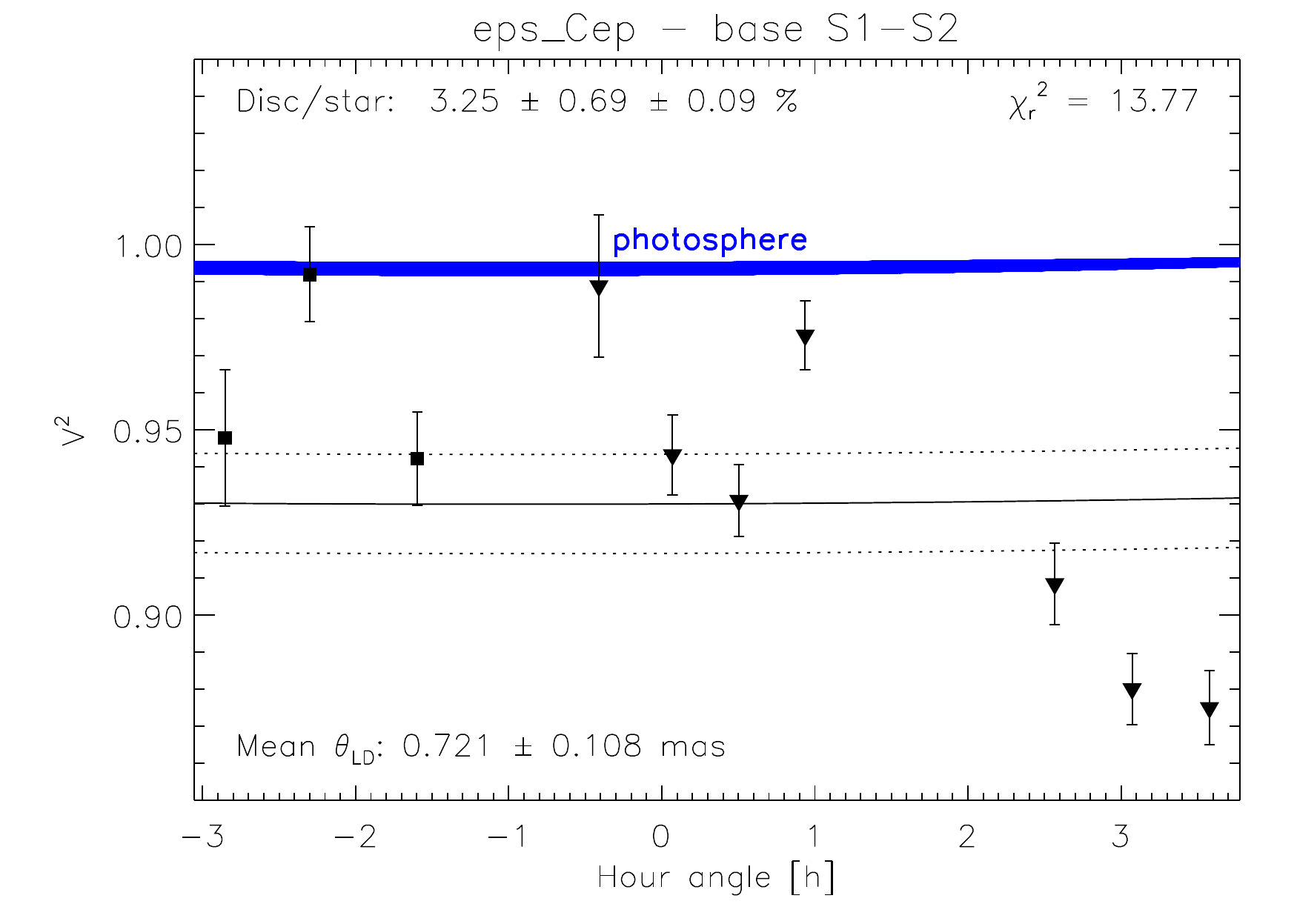}
\end{tabular}
\caption{(Continued)}
\end{figure*}

We estimate the level of circumstellar emission within the FLUOR field-of-view by fitting to each data set a model consisting of an oblate limb-darkened photosphere surrounded by a uniform emission filling the whole field-of-view.\footnote{For the sake of conciseness, we use the term ``disc'' to refer to this uniform emission in the rest of the paper.} The only free parameter in this fit is the disc/star flux ratio (considering disc flux within the field-of-view only), while the photospheric diameter and oblateness are based on Table~\ref{tab:sample} and the linear limb-darkening coefficient is taken from the tables of \citet{Claret00}.\footnote{except in the case of Vega, for which we take the contribution of gravity darkening into account in the linear limb-darkening coefficient \citep[see][for details]{Absil08}.} Although they are non-physical and can result in visibilities greater than one, negative values of the disc/star flux ratio were accepted and kept because their consideration is crucial in the statistical analysis. The results of the fitting procedure are given in Table~\ref{tab:result} and illustrated in Fig.~\ref{fig:result}, where squared visibilities are represented by different symbols depending on the year during which they were obtained. The stars already discussed in \citetalias{DiFolco07} and \citetalias{Absil08} are not included in Fig.~\ref{fig:result}, because the results have not changed for these stars. Conversely, the three stars discussed in \citet{Akeson09} have been further observed and/or re-analysed, so are included in Fig.~\ref{fig:result}. In each plot, the expected visibility of the photosphere is represented by a blue shaded region, whose width takes both the uncertainty on the median photospheric diameter and on the position angle of the oblate apparent photosphere into account. We chose to display the squared visibilities as a function of hour angle in Fig.~\ref{fig:result} for two main reasons: (i) the projected baseline length generally does not vary much on the north-south-oriented S1--S2 baseline of the CHARA array, and (ii) this allows easier diagnostics of possible binaries in our sample. Indeed, a (sub-)stellar companion within the field-of-view is expected to produce periodic variations in the squared visibility as a function of hour angle, on top of a global drop in the visibility level. Although the presence of a faint companion can generally not be rejected based on our interferometric data alone \citepalias[see][]{Absil08}, a background object producing $\sim 1\%$ of the photospheric flux within the $0\farcs8$ FLUOR field-of-view is highly unlikely around any of our target stars, as discussed in \citetalias{DiFolco07} (where the probability is shown to be less than $10^{-6}$ in the cases of tau~Cet and eps~Eri). We are therefore left with bound companions as the only possible origin of point-like source(s) within the FLUOR field-of-view.

	\subsection{Error bars and detection threshold} \label{sec:errors}

The final error bar $\sigma_f$ on the disc/star flux ratio is computed as the root square sum of two independent contributions: (i) the statistical error bar related to the statistical dispersion of the data points and (ii) the systematic error bar related to the uncertainty on the stellar model and on the calibrator diameters. The respective contributions of the two error bars are given as an inset in the plots of Fig.~\ref{fig:result}. The statistical error bar is directly taken from the outputs of the MPFIT routine in the IDL interpreted language \citep{Markwardt09}. In the cases where the reduced chi square ($\chi^2_r$) of the fit is larger than 1, we have considered that the statistical error bars on the individual data points could have been underestimated by the DRS. To prevent a possible underestimation of the final error bar on the star/disc contrast, the individual error bars were rescaled so that $\chi^2_r=1$. Our final error bars on the disc/star contrast can therefore be regarded as pessimistic. We note that most of the $\chi^2_r$ in our fits range between reasonable limits of $\sim0.3$ and $\sim3$. Only two stars show significantly higher $\chi^2_r$, of 5.5 (bet Leo) and 13.8 (eps Cep). Four other stars show very low $\chi^2_r$, 0.11 for eta Cas A and gam Ser, 0.13 for HD~69830, and 0.21 for ksi Boo. The reasons for these extreme values are discussed in Sect.~\ref{sec:individuals}.

\begin{figure}[t]
\centering
\includegraphics[width=\linewidth]{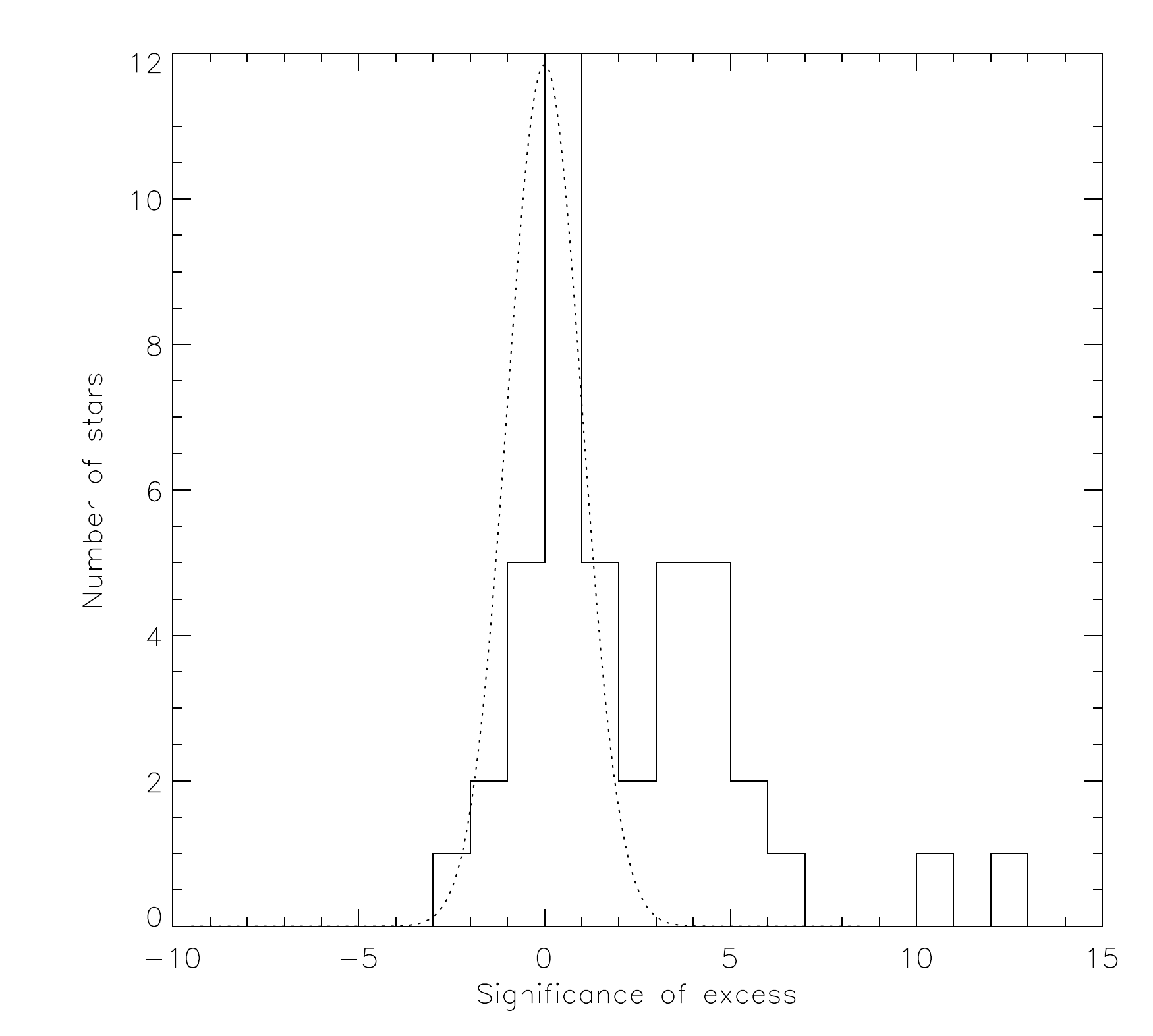}
\caption{Histogram of the excess significance ($\chi_f=f_{\rm CSE}/\sigma_f$) in our sample, compared to a centred Gaussian distribution (dotted line).} \label{fig:histo}
\end{figure}

In Fig.~\ref{fig:histo}, we plot the histogram of the significance level $\chi_f=f_{\rm CSE}/\sigma_f$ for all our excess measurements and compare it to a normal Gaussian distribution. This figure suggests that our measurements follow a bimodal distribution, with a set of non-detections more or less following a centred Gaussian distribution with a standard deviation of 1, and a set of detections located above $\chi_f=3$. A slight offset in the peak of the no-excess distribution (mean significance of $0.58\sigma$) suggests that an underlying population of non-significant excesses could be present. Alternately, the offset could be related to a small positive bias in our measurements. The most negative significance level in our sample is precisely $\chi_f=-3.0$, which confirms that events located outside the $[-3,3]$ interval most probably come from a true astrophysical signal rather than noise. A standard $3\sigma$ detection threshold is therefore considered as appropriate for our sample, corresponding to a probability of 0.27\% to be drawn from a Gaussian distribution with zero mean. To take the possible positive bias in our measurements into account, we add $0.58\sigma$ to the standard $3\sigma$ detection threshold, resulting in a formal threshold at $3.58\sigma$. Events with significance levels between 3 and 4 will generally be treated with care, on an individual basis.

	\subsection{Discussion of individual targets} \label{sec:individuals}

Here, we discuss some interesting or peculiar targets individually in a more detailed way. Among others, a short discussion is given for all the significant excess detections, including the possible presence of a faint point-like companion within the FLUOR field-of-view as the source of the measured excess. Ancillary data obtained as follow-up of excess detections are also presented here.

\paragraph{eta Cas A (HD 4614).} The fit to our data set yields a very low $\chi^2_r$ ($=0.11$) in this case, for which we have no convincing explanation. The most probable cause is a combination of overestimated error bars and chance alignment of these high-quality data points (obtained under excellent seeing conditions) along the best-fit model.

\paragraph{ups And (HD 9826).} The measured disc/star flux ratio ($0.53\pm0.17\%$) is at the limit of significance ($\chi_f=3.0$). The level of circumstellar excess is significantly less than most of the other detections, which have flux ratios generally around 1\% or larger. Taking the slight shift in the distribution of non-excess stars with respect to $\chi_f=0$ into account, we therefore consider this as a non-detection, even though the measured excess seems robust to DRS parameters, and the data set is of good quality (despite the relatively large $\chi^2_r$).

\paragraph{tau Cet (HD 10700).} This detection was first published in \citetalias{DiFolco07}. Since then, we have confirmed that the excess emission is not related to the presence of a close companion, using closure phase measurements obtained with the VLTI/PIONIER instrument \citep{Absil11}. The K-band excess must therefore be associated with an extended source around the star.

\paragraph{10 Tau (HD 22484).} This is one of our most significant detections ($\chi_f=11$), thanks to an excellent data quality that allowed us to reduce the uncertainty on the disc/star flux ratio down to $0.11\%$. The absence of large fluctuations in the squared visibilities, combined with the absence of radial velocity variations \citep{Nidever02}, suggest that the measured excess is not associated with a point-like companion. The low residuals and excellent $\chi^2_r$ in the fit of a limb-darkened stellar model to the long-baseline observations of \citet{Boyajian12a} also point towards the absence of a nearby faint companion. A definitive conclusion can, however, not be given concerning the (point-like or extended) nature of the excess.

\paragraph{zet Lep (HD 38678).} This star was suggested by \citet{Akeson09} to have a K-band circumstellar excess based on FLUOR data collected in 2006, although the measured excess (1.5\%) was considered as marginally significant by the authors. The additional data collected in 2008 within the present survey, of better quality than the 2006 data, show that this detection was spurious.

\begin{figure}[t]
\centering
\includegraphics[width=\linewidth]{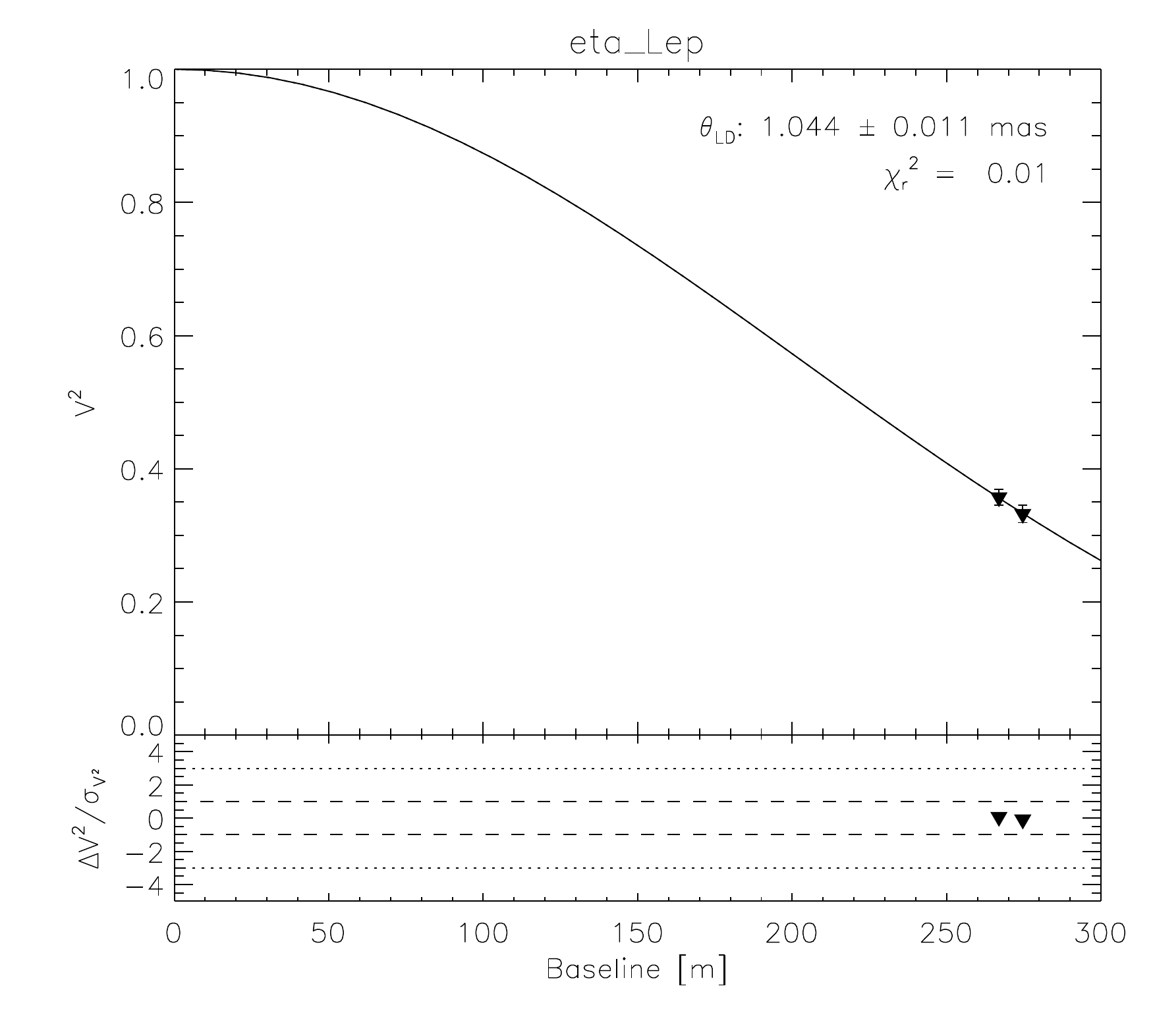}
\caption{Best fit of a limb-darkened photospheric model to the FLUOR data obtained on the CHARA S1-W1 baseline in the case of eta Lep. The residuals of the fit are shown at the bottom.} \label{fig:etaLep}
\end{figure}

\paragraph{eta Lep (HD 40136).} This star shows a K-band excess of $0.89\pm0.21\%$. This is the only excess detection around a single main sequence star for which a published direct diameter measurement was not available. To confirm the validity of our SBR diameter, we obtained additional long-baseline observations with the FLUOR interferometer. Two OBs were obtained on the long CHARA S1-W1 baseline (278\,m) on 28 October 2008 under excellent seeing conditions. Calibration was performed with interleaved observations of K giants from the catalogue of \citet{Merand05}. A limb-darkened photosphere was then fitted to the data, using the linear limb-darkening coefficient found in \citet{Claret00}. The result is shown in Fig.~\ref{fig:etaLep}, with a best-fit diameter $\theta_{\rm LD}= 1.044\pm0.011$\,mas. This diameter was then used as a reference for the analysis of short-baseline data. The short-baseline data are of good quality and do not show any evident sign of a point-like companion. Combined with the radial velocity stability of this star \citep{Lagrange09}, we conclude that the detected excess most probably originates in an extended circumstellar source, although no definitive conclusion can be given regarding the (point-like or extended) nature of the excess.

\paragraph{lam Gem (HD 56537).} This star, showing a $4.3\sigma$ detection ($f_{\rm CSE} = 0.74\pm0.17\%$), is somewhat peculiar, because it is one of the very few stars in our sample where the angular diameter based on surface-brightness relationships ($0.65\pm0.08$~mas) largely differs from the measured angular diameter based on long-baseline interferometric observations \citep[$\theta_{\rm LD}=0.835\pm0.013$~mas,][]{Boyajian12a}---although this difference is within $3\sigma$. In the context of our survey, we performed additional long-baseline visibility measurements on 28 October 2008 using the FLUOR beam-combiner on one of the longest CHARA baselines (S1-W1, 278~m). The three OBs obtained on this baseline yield an angular diameter $\theta_{\rm LD}=0.807\pm0.018$~mas, which confirms the \citet{Boyajian12a} diameter within error bars. In our circumstellar excess analysis presented above (Fig.~\ref{fig:result} and Table~\ref{tab:result}), we used the \citet{Boyajian12a} diameter. Using the SBR diameter would increase the detected excess to $0.91\pm0.18\%$ (within $1\sigma$ of our final result). Another peculiarity is that, based on lunar occultations \citep{Dunham77}, this star was suggested to have a faint companion, which would be much closer than the faint visual companion imaged at $9\farcs7$ \citep{DeRosa11}. This close-in companion was not confirmed by some other high-angular resolution studies \citep{Richichi96,Tokovinin10,Moerchen10}. The long-baseline data obtained by \citet{Boyajian12a} and in the present paper do not support the binary hypothesis either, as in both cases the visibilities seem to nicely follow a single limb-darkened disc model. Furthermore, radial velocity observations of lam Gem revealed the absence of companions at short orbital periods down to a few Jupiter masses \citep{Lagrange09}. We therefore conclude that the significant K-band excess detected in our data is most probably associated with an extended circumstellar structure.

\paragraph{HD 69830.} The low $\chi^2_r$ ($=0.13$) in the fit of this data set is due to the very small amount of data points and to the large error bars on individual points, which are directly related to the faintness of the source.

\paragraph{del Leo (HD 97603).} This target was originally used as a check star in the study of \citet{Akeson09}. No additional data was obtained on this target within our survey, so that we only use here the three already published visibilities to assess the disc/star flux ratio. The poor quality of this data set leads to a very large uncertainty on the circumstellar emission ($\sigma_f = 0.77\%$). We note that the limb-darkened diameters published by \citet{Akeson09} and \citet{Boyajian12a} are not at all in agreement for this target ($1.166\pm0.022$~mas versus $1.328\pm0.009$~mas, see Sect.~\ref{sec:method}). This discrepancy could be (partly) due to the non-negligible oblateness of the photosphere ($\rho=1.11$). In our analysis, we used the SBR diameter ($1.223\pm0.017$~mas), which falls in between the two interferometric measurements. Our model takes into account the unknown position angle of the photosphere on the sky, which explains the width of the photospheric model in Fig.~\ref{fig:result} for this star. Choosing one of the two interferometric diameter estimations would not have changed the final conclusion concerning this target, i.e., the absence of significant circumstellar excess emission.

\paragraph{bet Leo (HD 102647).} This target was already shown by \citet{Akeson09} to have a significant K-band excess, based on 2006 FLUOR data. We have expanded this data set, with additional data collected in 2009. A fit to the entire data set confirms the detection, although with a reduced magnitude of the excess and a significance just above our formal detection threshold of $3.58\sigma$. Our fit also features a large $\chi^2_r$ ($=5.5$) for a reason that is not well understood. Individual data points seem indeed of good quality, although the 2009 data set (represented by squares in Fig.~\ref{fig:result}) is affected by fluctuations in the photometric balance, which could lead to an underestimation of the error bars in the DRS. Nevertheless, the final result is robust to DRS parameters and the significance well above 3, so that we consider the excess detection to be significant. Because the variability seen in the squared visibility could be associated with a point-like companion, we performed dedicated aperture masking observations with the 18-hole mask within the NIRC2 camera on the Keck-II telescope. Observations were obtained in the K band on 1 June 2009 behind the Keck AO system \citep[see][for details about AO-assisted aperture masking]{Tuthill06}. No closure phase signal was detected in these observations, and we derived upper limits on the contrast of possible companions using the same method as in \citet{Lacour11}. The results are summarised in Table~\ref{tab:betleo}, and show that companions with flux ratios larger than $1\%$ cannot be located in the 20--320\,mas region. At larger angular separation, adaptive optics imaging has shown the absence of companions around bet Leo with a dynamic range of at least 5 magnitudes on the companion contrast at K band \citep{Leconte10,DeRosa11}. Combined with the absence of radial velocity variations \citep{Lagrange09} to cover shorter angular separations, these results strongly suggest that the circumstellar excess is associated with an extended source. Further interferometric observations are needed to assess whether the high $\chi^2_r$ could be due to variability in the excess.

\begin{table}[!t]
\caption{Upper limit on the contrast of companions around bet Leo expressed in terms of K-band magnitude, based on aperture masking observations with Keck/NIRC2.}
\label{tab:betleo}
\centering
\begin{tabular}{c c}
\hline\hline
Separation (mas) & $\Delta K$ \\
\hline
10--20 & 3.20 \\
20--40 & 5.40 \\
40--80 & 5.78 \\
80--160 & 5.67 \\
160--240 & 5.51 \\
240--320 & 5.19 \\
\hline
\end{tabular}
\end{table}

\paragraph{ksi Boo (HD 131156).} This target shows a relatively small $\chi_r^2$, which points towards an overestimation of the error bars. The significance of the detected excess ($\chi_f=3.7$, one of the smallest among the ``significant excess'' category) can therefore be considered as pessimistic. We also note that the measured excess is robust with respect to DRS parameters, which gives us a high confidence in its detection. Being a very nearby G-type star, ksi Boo has been searched for the presence of companions by various radial velocity survey programs over long periods of time \citep[e.g.,][]{Wittenmyer06}, showing the absence of Jupiter-sized companion with orbital periods up to about 10 years. Such period corresponds to an orbital semi-major axis around 5\,AU, i.e., an angular distance of about $0\farcs75$, thereby covering most of the FLUOR field-of-view. Together with the stability of the measured visibilities as a function of hour angle, this suggests that the detected excess is associated with an extended circumstellar emission. The low residuals and excellent $\chi^2_r$ in the fit of a limb-darkened stellar model to the long-baseline observations of \citet{Boyajian12a} also point towards the absence of a nearby faint companion. A definitive conclusion can, however, not be given concerning the (point-like or extended) nature of the excess.

\paragraph{kap CrB (HD 142091).} This is one of the six stars in our sample with a luminosity class IV (i.e., classified as a sub-giant), but the only one to have significantly moved away from the main sequence track. Indeed, with its mass of $1.8M_{\odot}$, kap CrB was originally an A-type star \citep{Johnson08}. Its evolutionary status has almost reached the base of the red giant branch, although its linear diameter is only about 5 time larger than the Sun's. It is orbited by a Jupiter-sized planet at 2.6\,AU and most probably by another low-mass companion on a longer orbital period, whose mass and orbital parameters are not precisely known \citep{Bonsor13a}. Here, we detect a significant K-band excess of $1.18\% \pm 0.19\%$, which we associate most likely with an extended emission owing to the absence of fluctuations in the measured visibilities. The possibility that the K-band excess is associated with a low-mass companion located within the FLUOR field-of-view can, however, not be excluded, since the high-contrast adaptive-optics images obtained by \citet{Bonsor13a} do not cover the region within $0\farcs3$. Because kap CrB is close to the red giant branch, our FLUOR observations could trace different dust populations or physical phenomena than for the rest of our sample. Therefore, we will exclude kap CrB from the statistical study of Sect.~\ref{sec:stat} (which focuses specifically on main sequence stars).

\paragraph{gam Ser (HD 142860).} As for HD~69830, the very low $\chi^2_r$ ($=0.11$) obtained when fitting this data set is due to the small number of data points and to the large error bars.

\paragraph{mu Her (HD 161797).} Although the detection has a formal significance $\chi_f=3.1$, this target will be classified in the non-detection category for two reasons. First, the data are mostly of mediocre quality, and show a relatively large $\chi^2_r$. Second, this is one of the rare stars for which the final result depends significantly on the DRS parameters and on OB selection process. Indeed, small changes in the standard DRS parameters will often lead to significances below $3\sigma$ for the excess. This behaviour can be related to the fact that the best fit is mostly constrained by the two right-most data points in the plot (see Fig.~\ref{fig:result}), which are not in good agreement with each other.

\paragraph{alf Lyr (HD 172167).} The detection of a near-infrared circumstellar excess around Vega was first presented by \citet{Absil06}, and then confirmed at H band by \citet{Defrere11}. In these papers, it was argued that the origin of the excess is most probably associated with an extended emission. This is now backed up by the absence of closure phase signal in the MIRC data set obtained by \citet{Monnier12} at the CHARA array.

\paragraph{110 Her (HD 173667).} A circumstellar emission is detected at $3.8\sigma$ around this star. The data quality is good and the presence of the excess is robust to the DRS parameters, so that we are confident in adding this star to the near-infrared excess category. The absence of large fluctuations in the squared visibilities, combined with the absence of radial velocity variations \citep{Nidever02}, suggest that the measured excess is not associated with a point-like companion. The low residuals and excellent $\chi^2_r$ in the fit of a limb-darkened stellar model to the long-baseline observations of \citet{Boyajian12a} also point towards the absence of a nearby faint companion. A definitive conclusion can, however, not be given concerning the (point-like or extended) nature of the excess.

\begin{figure}[t]
\centering
\includegraphics[width=\linewidth]{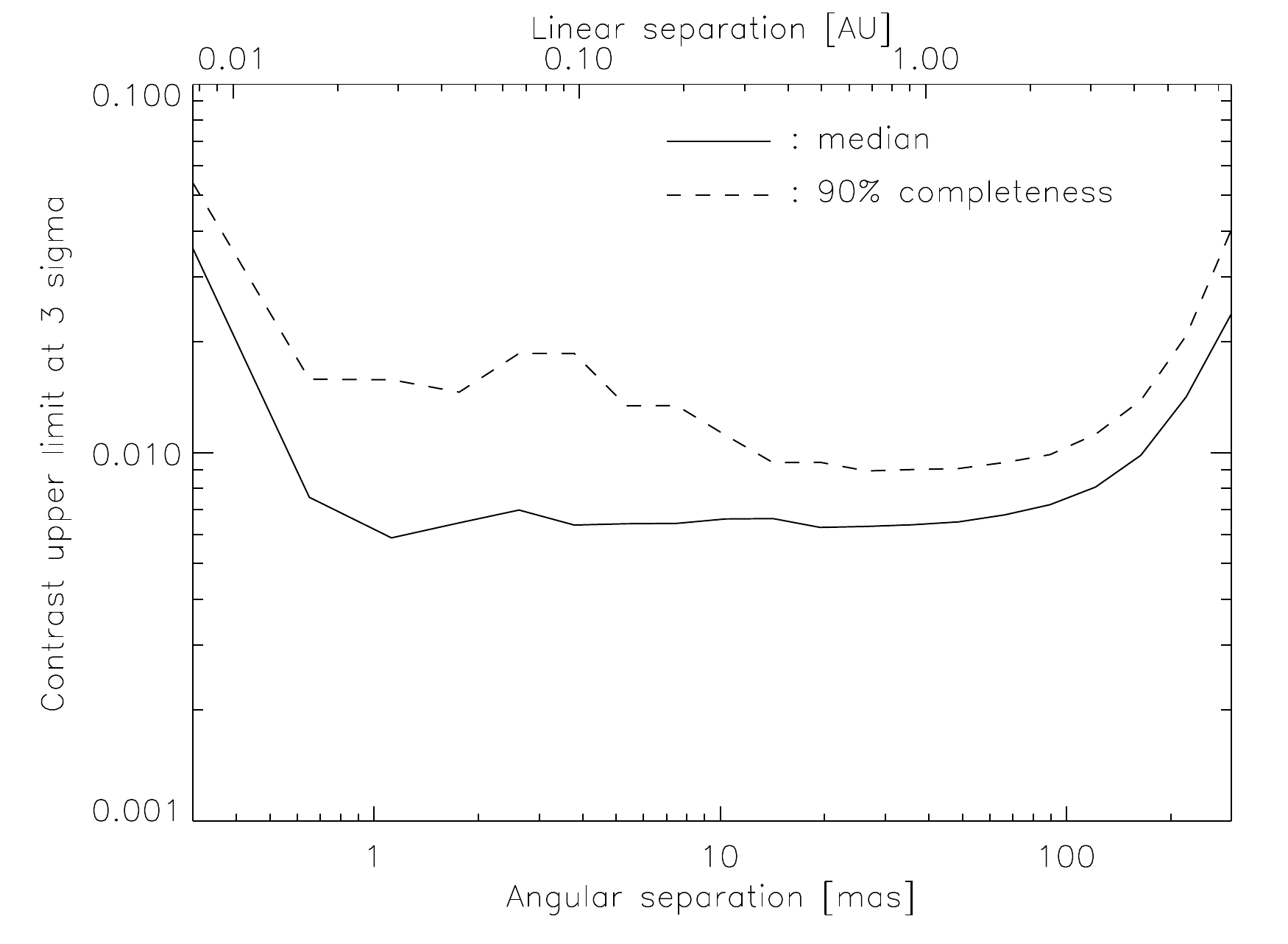}
\caption{Sensitivity of our CHARA/MIRC closure phases to faint companions around zet Aql, expressed in terms of upper limit on the companion contrast. The median sensitivity is depicted with a solid line, while a dashed line shows the sensitivity at 90\% completeness.} \label{fig:zetAql}
\end{figure}

\paragraph{zet Aql (HD 177724).} The detection of a significant circumstellar excess ($1.69\% \pm 0.31\%$) around zet Aql was already described in \citetalias{Absil08}. In that paper, it was argued that the excess could be due to an M-type main sequence companion. Since then, we obtained closure phase measurements with the MIRC interferometer at the CHARA array. Two observations of zet Aql were obtained on 16 and 17 August 2010, using MIRC on a 4T configuration in the H band with a low spectral dispersion. Calibration was performed with bracketing observations of gam Lyr, sig Cyg, and/or eps Aql. The data were reduced with the standard MIRC pipeline \citep{Monnier07}, making use of simultaneous photometric measurements to enhance data quality \citep{Che10}. We then applied a closure phase analysis similar to the one described by \citet{Absil11}, and found a $3\sigma$ upper limit ranging between 1\% and 2\% (at 90\% completeness) on the contrast of any companion located between 1\,mas and a 200\,mas ($\sim5$\,AU), as illustrated in Fig.~\ref{fig:zetAql}. Beyond this angular separation, the sensitivity to companions degrades significantly owing to the poor coupling of off-axis sources into single-mode fibres. The MIRC observations mostly rule out the presence of a faint companion as the source of the near-infrared excess around zet Aql up to a distance of about 5\,AU, although it cannot be excluded that a faint companion hides in the parts of the search region where the MIRC sensitivity is larger than 1.69\%. We cannot exclude either that the companion colours are extremely red (e.g., contrast lower than 1\% at H band while it is of 1.69\% at K band). These two possibilities are, however, considered as unlikely. It was shown in \citetalias{Absil08} that companions beyond 8\,AU can also be ruled out, thanks to high-contrast adaptive optics observations obtained with CFHT/PUEO. This reduces the possible location of a companion to the 5--8\,AU region. Recent observations with the Keck Interferometer Nuller (KIN) show that a companion accounting for $\ge 1\%$ of the photospheric flux in the mid-infrared is not likely in that region either, although KIN observations are not sensitive in all parts of the field-of-view because of the nulling instrument transmission pattern (B.~Mennesson et al., in prep.). We conclude that the detected circumstellar excess is most likely associated with an extended emission.

\paragraph{alf Aql (HD 187642).} The oblate limb-darkened stellar model that we use for Altair was derived from the works of \citet{Peterson06} and \citet{Monnier07}. It does not take into account the details of the intensity map derived by means of interferometric image reconstruction in \citet{Monnier07}, but is sufficiently realistic in order not to significantly bias the estimation of the disc/star flux ratio at short baselines. The observations show a very significant drop in visibility, leading to an estimated disc/star flux ratio of $3.07\% \pm 0.24\%$, the largest in our sample. We are confident that this circumstellar excess is not related to a faint companion, because its signature would have prominently shown up in the closure phases obtained at long baselines by \citet{Peterson06} and \citet{Monnier07}. Indeed, when the visibility is close to zero (as was the case in these studies), the effect of a faint off-axis companion is strongly enhanced in the closure phases \citep{Chelli09}. The excess emission is thus most probably associated with an extended circumstellar source. Assessing whether the detected K-band excess is related to the marginal detection of a mid-infrared excess using the Keck Interferometer Nuller \citep{MillanGabet11} is beyond the scope of the present paper.

\paragraph{61 Cyg A\&B (HD 201091, HD 201092).} The data used here were first presented in \citet{Kervella08}, where the circumstellar environment was not discussed. We re-reduced and re-calibrated this data set to reach the best possible accuracy and robustness on the disc/star flux ratio. No excess emission is found around either star.

\paragraph{alf Cep (HD 203280).} The oblate limb-darkened photospheric model used for this star is derived from the interferometric observations and modelling of \citet{Zhao09}. The excess emission at K band found around this star is most probably related to an extended circumstellar emission, since, following the same argument as for alf Aql, no hint for a faint companion was found in the long-baseline closure phase measurements of \citet{Zhao09}.

\paragraph{eps Cep (HD 211336).} This target features the largest $\chi^2_r$ ($=13.8$) in our survey and displays significant fluctuations in its squared visibilities. The shape of these fluctuations suggests that the circumstellar excess could be due to a point-like source rather than an extended disc. We followed-up this target with single-pupil coronagraphic imaging, which confirmed there is of a stellar companion \citep{Mawet11}. While fitting a detailed binary model to our data set is beyond the scope of this paper, we note that the estimated circumstellar excess ($3.25\pm0.69\%$) is within $2\sigma$ from the flux ratio measured in the coronagraphic images ($2.0\pm0.5\%$, also at K band), which suggests that the companion is the sole responsible for the visibility drop. Due to its binary nature, we remove eps Cep from our statistical sample in Sect.~\ref{sec:stat}.

\section{Statistical analysis} \label{sec:stat}

\begin{table}[!t]
\caption{Summary of the detected K-band excesses in the final sample, from which eps Cep and kap CrB have been removed (see Sect.~\ref{sec:individuals}).}
\label{tab:stat}
\centering
\begin{tabular}{c c c c c}
\hline\hline
 & A & F & G-K & Total\\
\hline
Outer reservoir & 2/7 & 3/7 & 1/4 & 6/18 \\
No outer reservoir & 4/5 & 0/7 & 1/10 & 5/22 \\
\hline
Total & 6/12 & 3/14 & 2/14 & 11/40 \\
\hline
\end{tabular}
\end{table}

Considering the above discussion, our final statistical sample consists in 40 stars (after removing kap CrB and eps Cep), of which 11 have a significant K-band excess emission (see Table~\ref{tab:stat}), resulting in an overall occurrence rate of $28^{+8}_{-6}\%$. The (asymmetric) statistical uncertainty on the occurrence rate results from a numerical integration of the binomial distribution, as described in the Appendix of \citet{Burgasser03}. Out of these eleven excesses, we are confident that at least seven are associated with an extended circumstellar emission. The remaining four might be related to faint companions, although we argue in Sect.~\ref{sec:individuals} that this scenario is less likely than the extended emission scenario. The nature of the extended circumstellar emission cannot be derived directly from our interferometric data. It can only be addressed by more high-angular resolution observations and by dedicated radiative transfer modelling \citep[see e.g.,][]{Lebreton13}. It is the purpose of the present study to attack this problem from another point of view, that is, by means of a statistical analysis. By examining the possible correlations between resolved near-infrared excesses and various stellar parameters, including the presence of outer dust reservoirs, we aim to shed new light on the origin of these excesses. We do not include the stars studied in the same way by other interferometers, such as alf PsA \citep{Absil09} and bet Pic \citep{Defrere12}. Their case will be briefly discussed in Sect.~\ref{sec:discussion} in light of the statistical trends found in the FLUOR sample.

\begin{figure}[t]
\centering
\includegraphics[width=\linewidth]{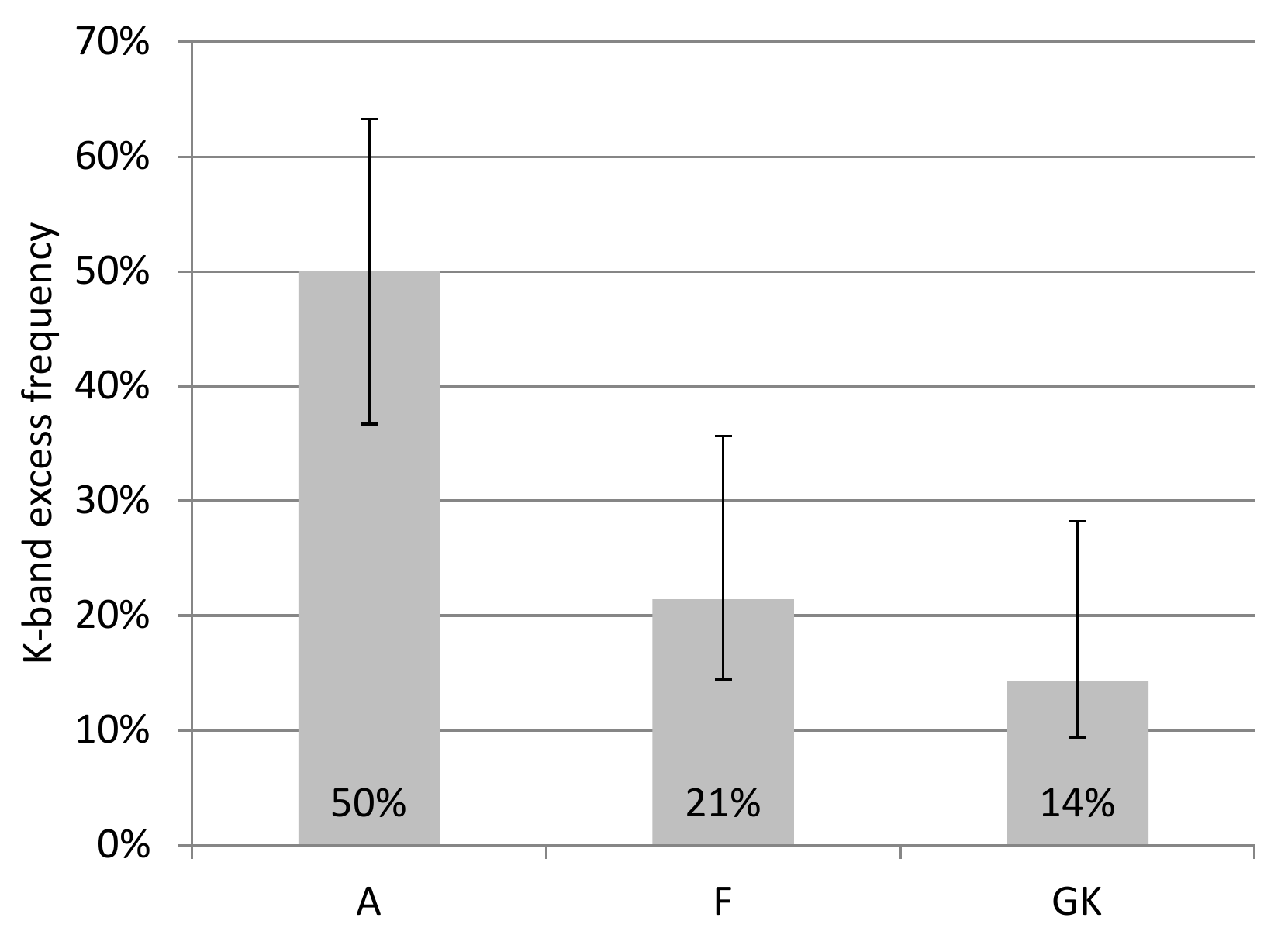}
\caption{Occurrence rate of significant K-band excess emission as a function of the spectral type of the target stars.} \label{fig:sptype}
\end{figure}

\begin{figure}[t]
\centering
\includegraphics[width=\linewidth]{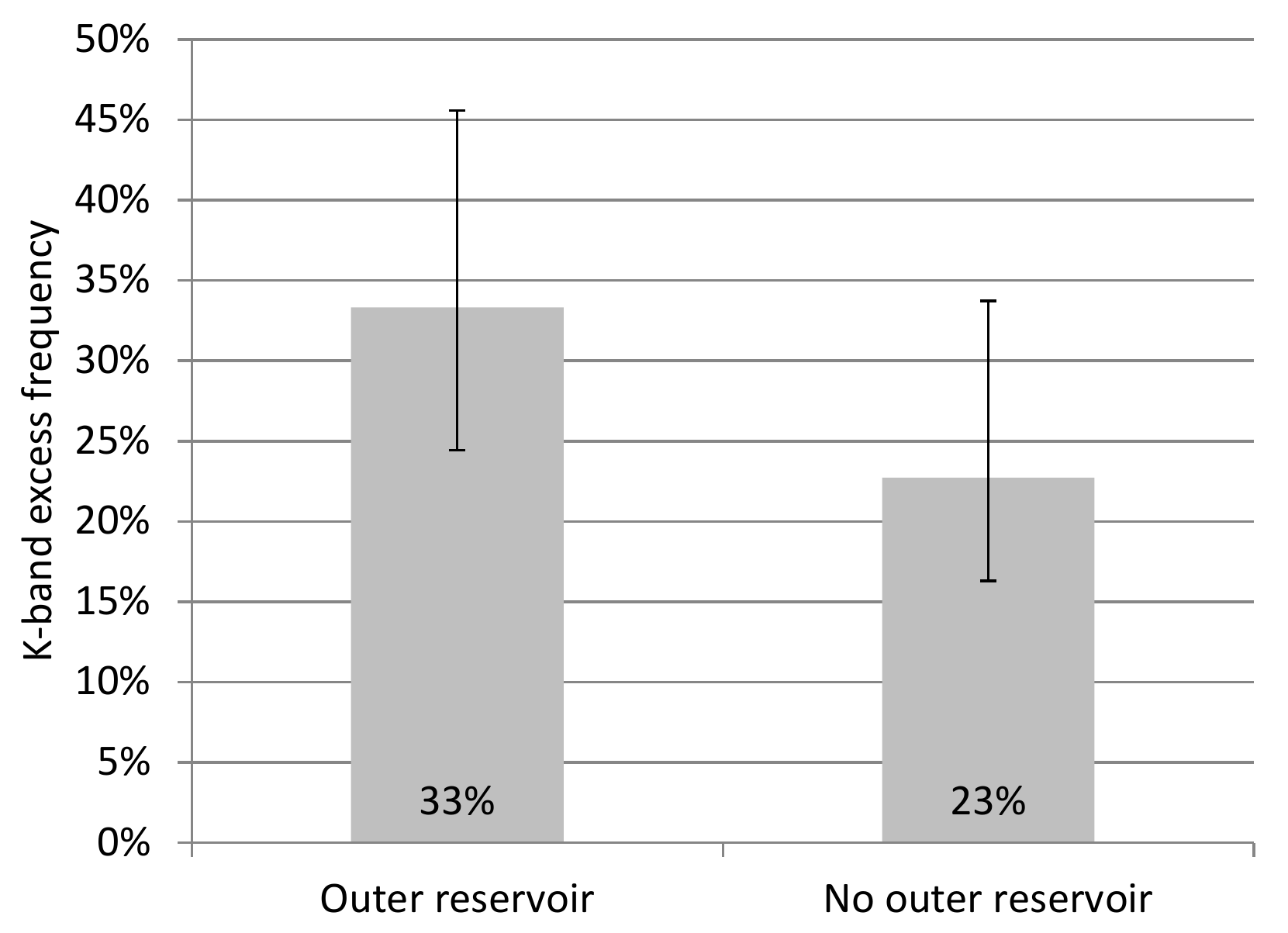}
\caption{Same as Fig.~\ref{fig:sptype} as a function of the presence of outer dust reservoirs around the target stars.} \label{fig:dust}
\end{figure}

\begin{figure}[t]
\centering
\includegraphics[width=\linewidth]{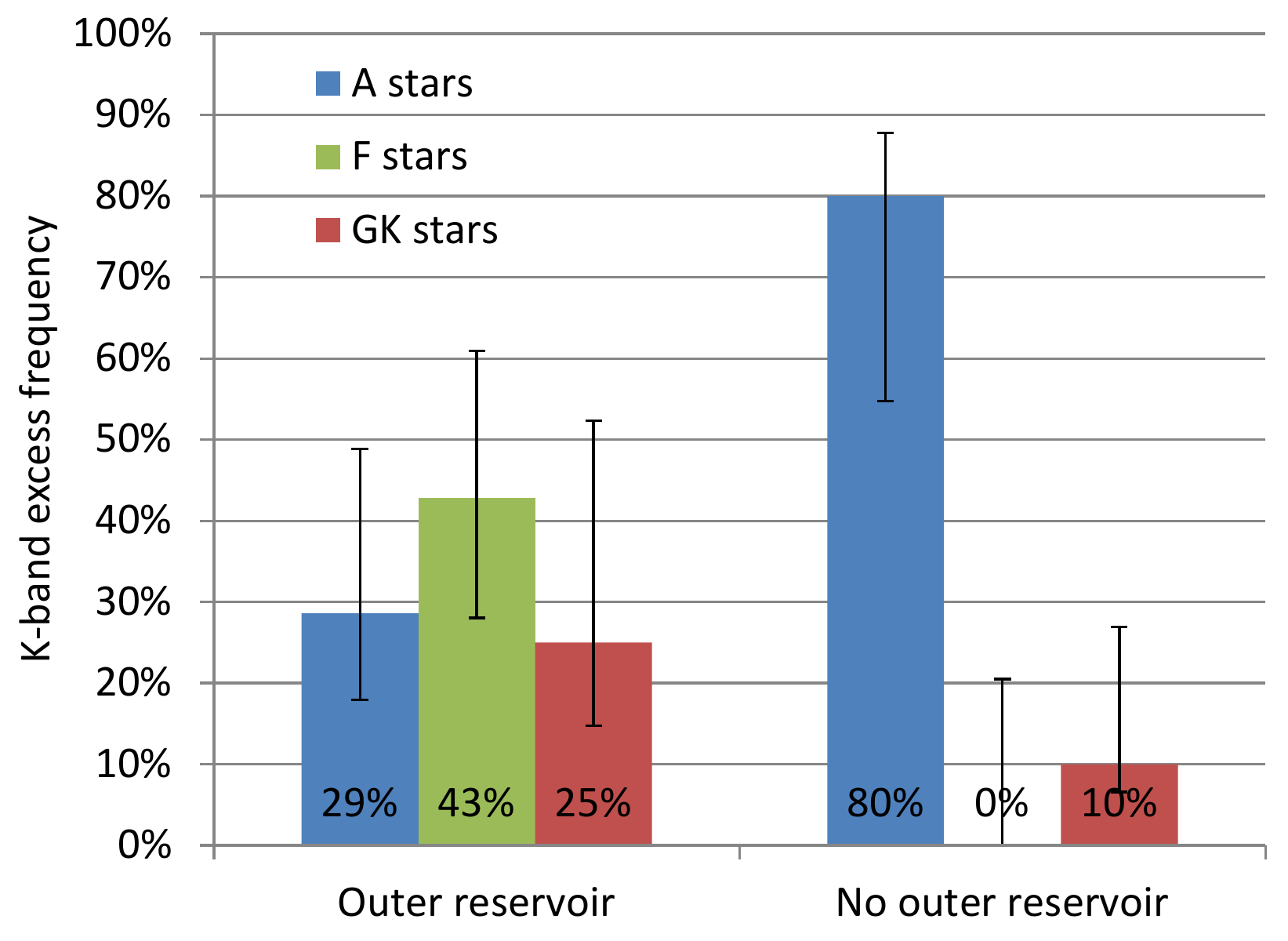}
\caption{Same as Fig.~\ref{fig:sptype} after separating the target stars based on both their spectral type and the presence of outer dust reservoirs.} \label{fig:dusttype}
\end{figure}

In Fig.~\ref{fig:sptype}, we present the K-band excess detection frequency as a function of the spectral type of the target stars. An occurrence rate of $50^{+13}_{-13}\%$ is found around A-type stars, while FGK-type stars (referred to as solar-type stars in the following) have an occurrence rate of $18^{+9}_{-5}\%$. The probability (or p-value) that these two samples are taken from the same population is only 0.037 based on a $\chi^2$ test, so that this difference in occurrence rate can be considered as significant. Figure~\ref{fig:dust} shows the occurrence rate as a function of the presence of outer dust reservoirs around the target stars (to within the sensitivity of previous mid- to far-infrared spectro-photometric surveys). Stars harbouring outer dust reservoirs show more frequently the sign of abundant hot dust, with an occurrence rate of $33^{+12}_{-9}\%$ compared to $23^{+11}_{-6}\%$ for stars without known outer dust reservoirs. This difference is, however, not significant (p-value of 0.45 in the $\chi^2$ test).

A more instructive way to present these two results is proposed in Fig.~\ref{fig:dusttype}, where we separate the target stars according to both spectral type and presence of outer dust reservoirs. This figure shows that stars with outer reservoirs have similar near-infrared excess occurrence rate regardless of their spectral type: $29^{+20}_{-11}\%$ for A-type stars and $36^{+16}_{-11}\%$ for solar-type stars. Stars without known outer dust, on the other hand, have very different behaviours with respect to K-band excess emission depending on their spectral type, with only one K-band excess found around solar-type stars (global occurrence rate of $6^{+11}_{-2}\%$), while $80^{+8}_{-25}\%$ of A-type stars show excess emission. This difference is very significant, with an associated p-value of $5\times10^{-4}$ in the $\chi^2$ test. We note, however, that only five A-type stars without outer reservoirs are present in our sample, so that this result is based on small number statistics and may not be representative of A-type stars in general. Another interesting trend found in Fig.~\ref{fig:dusttype} is the difference in the K-band excess occurrence rate for solar-type stars as a function of the presence of outer dust reservoirs. A p-value of 0.04 is found when comparing the K-band  excess occurrence rate for the two samples, which confirms the significance of this trend.

\begin{figure}[t]
\centering
\includegraphics[width=\linewidth]{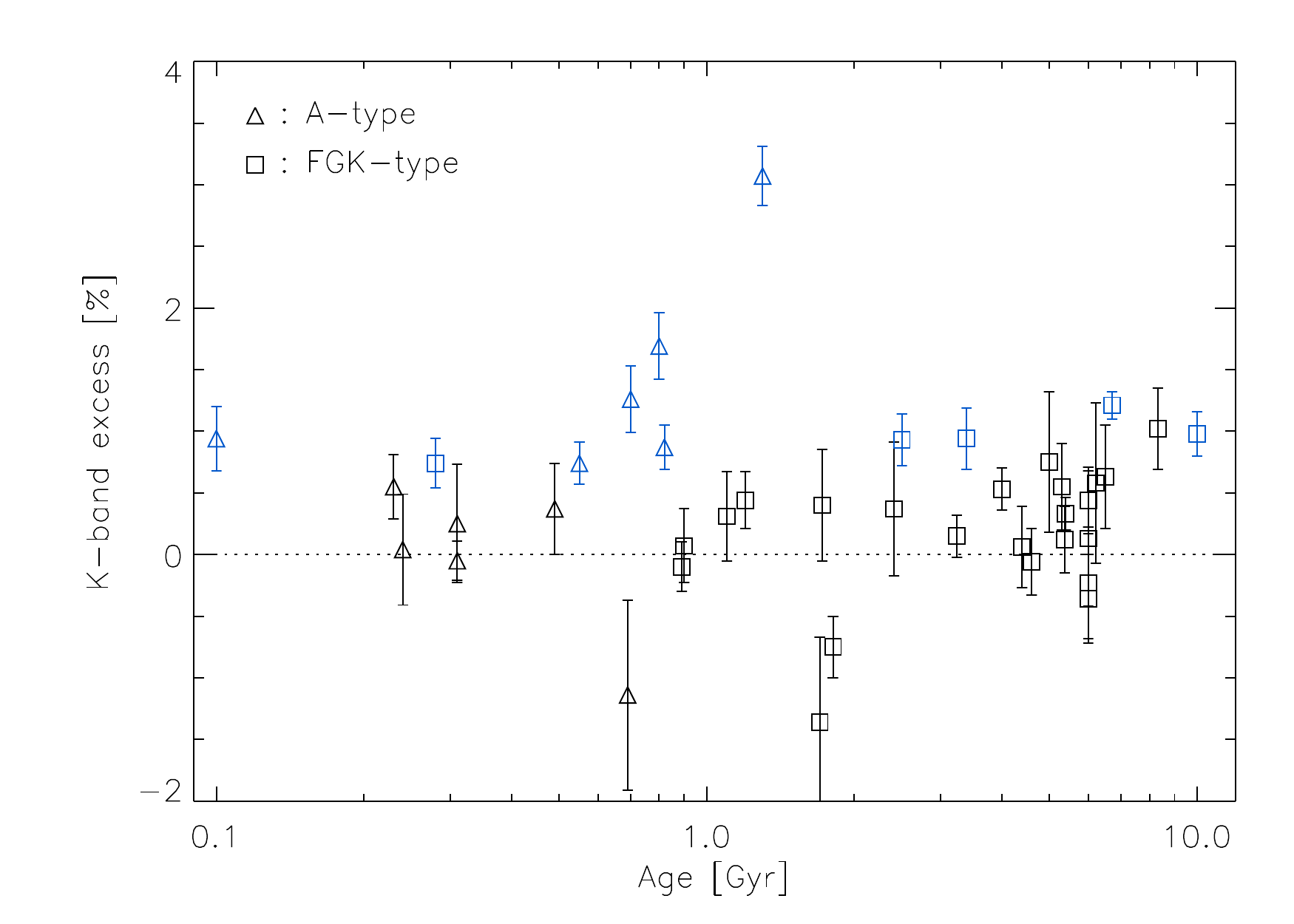}
\caption{Measured K-band disc/star flux ratio as a function of the age of the target stars. The error bars on the ages are generally huge and have been omitted for the sake of clarity. Triangles (resp.\ squares) are used for A-type (resp.\ solar-type) stars. Blue (resp.\ black) symbols are used for stars with (resp.\ without) significant K-band excess.} \label{fig:age}
\end{figure}

\begin{figure}[t]
\centering
\includegraphics[width=\linewidth]{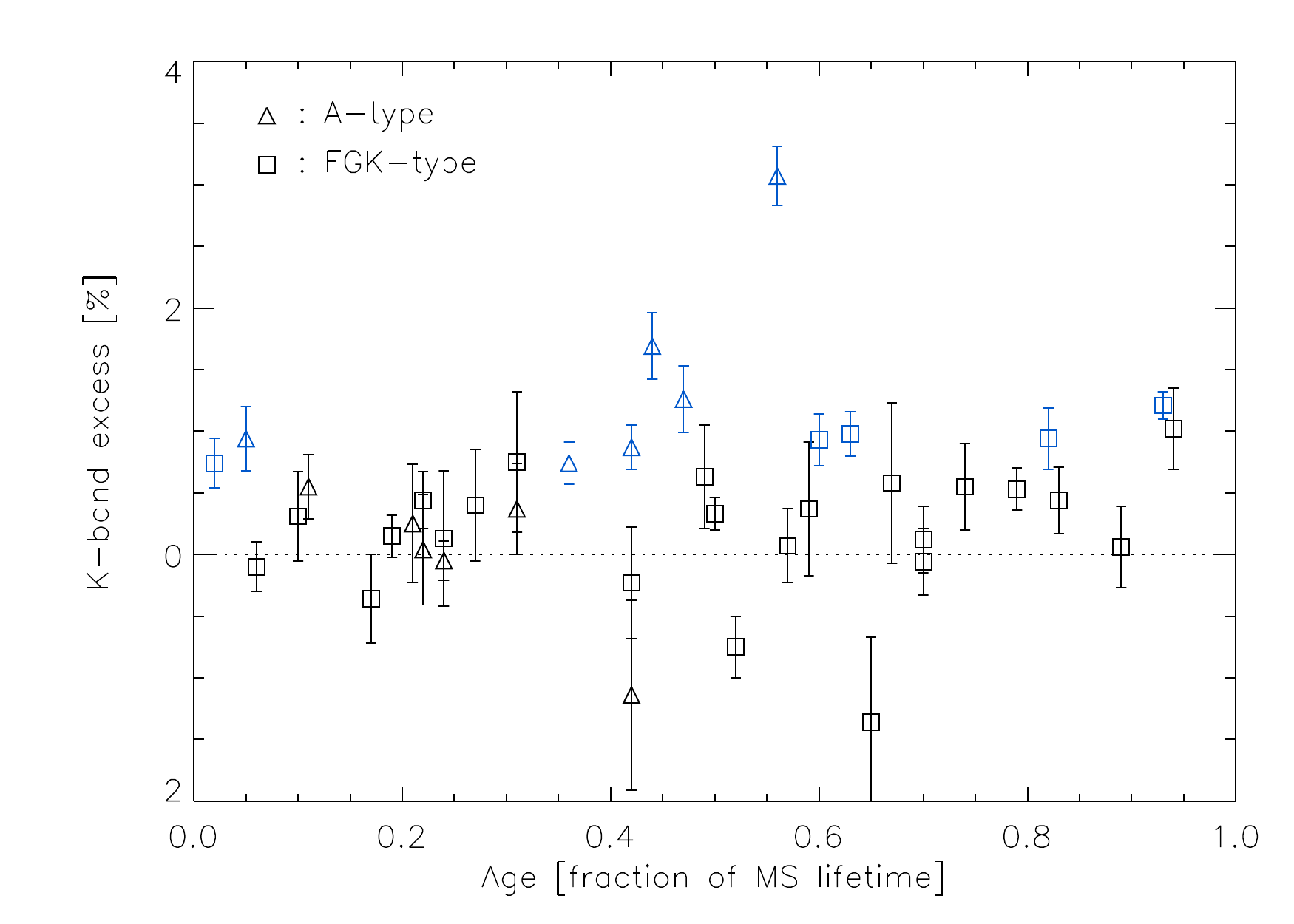}
\caption{Same as Fig.~\ref{fig:age} with the age expressed as a fraction of the main sequence lifetime.} \label{fig:agefrac}
\end{figure}

\begin{figure}[t]
\centering
\includegraphics[width=\linewidth]{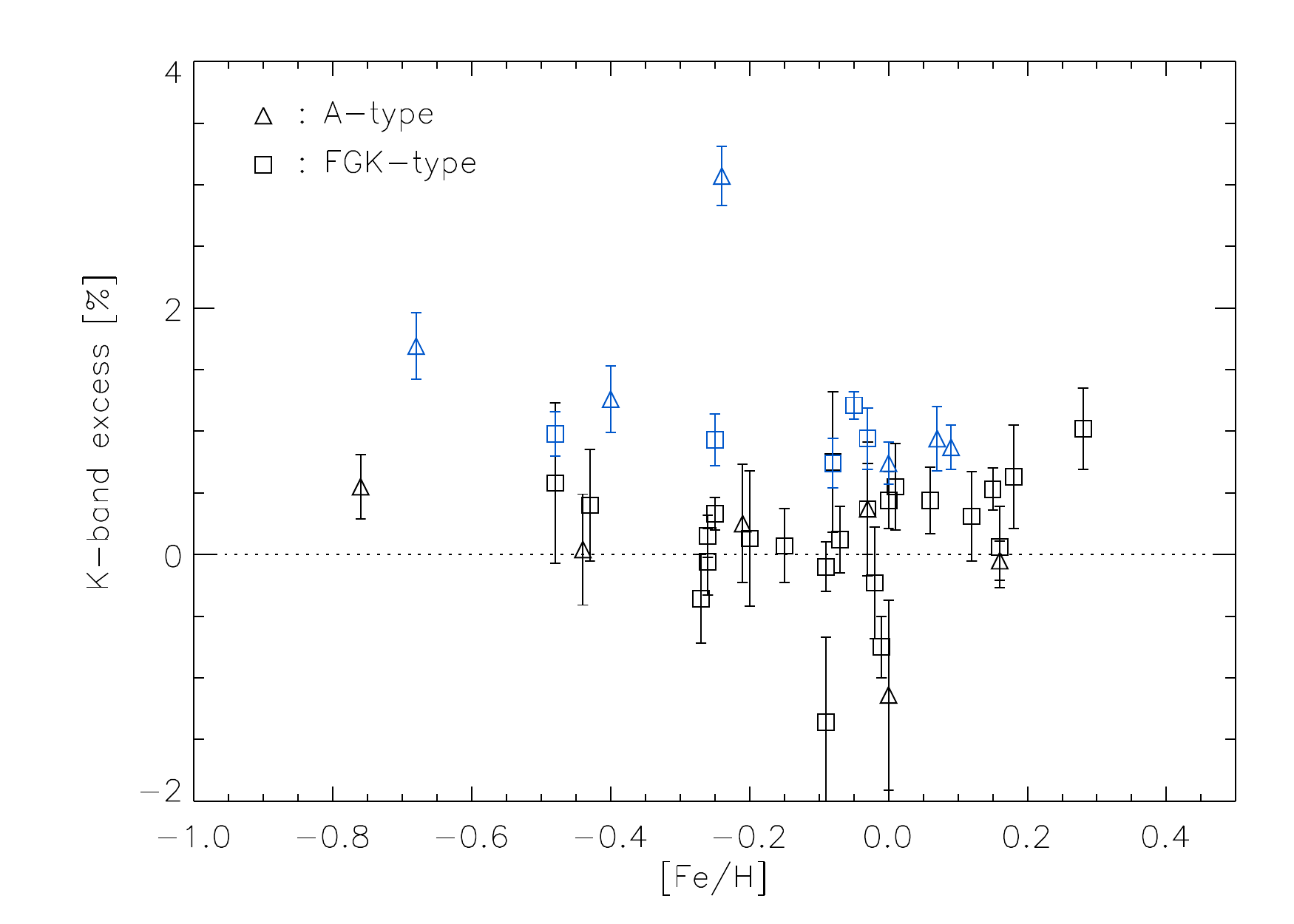}
\caption{Same as Fig.~\ref{fig:age} as a function of stellar metallicity.} \label{fig:metal}
\end{figure}

Correlations with other stellar parameters than just spectral type and the presence of outer dust reservoirs are worth investigating. Stellar age in particular could indicate the time dependence of the K-band excesses. Figure~\ref{fig:age} shows the absence of correlation between stellar age and K-band excess in our sample, although there is a trend for A-type stars to have larger K-band excesses at older ages. A similar trend does not seem to appear in the case of solar-type stars, for which both the youngest (ksi Boo, 280\,Myr) and the oldest (tau Cet, 10 Gyr) targets show a significant K-band excess. To check whether the excesses could be associated with a particular period in the stellar evolution, we plot the measured K-band excess in Fig.~\ref{fig:agefrac} as a function of the evolution of the star on the main sequence, represented by the fractional age relative to the main sequence lifetime. Here again, no significant correlation is found. Finally, another interesting stellar parameter to investigate is metallicity, which is known to be correlated to the presence of high-mass extrasolar planets, but not to low-mass planets \citep[e.g.,][]{Santos01}. Furthermore, metallicity could be slightly correlated to the presence of cold dust \citep{Maldonado12}. We collected metallicities in the literature for most of our stars\footnote{Only lam Gem was found to have no metallicity estimation in the literature.}, using the catalogues of \citet{Soubiran10} and \citet{Gray03,Gray06}. The result is shown in Fig.~\ref{fig:metal} and suggests there is no correlation.

\section{Discussion} \label{sec:discussion}

In this section, we discuss the implications of the statistical trends found above on the nature and origin of the detected K-band excesses. Because A-type and solar-type stars show significantly different behaviour in terms of correlation between cold dust and K-band excess, they are discussed separately.

	\subsection{Solar-type stars}

In the case of solar-type stars, we found in Sect.~\ref{sec:stat} a significant correlation between the detection of a K-band excess and the presence of outer dust reservoir(s). This correlation suggests that K-band excesses could indeed come from debris dust and that they could be directly connected to massive outer reservoirs of dust and planetesimals. Among the five solar-type stars showing significant K-band excess, only ksi Boo has no detectable outer debris disc, at the sensitivity level of \textsc{Spitzer}/MIPS and \textsc{Herschel}/PACS. This star is the only ``young'' solar-type star in our sample, with an estimated age of 280\,Myr \citep{Gaspar13}. Its peculiar behaviour compared to the rest of the solar-type sample might therefore be related to its evolutionary status, although such an assertion needs to be backed up with more observations of young solar-type stars. Among the other four solar-type stars with K-band excess, only eta Lep has a spectro-photometric excess detected at both mid-and far-infrared wavelengths, while the other three only show a far-infrared excess. The outer dust reservoirs that we are facing here are thus generally located far from the inner regions where the K-band excesses originate. Recall that the FLUOR field-of-view is only $0\farcs8$ in radius, which translates to 8\,AU for a star at 10\,pc.

\begin{figure}[t]
\centering
\includegraphics[width=\linewidth]{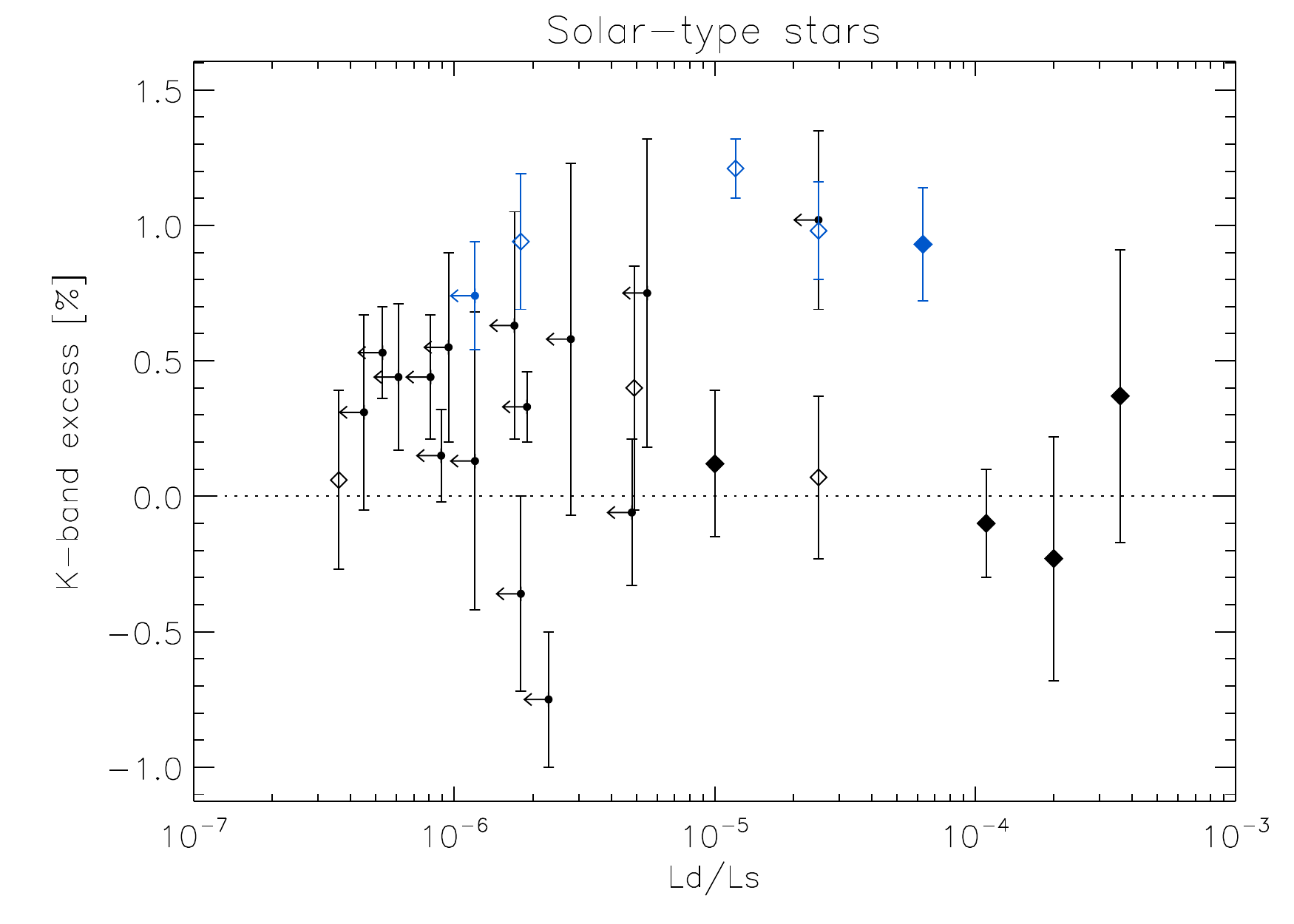}
\caption{Measured K-band disc/star flux ratio as function of the fractional luminosity associated with the outer dust reservoir, in the case of solar-type stars. When no outer reservoir is detected, $3\sigma$ upper limits are used (dots with arrows instead of diamonds for actually measured fractional luminosities). Stars showing a far-infrared excess only are represented by empty diamond, while filled diamonds correspond to stars showing a mid-infrared excess (disregarding the presence or not of a far-infrared excess).} \label{fig:lum_fgk}
\end{figure}

To investigate the inner/outer disc correlation further, we plot in Fig.~\ref{fig:lum_fgk} the measured K-band excesses as a function of the fractional luminosity of the outer dust for our solar-type stars. This figure suggests that the level of K-band emission is not directly related to the amount of dust present in the outer reservoir (provided that outer dust is present of course). This lack of a direct correlation between the luminosity levels of outer and inner dust discs might be real, or could be because the actual fractional luminosity of cold dust around stars with no detected far-infrared excess is much lower than the $3\sigma$ upper limits derived from far-infrared observations. In Fig.~\ref{fig:lum_fgk}, we note another intriguing trend, that stars with mid-infrared spectro-photometric excesses have less frequently significant K-band excesses. Indeed, three of the six stars featuring only a far-infrared excess emission (i.e., $50\pm18\%$) show a K-band excess, while only one star out of five showing mid-infrared excess (i.e., $20^{+25}_{-8}\%$) has a K-band excess. While this trend is far from significant due to the small number statistics, it could indicate that the K-band and mid-infrared excesses are actually two different manifestations of a common phenomenon, which relates to the outer dust reservoir. The correlation between the presence of mid-infrared and far-infrared excess was indeed demonstrated by \citet{Bryden09}, with 2/13 far-infrared excess stars showing a mid-infrared excess, while only 1/133 stars without far-infrared excess show a mid-infrared excess. If real, the trend for a mutual exclusion between near- and mid-infrared excess could be related to the architecture of the inner planetary system, with planets trapping dust and preventing it from going closer to the star in some cases (e.g., eps Eri, HD~69830). It is, however, premature to draw any conclusion on this matter.

Notwithstanding, the correlation between the presence of outer and inner discs remains, and its origin must be investigated. A possibility would be that the presence of inner parent body populations correlates with the presence of outer parent body populations. However, because of the very short lifetime of parent populations at orbital distances of a few AUs, the detected K-band excesses cannot be due to the in-situ steady-state evolution of an asteroid belt \citep{Wyatt07}. Two main possibilities then remain: (i) a transient local event, or (ii) an inward transport of material from the outer disc. 

Witnessing the aftermath of (supposedly rare) major collisions between large planetesimals for 5 out of 28 surveyed solar-type stars would be a likely scenario only if the associated K-band emission level remains high during a significant fraction of the stellar age, i.e., at least hundreds of Myr. According to \citet{Jackson12}, the mid-infrared emission level associated with the Moon-forming collision would remain detectable at most for a few tens of Myr. We expect that the same would hold true for the K-band excess in case of a collision happening closer to the star (so that the dust can be heated up to an appropriate temperature to produce significant K-band emission). Furthermore, such collisions are only expected to happen during the final stages of terrestrial planet formation, that is, at ages of a few hundred Myr at most, which is at odds with the ages of most solar-type stars in our survey. This scenario is therefore deemed unlikely.

Transport mechanisms from the outer towards the inner disc have already been discussed by \citet{Bonsor12}. The authors show that scattering by a chain of tightly packed, low-mass planets could potentially produce a sufficient dust replenishment rate to retain the K-band excess at the observed level of around 1\%. However, this scenario only works for young systems, and the lack of time dependence seen in our survey is a critical piece of evidence against this scenario (although it could potentially be at work for younger systems). Furthermore, the likelihood of having such a tightly packed planetary system around 5 of the 28 surveyed solar-type stars is not considered as very high, since radial velocity measurements have only (tentatively) detected planets around one of these five stars \citep[tau Cet,][]{Tuomi13}. Besides scattering, major dynamical instabilities akin to the solar system's Late Heavy Bombardment may also be responsible for the presence of large amounts of dust in the innermost regions. Based on N-body simulations, \citet{Bonsor13b} conclude, however, that such dynamical instabilities occur too rarely and that their aftermath does not last long enough to be a probable explanation for the large fraction of bright exozodiacal discs observed within the present survey. 

Because none of these two production scenarios seem sufficient to explain the amount of dust in the detected exozodiacal discs, we propose that dust trapping mechanisms could be at work in the innermost parts of the planetary systems. Some possible trapping mechanisms acting close to the sublimation radius are the following: (i) the pile-up of grains due to the interplay among sublimation, Poynting-Robertson drag, and radiation pressure \citep{Kobayashi09}; (ii) the trapping of charged nano-grains in the stellar magnetic field \citep{Czechowski10}; (iii) the resonant trapping by a planet, although the presence of short-period planets is unlikely for four of the five systems based on radial velocity measurements. Such trapping mechanisms could work in concert with both production mechanisms (major collision or transport).

	\subsection{A-type stars}

\begin{figure}[t]
\centering
\includegraphics[width=\linewidth]{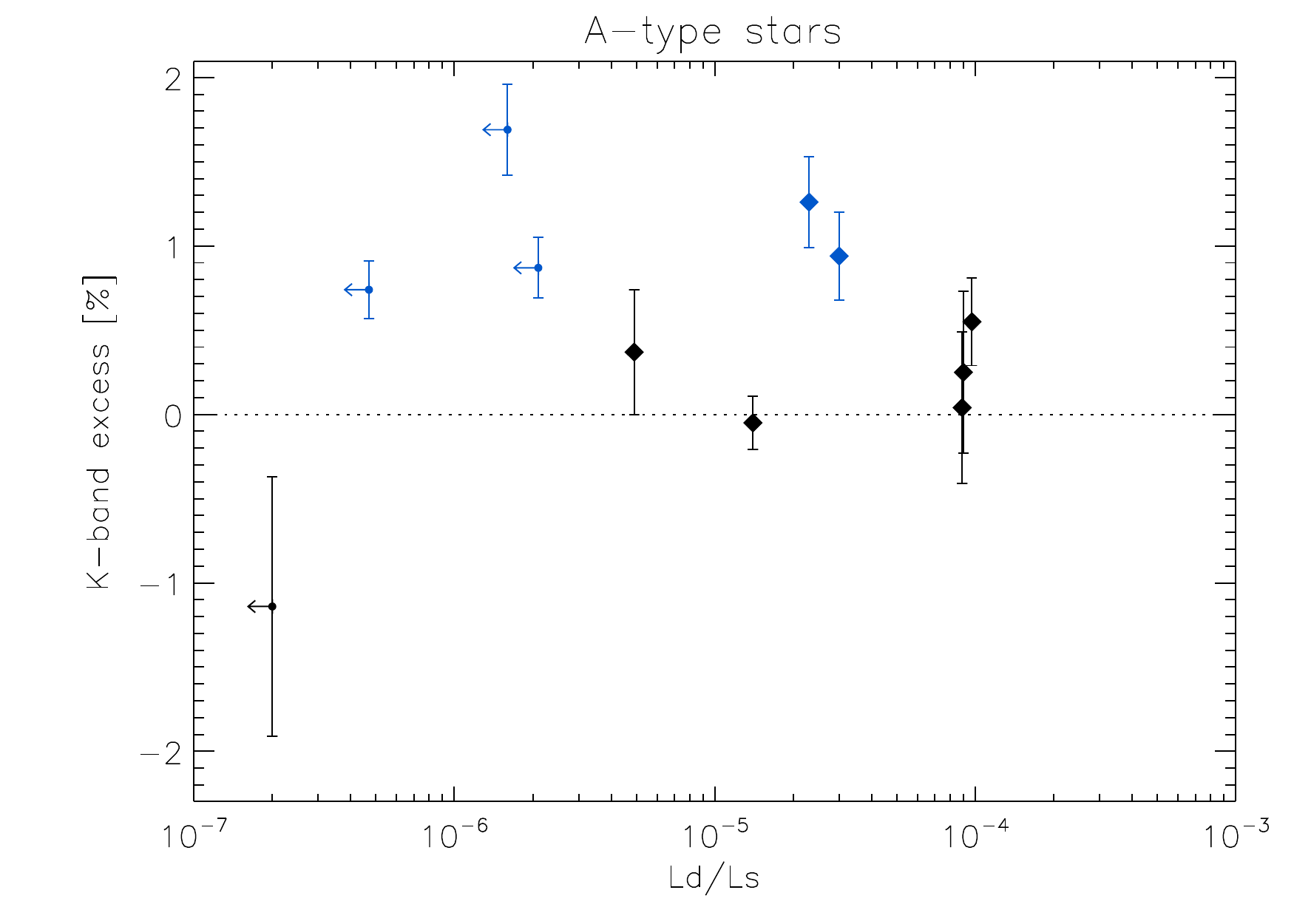}
\caption{Same as Fig.~\ref{fig:lum_fgk} in the case of A-type stars.} \label{fig:lum_a}
\end{figure}

In \citetalias{Absil08} and \citet{Mennesson13}, we discussed the various possible origins for a K-band excess around A-type stars, and concluded that circumstellar dust is much more plausible than photosphere-related effects. The absence of correlation between the presence of K-band excesses and outer dust reservoirs urges us to reconsider all possibilities. In Fig.~\ref{fig:lum_a}, we plot the measured K-band excesses as a function of the fractional luminosity of dust in the outer reservoir for the A-type stars. This figure suggests a possible anti-correlation between these two quantities, which is in line with the high K-band excess occurrence rate for A-type stars without cold dust reservoirs revealed in Fig.~\ref{fig:dusttype}. This correlation, which suggests that the presence of outer dust reservoirs could prevent the formation of bright exozodiacal discs in the case of A-type stars, is, however, not significant (p-value of 0.08 when comparing the two populations) and will therefore not be considered as a valuable clue in our investigation of the origin of near-infrared excesses.

\begin{figure}[t]
\centering
\includegraphics[width=\linewidth]{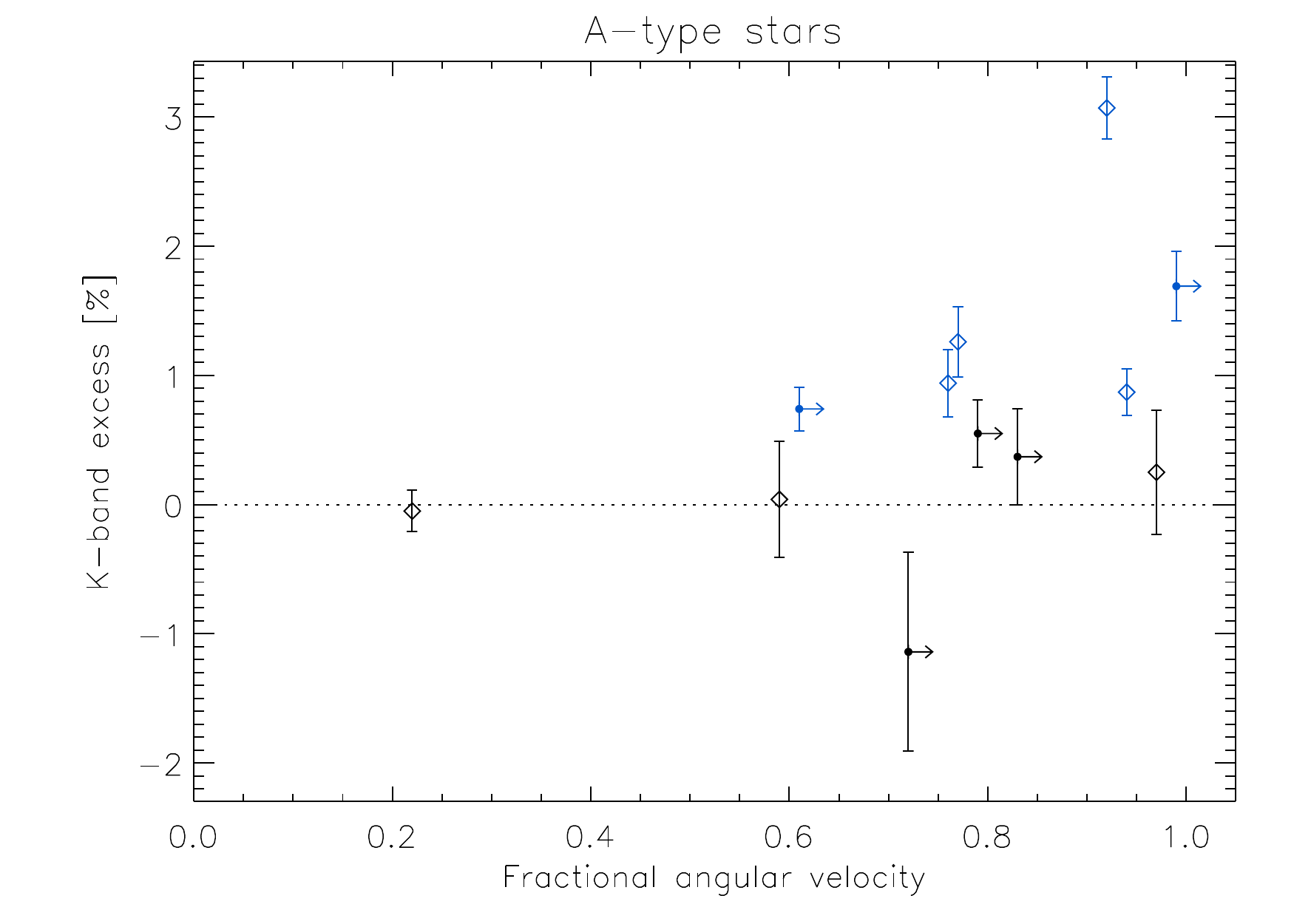}
\caption{Measured K-band disc/star flux ratio as function of fractional angular velocity (relative to the critical angular velocity) in the case of A-type stars. Diamond are used when the inclination of the photosphere spin axis is known. Lower limits are used otherwise (computed by assuming the star to be equator-on).} \label{fig:omega}
\end{figure}

To investigate whether the measured K-band excesses could be related to mass loss, we plot their level in Fig.~\ref{fig:omega} as a function of the fractional angular velocity $\omega$ of the A-type stars relative to their break-up velocity. The fractional angular velocity is computed as in the Appendix of \citet{Owocki94}, taking the photospheric inclination to the line-of-sight into account when available. When no information is available on the inclination, we compute a lower limit on $\omega$ by assuming the star to be seen equator-on, except in the cases where a debris disc has been resolved by direct imaging. In the latter case, we assume that the disc and star spin axes are aligned, as suggested by \citet{LeBouquin09}, \citet{Watson11}, and J.\ Greaves et al.\ (in prep.). Figure~\ref{fig:omega} suggests that the K-band excess level correlates with $\omega$. In particular, three of the four stars with fractional angular velocity higher than 0.9 feature a K-band excess. Furthermore, these three stars (zet Aql, alf Aql and alf Cep) do not have a detected cold dust reservoir, which reinforces a possible photospheric origin of the excess. The correlation found here is, however, subject to the fact that only lower limits on $\omega$ are available for most stars. It could turn out that many of them rotate close to the break-up velocity, and the correlation would then disappear. We also note that rapid rotators do not all have a K-band excess, gam Oph ($\omega=0.97$, assuming the photosphere to have a $40\degr$ inclination as the disc) being the most striking example.

To be a plausible origin for the detected K-band excess, mass loss from rapidly rotating stars should be fully resolved by our $\sim30$-m baseline, which translates into an angular radius of at least 8\,mas at K band. This corresponds to linear distances of about 5, 10, and 18 stellar radii in the cases of alf Aql, alf Cep and zet Aql, respectively, while theory predicts the near-infrared emission associated with winds from hot stars to originate mostly in a region $<1.5 R_{\star}$ \citep{Lamers99}. Furthermore, we discussed in \citetalias{Absil08} that the mass-loss rates associated with A-type stars are generally very low and should not give birth to significant near-infrared emission. The possibility that some of the detected K-band excesses are actually associated with mass loss from the photosphere should, however, be investigated further, using for instance spectro-interferometric observations at wavelengths corresponding to expected winds emission/absorption lines.

Finally, we note that two other A-type stars have been observed on other interferometers with the same aim and using the same strategy: alf PsA on VLTI/VINCI \citep{Absil09} and bet Pic on VLTI/PIONIER \citep{Defrere12}. Both have a K-band and a far-infrared excess. Including them in our sample would have increased the occurrence rate of K-band excesses for stars with known cold dust, to $44^{+16}_{-14}\%$. This occurrence rate would then be slightly higher than for dusty solar-type stars, but the discussion given above would remain the same.

	\subsection{Sensitivity of FLUOR and comparison with other studies}

Excluding eps Cep from Table~\ref{tab:result}, the median error bar on the disc/star flux ratio measured by CHARA/FLUOR at K band is $0.27\%$, which results in a median $3\sigma$ sensitivity of about $0.8\%$. This figure has to be compared to the K-band excess that would be measured when looking at our Sun from outside the solar system, which amounts to $2.2\times10^{-6}$ using the \citet{Kelsall98} empirical model for the zodiacal cloud implemented in the ZODIPIC package.\footnote{ZODIPIC is an IDL program developed and maintained by Mark Kuchner for synthesizing images of exozodiacal clouds (see http://asd.gsfc.nasa.gov/Marc.Kuchner/home.html).} The typical $3\sigma$ sensitivity limit of our CHARA/FLUOR observations would then be equivalent to about 3600 times the density of the solar zodiacal disc. The excess associated with such a bright exozodiacal disc at a wavelength of 10\,$\mu$m would amount to almost 20\% of the photospheric flux, which would be readily detected by mid-infrared spectro-photometry \citep[see e.g.][]{Schutz09,Lawler09,Morales09}. Although accurate mid-infrared spectro-photometry is not available for all our sources, the general absence of large excess at 10\,$\mu$m suggests that the discs detected by FLUOR are significantly different from the solar zodiacal disc, with much higher dust temperatures. This conclusion has already been made by various authors based on a more thorough modelling of exozodiacal discs using multi-colour interferometric observations \citep{Absil06,Akeson09,Defrere11,Defrere12,Mennesson13,Lebreton13}.

The most comprehensive interferometric survey of exozodiacal dust to date was performed by \citet{MillanGabet11} using the Keck Interferometer Nuller (KIN) in the mid-infrared domain. With an angular resolution $\lambda/2B=10$~mas and a field-of-view radius of about $0\farcs5$, the KIN probes circumstellar dust on spatial scales similar to our survey. While the KIN sensitivity level is about eight times better than in our survey in terms of equivalent zodiacal dust density (average individual uncertainty of 160 zodis compared to 1200 zodis in our case), the excess detection rate in the KIN survey was much lower than in ours, with an overall detection rate of 3/25 ($12^{+9}_{-4}\%$), taking two marginal detections into account. This figure must be compared to the overall occurrence rate of $28^{+8}_{-6}\%$ found in the present study. Among the 25 stars observed by \citet{MillanGabet11}, 11 have also been observed with CHARA/FLUOR. One of these eleven stars shows a significant mid-infrared excess in KIN data (eta Crv), and two other show a marginal excess (gam Oph, alf Aql). Out of the same eleven stars, only one shows a significant K-band excess (alf Aql). The difference in detection rates between the KIN and CHARA/FLUOR surveys could therefore be entirely due to the particular choice of targets, with only two A-type stars in the KIN sample for instance. That \citet{MillanGabet11} did not specifically focus on stars with outer dust reservoirs may also explain the discrepancy in excess occurrence rate. We leave it to a future study, including new KIN observations, to investigate the correlation between near-infrared and mid-infrared resolved circumstellar excesses (B.\ Mennesson et al., in prep.).

Other interferometric studies have specifically targeted exozodiacal discs in the mid-infrared, using the VLTI/MIDI instrument. \citet{Smith09} obtained the first mid-infrared interferometric detection of exozodiacal dust around HD\,69830 and eta Crv. These two stars were already known to possess a large mid-infrared excess from spectro-photometric measurements. \citet{Smith12} then resolved the dust located in the terrestrial planet forming region of two young main sequence stars (HD\,113766 and HD\,172555) with the same instrument. The typical error bar on individual visibility measurements is around 5\% in these observations, which would give a typical $3\sigma$ sensitivity of a few percent on the disc/star flux ratio after averaging five to ten independent visibility measurements. This would correspond to a few hundred zodi of dust around a solar-type star, which falls in the same ballpark as the KIN sensitivity. We note that both HD\,69830 and eta Crv were observed within the present study, without any sign of K-band excess. This emphasises, once again, that prominent near-infrared and mid-infrared excesses are not necessarily associated.

Besides interferometric studies, spectro-photometric surveys for exozodiacal dust have also been performed from the ground and from space. The most sensitive search around solar-type stars was performed by \citet{Lawler09} using the \textsc{Spitzer}/IRS mid-infrared spectrograph. At a $3\sigma$ sensitivity level of about 1000 times the level of the zodiacal emission in the short IRS wavelength band (8.5--12\,$\mu$m), the authors found a success rate for excesses of about 1\%. This occurrence rate, recently confirmed by WISE mid-infrared photometry \citep{Morales12,Padgett13}, is in stark contrast to the present study, and points again to a non-solar system morphology for the exozodiacal discs detected at K band.

It is also interesting to compare the results of our study with the occurrence rate of cold dust belts found with \textsc{Spitzer} or \textsc{Herschel} far-infrared photometry. In the case of solar-type stars, the \textsc{Herschel} far-infrared excess frequency is of $20.2\% \pm 2\%$ for an unbiased sample \citep{Eiroa13}. This frequency cannot be directly compared with the occurrence rate of $18^{+9}_{-5}\%$ found around solar-type stars in our survey, because of the bias in our sample. However, if we discard the ``young'' solar-type star ksi Boo from our sample and accept that the presence of bright exozodiacal discs around mature solar-type stars is directly related to the presence of cold dust populations, we could tentatively de-bias our result based on the \textsc{Herschel} occurrence rate. Making the further (very tentative) assumption that faint cold dust populations not detectable by \textsc{Herschel} cannot be at the origin of detectable exozodiacal emission at K band, this would yield an unbiased occurrence rate of about 8\% for bright exozodiacal discs around solar-type stars. This rate is still much higher than what was found in mid-infrared spectro-photometric surveys. In the case of A-type stars, the \textsc{Spitzer} far-infrared excess rate of about 32\% \citep{Su06} is lower than the $50\% \pm 13\%$ found in our sample, but here again a direct comparison is not possible due to the biases in our sample.

\section{Conclusions}

We have conducted a survey using the FLUOR interferometer at the CHARA array to search for resolved near-infrared emission at sub-percent-level accuracy around a sample of 42 stars. About half of our targets were chosen based on the presence of ``outer'' dust populations at separations larger than a few AU. A significant K-band excess was found around 13 of these 42 stars. Since the presence of a faint companion within the $0\farcs8$ field-of-view of FLUOR cannot be ruled out from our interferometric observations, we used ancillary data to show that the excess is associated with an extended source in most cases. We argued that hot circumstellar dust is the most likely source for these near-infrared excesses. Only one star (eps Cep) showed clear signs of binarity in the measured visibilities. The binarity of this source was confirmed by coronagraphic observations \citep{Mawet11}, and the star was therefore removed from our statistical sample. Considering the fact that one of our targets turned out to be a giant (kap CrB), we are left with 11 excesses out 40 single main sequence stars, i.e., an occurrence rate of $28^{+8}_{-6}\%$. 

The typical $1\sigma$ error bar on the measured disc/star flux ratio in our sample is 0.27\%, which for solar-type stars gives a typical $3\sigma$ sensitivity of about 3600 times the density of the solar zodiacal cloud (i.e., 3600 \emph{zodi}). At this sensitivity level, we found significantly more resolved excesses in the near-infrared than was found in the mid-infrared by \citet{MillanGabet11} with eight times better sensitivity in terms of \emph{zodi}. This suggests that the detected dust discs have a morphology that is very different from the solar zodiacal cloud, a conclusion that has already been reached by several authors based on radiative transfer modelling of these resolved exozodiacal discs.

An investigation of the near-infrared excess occurrence rate as a function of stellar properties brings two interesting correlations to light: (i) A-type stars show significantly more positive detections than solar-type (F,G,K) stars, and (ii) the presence of a near-infrared excess correlates with the presence of outer dust reservoir(s) in the case of solar-type stars. We suggested a few possible scenarios where the inner dust content of planetary systems around solar-type stars could be directly related to outer reservoir(s). Conversely, the origin of inner dust around A-type stars remains mysterious, although we showed a possible correlation with the rotational velocity, which may hint at mass-loss effects. Enlarging the stellar sample is required to confirm these statistical trends. This will be the subject of future papers exploiting the capabilities of the VLTI/PIONIER interferometer in the Southern hemisphere and of a refurbished version of FLUOR at the CHARA array.

\begin{acknowledgements}
The authors thank the French National Research Agency (ANR) for financial support through contract ANR-2010 BLAN-0505-01 (EXOZODI). O.A.\ acknowledges financial support from the European Commission's Sixth Framework Programme as a Marie Curie Intra-European Fellow and from an F.R.S.-FNRS Postdoctoral Fellowship during part(s) of this work. The authors also thank B.~Matthews, J.~Greaves, C.~Eiroa, and J.~Marshall for sharing early \textsc{Herschel}/PACS results; G.~Bryden for sharing unpublished \textsc{Spitzer}/MIPS data; T.~Boyajian for making her interferometric stellar diameters available to us; A.~G\'asp\'ar, G.~Rieke, and J.~Maldonado for useful information on stellar ages; D.~Mawet and B.~Mennesson for follow-up observations with vortex coronagraphy and/or nulling interferometry; M.~Ireland and A.~Kraus for carrying out the aperture masking observations of bet Leo; E.~Pedretti for his contribution to the MIRC observations; R.~Soummer and J.~Leconte for information on AEOS coronagraphic observations; and A.~Bonsor for useful discussions on planet scattering and on kap CrB. The CHARA Array is currently funded by the National Science Foundation through AST-1211929 and by Georgia State University. This work made use of the Smithsonian/NASA Astrophysics Data System (ADS) and of the Centre de Donn\'ees astronomiques de Strasbourg (CDS).
\end{acknowledgements}

\bibliographystyle{aa} 
\bibliography{fluor_survey} 

\begin{thebibliography}{117}
\expandafter\ifx\csname natexlab\endcsname\relax\def\natexlab#1{#1}\fi

\bibitem[{{Absil} {et~al.}(2006){Absil}, {Di Folco}, {M{\'e}rand}, {Augereau},
  {Coud{\'e} du Foresto}, {Aufdenberg}, {Kervella}, {Ridgway}, {Berger}, {ten
  Brummelaar}, {Sturmann}, {Sturmann}, {Turner}, \& {McAlister}}]{Absil06}
{Absil}, O., {Di Folco}, E., {M{\'e}rand}, A., {et~al.} 2006, \aap, 452, 237

\bibitem[{{Absil} {et~al.}(2008){Absil}, {Di Folco}, {M{\'e}rand}, {Augereau},
  {Coud{\'e} du Foresto}, {Defr{\`e}re}, {Kervella}, {Aufdenberg}, {Desort},
  {Ehrenreich}, {Lagrange}, {Montagnier}, {Olofsson}, {ten Brummelaar},
  {McAlister}, {Sturmann}, {Sturmann}, \& {Turner}}]{Absil08}
{Absil}, O., {Di Folco}, E., {M{\'e}rand}, A., {et~al.} 2008, \aap, 487, 1041

\bibitem[{{Absil} {et~al.}(2011){Absil}, {Le Bouquin}, {Berger}, {Lagrange},
  {Chauvin}, {Lazareff}, {Zins}, {Haguenauer}, {Jocou}, {Kern}, {Millan-Gabet},
  {Rochat}, \& {Traub}}]{Absil11}
{Absil}, O., {Le Bouquin}, J.-B., {Berger}, J.-P., {et~al.} 2011, \aap, 535,
  A68

\bibitem[{{Absil} {et~al.}(2009){Absil}, {Mennesson}, {Le Bouquin}, {Di Folco},
  {Kervella}, \& {Augereau}}]{Absil09}
{Absil}, O., {Mennesson}, B., {Le Bouquin}, J.-B., {et~al.} 2009, \apj, 704,
  150

\bibitem[{{Akeson} {et~al.}(2009){Akeson}, {Ciardi}, {Millan-Gabet}, {Merand},
  {Di Folco}, {Monnier}, {Beichman}, {Absil}, {Aufdenberg}, {McAlister},
  {Brummelaar}, {Sturmann}, {Sturmann}, \& {Turner}}]{Akeson09}
{Akeson}, R.~L., {Ciardi}, D.~R., {Millan-Gabet}, R., {et~al.} 2009, \apj, 691,
  1896

\bibitem[{{Arribas} \& {Martinez Roger}(1989)}]{Arribas89}
{Arribas}, S. \& {Martinez Roger}, C. 1989, \aap, 215, 305

\bibitem[{{Aufdenberg} {et~al.}(2006){Aufdenberg}, {M\'erand}, {Coud\'e du
  Foresto}, {Absil}, {Di Folco}, {Kervella}, {Ridgway}, {Berger}, {ten
  Brummelaar}, {McAlister}, {Sturmann}, {Sturmann}, \& {Turner}}]{Aufdenberg06}
{Aufdenberg}, J.~A., {M\'erand}, A., {Coud\'e du Foresto}, V., {et~al.} 2006,
  \apj, 645, 664

\bibitem[{{Augereau} {et~al.}(1999){Augereau}, {Lagrange}, {Mouillet},
  {Papaloizou}, \& {Grorod}}]{Augereau99}
{Augereau}, J.~C., {Lagrange}, A.~M., {Mouillet}, D., {Papaloizou}, J.~C.~B.,
  \& {Grorod}, P.~A. 1999, \aap, 348, 557

\bibitem[{{Aumann} \& {Probst}(1991)}]{Aumann91}
{Aumann}, H.~H. \& {Probst}, R.~G. 1991, \apj, 368, 264

\bibitem[{{Backman} {et~al.}(2009){Backman}, {Marengo}, {Stapelfeldt}, {Su},
  {Wilner}, {Dowell}, {Watson}, {Stansberry}, {Rieke}, {Megeath}, {Fazio}, \&
  {Werner}}]{Backman09}
{Backman}, D., {Marengo}, M., {Stapelfeldt}, K., {et~al.} 2009, \apj, 690, 1522

\bibitem[{{Baines} {et~al.}(2008){Baines}, {McAlister}, {ten Brummelaar},
  {Turner}, {Sturmann}, {Sturmann}, {Goldfinger}, \& {Ridgway}}]{Baines08}
{Baines}, E.~K., {McAlister}, H.~A., {ten Brummelaar}, T.~A., {et~al.} 2008,
  \apj, 680, 728

\bibitem[{{Baliunas} {et~al.}(1996){Baliunas}, {Nesme-Ribes}, {Sokoloff}, \&
  {Soon}}]{Baliunas96}
{Baliunas}, S.~L., {Nesme-Ribes}, E., {Sokoloff}, D., \& {Soon}, W.~H. 1996,
  \apj, 460, 848

\bibitem[{{Beichman} {et~al.}(2005){Beichman}, {Bryden}, {Gautier},
  {Stapelfeldt}, {Werner}, {Misselt}, {Rieke}, {Stansberry}, \&
  {Trilling}}]{Beichman05}
{Beichman}, C.~A., {Bryden}, G., {Gautier}, T.~N., {et~al.} 2005, \apj, 626,
  1061

\bibitem[{{Beichman} {et~al.}(2006){Beichman}, {Tanner}, {Bryden},
  {Stapelfeldt}, {Werner}, {Rieke}, {Trilling}, {Lawler}, \&
  {Gautier}}]{Beichman06}
{Beichman}, C.~A., {Tanner}, A., {Bryden}, G., {et~al.} 2006, \apj, 639, 1166

\bibitem[{{Blackwell} {et~al.}(1979){Blackwell}, {Shallis}, \&
  {Selby}}]{Blackwell79}
{Blackwell}, D.~E., {Shallis}, M.~J., \& {Selby}, M.~J. 1979, \mnras, 188, 847

\bibitem[{{Bonsor} {et~al.}(2012){Bonsor}, {Augereau}, \&
  {Th{\'e}bault}}]{Bonsor12}
{Bonsor}, A., {Augereau}, J.-C., \& {Th{\'e}bault}, P. 2012, \aap, 548, A104

\bibitem[{{Bonsor} {et~al.}(2013){Bonsor}, {Kennedy}, {Crepp}, {Johnson},
  {Wyatt}, {Sibthorpe}, \& {Su}}]{Bonsor13a}
{Bonsor}, A., {Kennedy}, G.~M., {Crepp}, J.~R., {et~al.} 2013, \mnras, 431,
  3025

\bibitem[{{Bonsor} \& {Raymond}(2013)}]{Bonsor13b}
{Bonsor}, A. \& {Raymond}, S. 2013, \mnras, submitted

\bibitem[{{Bord{\'e}} {et~al.}(2002){Bord{\'e}}, {Coud{\'e} du Foresto},
  {Chagnon}, \& {Perrin}}]{Borde02}
{Bord{\'e}}, P., {Coud{\'e} du Foresto}, V., {Chagnon}, G., \& {Perrin}, G.
  2002, \aap, 393, 183

\bibitem[{{Boyajian} {et~al.}(2012){Boyajian}, {McAlister}, {van Belle},
  {Gies}, {ten Brummelaar}, {von Braun}, {Farrington}, {Goldfinger}, {O'Brien},
  {Parks}, {Richardson}, {Ridgway}, {Schaefer}, {Sturmann}, {Sturmann},
  {Touhami}, {Turner}, \& {White}}]{Boyajian12a}
{Boyajian}, T.~S., {McAlister}, H.~A., {van Belle}, G., {et~al.} 2012, \apj,
  746, 101

\bibitem[{{Bryden} {et~al.}(2009){Bryden}, {Beichman}, {Carpenter}, {Rieke},
  {Stapelfeldt}, {Werner}, {Tanner}, {Lawler}, {Wyatt}, {Trilling}, {Su},
  {Blaylock}, \& {Stansberry}}]{Bryden09}
{Bryden}, G., {Beichman}, C.~A., {Carpenter}, J.~M., {et~al.} 2009, \apj, 705,
  1226

\bibitem[{{Burgasser} {et~al.}(2003){Burgasser}, {Kirkpatrick}, {Reid},
  {Brown}, {Miskey}, \& {Gizis}}]{Burgasser03}
{Burgasser}, A.~J., {Kirkpatrick}, J.~D., {Reid}, I.~N., {et~al.} 2003, \apj,
  586, 512

\bibitem[{{Carpenter} {et~al.}(2009){Carpenter}, {Bouwman}, {Mamajek}, {Meyer},
  {Hillenbrand}, {Backman}, {Henning}, {Hines}, {Hollenbach}, {Kim},
  {Moro-Martin}, {Pascucci}, {Silverstone}, {Stauffer}, \&
  {Wolf}}]{Carpenter09}
{Carpenter}, J.~M., {Bouwman}, J., {Mamajek}, E.~E., {et~al.} 2009, \apjs, 181,
  197

\bibitem[{{Che} {et~al.}(2010){Che}, {Monnier}, \& {Webster}}]{Che10}
{Che}, X., {Monnier}, J.~D., \& {Webster}, S. 2010, in Proc. SPIE, Vol. 7734,
  {Optical and Infrared Interferometry II}, 77342V

\bibitem[{{Che} {et~al.}(2011){Che}, {Monnier}, {Zhao}, {Pedretti}, {Thureau},
  {M{\'e}rand}, {ten Brummelaar}, {McAlister}, {Ridgway}, {Turner}, {Sturmann},
  \& {Sturmann}}]{Che11}
{Che}, X., {Monnier}, J.~D., {Zhao}, M., {et~al.} 2011, \apj, 732, 68

\bibitem[{{Chelli} {et~al.}(2009){Chelli}, {Duvert}, {Malbet}, \&
  {Kern}}]{Chelli09}
{Chelli}, A., {Duvert}, G., {Malbet}, F., \& {Kern}, P. 2009, \aap, 498, 321

\bibitem[{{Chen} {et~al.}(2006){Chen}, {Sargent}, {Bohac}, {Kim},
  {Leibensperger}, {Jura}, {Najita}, {Forrest}, {Watson}, {Sloan}, \&
  {Keller}}]{Chen06}
{Chen}, C.~H., {Sargent}, B.~A., {Bohac}, C., {et~al.} 2006, \apjs, 166, 351

\bibitem[{{Christou} \& {Drummond}(2006)}]{Christou06}
{Christou}, J.~C. \& {Drummond}, J.~D. 2006, \aj, 131, 3100

\bibitem[{{Churcher} {et~al.}(2011){Churcher}, {Wyatt}, {Duch{\^e}ne},
  {Sibthorpe}, {Kennedy}, {Matthews}, {Kalas}, {Greaves}, {Su}, \&
  {Rieke}}]{Churcher11}
{Churcher}, L.~J., {Wyatt}, M.~C., {Duch{\^e}ne}, G., {et~al.} 2011, \mnras,
  417, 1715

\bibitem[{{Claret}(2000)}]{Claret00}
{Claret}, A. 2000, \aap, 363, 1081

\bibitem[{{Coud\'e du Foresto} {et~al.}(2003){Coud\'e du Foresto}, {Bord\'e},
  {M\'erand}, {Baudouin}, {Remond}, {Perrin}, {Ridgway}, {ten Brummelaar}, \&
  {McAlister}}]{Coude03}
{Coud\'e du Foresto}, V., {Bord\'e}, P.~J., {M\'erand}, A., {et~al.} 2003, in
  Proc. SPIE, Vol. 4838, {Interferometry in Optical Astronomy II}, 280--285

\bibitem[{{Coud\'e du Foresto} {et~al.}(1997){Coud\'e du Foresto}, {Ridgway},
  \& {Mariotti}}]{Coude97}
{Coud\'e du Foresto}, V., {Ridgway}, S., \& {Mariotti}, J.-M. 1997, \aaps, 121,
  379

\bibitem[{{Czechowski} \& {Mann}(2010)}]{Czechowski10}
{Czechowski}, A. \& {Mann}, I. 2010, \apj, 714, 89

\bibitem[{{De Rosa} {et~al.}(2011){De Rosa}, {Bulger}, {Patience}, {Leland},
  {Macintosh}, {Schneider}, {Song}, {Marois}, {Graham}, {Bessell}, \&
  {Doyon}}]{DeRosa11}
{De Rosa}, R.~J., {Bulger}, J., {Patience}, J., {et~al.} 2011, \mnras, 415, 854

\bibitem[{{Defr{\`e}re} {et~al.}(2011){Defr{\`e}re}, {Absil}, {Augereau}, {di
  Folco}, {Berger}, {Coud{\'e} Du Foresto}, {Kervella}, {Le Bouquin},
  {Lebreton}, {Millan-Gabet}, {Monnier}, {Olofsson}, \& {Traub}}]{Defrere11}
{Defr{\`e}re}, D., {Absil}, O., {Augereau}, J.-C., {et~al.} 2011, \aap, 534, A5

\bibitem[{{Defr{\`e}re} {et~al.}(2012){Defr{\`e}re}, {Lebreton}, {Le Bouquin},
  {Lagrange}, {Absil}, {Augereau}, {Berger}, {di Folco}, {Ertel}, {Kluska},
  {Montagnier}, {Millan-Gabet}, {Traub}, \& {Zins}}]{Defrere12}
{Defr{\`e}re}, D., {Lebreton}, J., {Le Bouquin}, J.-B., {et~al.} 2012, \aap,
  546, L9

\bibitem[{{Di Folco} {et~al.}(2007){Di Folco}, {Absil}, {Augereau},
  {M{\'e}rand}, {Coud{\'e} du Foresto}, {Th{\'e}venin}, {Defr{\`e}re},
  {Kervella}, {ten Brummelaar}, {McAlister}, {Ridgway}, {Sturmann}, {Sturmann},
  \& {Turner}}]{DiFolco07}
{Di Folco}, E., {Absil}, O., {Augereau}, J.-C., {et~al.} 2007, \aap, 475, 243

\bibitem[{{Di Folco} {et~al.}(2004){Di Folco}, {Th{\'e}venin}, {Kervella},
  {Domiciano de Souza}, {Coud{\'e} du Foresto}, {S{\'e}gransan}, \&
  {Morel}}]{DiFolco04}
{Di Folco}, E., {Th{\'e}venin}, F., {Kervella}, P., {et~al.} 2004, \aap, 426,
  601

\bibitem[{{Dodson-Robinson} {et~al.}(2011){Dodson-Robinson}, {Beichman},
  {Carpenter}, \& {Bryden}}]{Dodson11}
{Dodson-Robinson}, S.~E., {Beichman}, C.~A., {Carpenter}, J.~M., \& {Bryden},
  G. 2011, \aj, 141, 11

\bibitem[{{Domiciano de Souza} {et~al.}(2005){Domiciano de Souza}, {Kervella},
  {Jankov}, {Vakili}, {Ohishi}, {Nordgren}, \& {Abe}}]{Domiciano05}
{Domiciano de Souza}, A., {Kervella}, P., {Jankov}, S., {et~al.} 2005, \aap,
  442, 567

\bibitem[{{Ducati}(2002)}]{Ducati02}
{Ducati}, J.~R. 2002, VizieR Online Data Catalog, 2237, 0

\bibitem[{{Dunham}(1977)}]{Dunham77}
{Dunham}, D.~W. 1977, Occultation Newsletter, International Occultation Timing
  Association (IOTA), 1, 119

\bibitem[{{Eiroa} {et~al.}(2010){Eiroa}, {Fedele}, {Maldonado},
  {Gonz{\'a}lez-Garc{\'{\i}}a}, {Rodmann}, {Heras}, {Pilbratt}, {Augereau},
  {Mora}, {Montesinos}, {Ardila}, {Bryden}, {Liseau}, {Stapelfeldt},
  {Launhardt}, {Solano}, {Bayo}, {Absil}, {Ar{\'e}valo}, {Barrado},
  {Beichmann}, {Danchi}, {Del Burgo}, {Ertel}, {Fridlund}, {Fukagawa},
  {Guti{\'e}rrez}, {Gr{\"u}n}, {Kamp}, {Krivov}, {Lebreton}, {L{\"o}hne},
  {Lorente}, {Marshall}, {Mart{\'{\i}}nez-Arn{\'a}iz}, {Meeus}, {Montes},
  {Morbidelli}, {M{\"u}ller}, {Mutschke}, {Nakagawa}, {Olofsson}, {Ribas},
  {Roberge}, {Sanz-Forcada}, {Th{\'e}bault}, {Walker}, {White}, \&
  {Wolf}}]{Eiroa10}
{Eiroa}, C., {Fedele}, D., {Maldonado}, J., {et~al.} 2010, \aap, 518, L131

\bibitem[{{Eiroa} {et~al.}(2013){Eiroa}, {Marshall}, {Mora}, {Montesinos},
  {Absil}, {Augereau}, \& A.}]{Eiroa13}
{Eiroa}, C., {Marshall}, J., {Mora}, A., {et~al.} 2013, \aap, in press

\bibitem[{{G{\'a}sp{\'a}r} {et~al.}(2013){G{\'a}sp{\'a}r}, {Rieke}, \&
  {Balog}}]{Gaspar13}
{G{\'a}sp{\'a}r}, A., {Rieke}, G.~H., \& {Balog}, Z. 2013, \apj, 768, 25

\bibitem[{{Gray} {et~al.}(2006){Gray}, {Corbally}, {Garrison}, {McFadden},
  {Bubar}, {McGahee}, {O'Donoghue}, \& {Knox}}]{Gray06}
{Gray}, R.~O., {Corbally}, C.~J., {Garrison}, R.~F., {et~al.} 2006, \aj, 132,
  161

\bibitem[{{Gray} {et~al.}(2003){Gray}, {Corbally}, {Garrison}, {McFadden}, \&
  {Robinson}}]{Gray03}
{Gray}, R.~O., {Corbally}, C.~J., {Garrison}, R.~F., {McFadden}, M.~T., \&
  {Robinson}, P.~E. 2003, \aj, 126, 2048

\bibitem[{{Habing} {et~al.}(2001){Habing}, {Dominik}, {Jourdain de Muizon},
  {Laureijs}, {Kessler}, {Leech}, {Metcalfe}, {Salama}, {Siebenmorgen},
  {Trams}, \& {Bouchet}}]{Habing01}
{Habing}, H.~J., {Dominik}, C., {Jourdain de Muizon}, M., {et~al.} 2001, \aap,
  365, 545

\bibitem[{{Holmberg} {et~al.}(2009){Holmberg}, {Nordstr{\"o}m}, \&
  {Andersen}}]{Holmberg09}
{Holmberg}, J., {Nordstr{\"o}m}, B., \& {Andersen}, J. 2009, \aap, 501, 941

\bibitem[{{Jackson} \& {Wyatt}(2012)}]{Jackson12}
{Jackson}, A.~P. \& {Wyatt}, M.~C. 2012, \mnras, 425, 657

\bibitem[{{Johnson} {et~al.}(2008){Johnson}, {Marcy}, {Fischer}, {Wright},
  {Reffert}, {Kregenow}, {Williams}, \& {Peek}}]{Johnson08}
{Johnson}, J.~A., {Marcy}, G.~W., {Fischer}, D.~A., {et~al.} 2008, \apj, 675,
  784

\bibitem[{{Kelsall} {et~al.}(1998){Kelsall}, {Weiland}, {Franz}, {Reach},
  {Arendt}, {Dwek}, {Freudenreich}, {Hauser}, {Moseley}, {Odegard},
  {Silverberg}, \& {Wright}}]{Kelsall98}
{Kelsall}, T., {Weiland}, J.~L., {Franz}, B.~A., {et~al.} 1998, \apj, 508, 44

\bibitem[{{Kervella} {et~al.}(2008){Kervella}, {M{\'e}rand}, {Pichon},
  {Th{\'e}venin}, {Heiter}, {Bigot}, {ten Brummelaar}, {McAlister}, {Ridgway},
  {Turner}, {Sturmann}, {Sturmann}, {Goldfinger}, \& {Farrington}}]{Kervella08}
{Kervella}, P., {M{\'e}rand}, A., {Pichon}, B., {et~al.} 2008, \aap, 488, 667

\bibitem[{{Kervella} {et~al.}(2004){Kervella}, {Th{\'e}venin}, {Di Folco}, \&
  {S{\'e}gransan}}]{Kervella04}
{Kervella}, P., {Th{\'e}venin}, F., {Di Folco}, E., \& {S{\'e}gransan}, D.
  2004, \aap, 426, 297

\bibitem[{{Kervella} {et~al.}(2003){Kervella}, {Th{\'e}venin}, {S{\'e}gransan},
  {Berthomieu}, {Lopez}, {Morel}, \& {Provost}}]{Kervella03}
{Kervella}, P., {Th{\'e}venin}, F., {S{\'e}gransan}, D., {et~al.} 2003, \aap,
  404, 1087

\bibitem[{{Kharchenko} \& {Roeser}(2009)}]{Kharchenko09}
{Kharchenko}, N.~V. \& {Roeser}, S. 2009, VizieR Online Data Catalog, 1280, 0

\bibitem[{{Kobayashi} {et~al.}(2009){Kobayashi}, {Watanabe}, {Kimura}, \&
  {Yamamoto}}]{Kobayashi09}
{Kobayashi}, H., {Watanabe}, S.-I., {Kimura}, H., \& {Yamamoto}, T. 2009,
  \icarus, 201, 395

\bibitem[{{Lacour} {et~al.}(2011){Lacour}, {Tuthill}, {Amico}, {Ireland},
  {Ehrenreich}, {Huelamo}, \& {Lagrange}}]{Lacour11}
{Lacour}, S., {Tuthill}, P., {Amico}, P., {et~al.} 2011, \aap, 532, A72

\bibitem[{{Lagrange} {et~al.}(2009){Lagrange}, {Desort}, {Galland}, {Udry}, \&
  {Mayor}}]{Lagrange09}
{Lagrange}, A.-M., {Desort}, M., {Galland}, F., {Udry}, S., \& {Mayor}, M.
  2009, \aap, 495, 335

\bibitem[{{Lamers} \& {Cassinelli}(1999)}]{Lamers99}
{Lamers}, H.~J.~G.~L.~M. \& {Cassinelli}, J.~P. 1999, {Introduction to Stellar
  Winds} (Cambridge University Press)

\bibitem[{{Lawler} {et~al.}(2009){Lawler}, {Beichman}, {Bryden}, {Ciardi},
  {Tanner}, {Su}, {Stapelfeldt}, {Lisse}, \& {Harker}}]{Lawler09}
{Lawler}, S.~M., {Beichman}, C.~A., {Bryden}, G., {et~al.} 2009, \apj, 705, 89

\bibitem[{{Le Bouquin} {et~al.}(2009){Le Bouquin}, {Absil}, {Benisty}, {Massi},
  {M{\'e}rand}, \& {Stefl}}]{LeBouquin09}
{Le Bouquin}, J.-B., {Absil}, O., {Benisty}, M., {et~al.} 2009, \aap, 498, L41

\bibitem[{{Le Bouquin} {et~al.}(2011){Le Bouquin}, {Berger}, {Lazareff},
  {Zins}, {Haguenauer}, {Jocou}, {Kern}, {Millan-Gabet}, {Traub}, {Absil},
  {Augereau}, {Benisty}, {Blind}, {Bonfils}, {Bourget}, {Delboulbe},
  {Feautrier}, {Germain}, {Gitton}, {Gillier}, {Kiekebusch}, {Kluska},
  {Knudstrup}, {Labeye}, {Lizon}, {Monin}, {Magnard}, {Malbet}, {Maurel},
  {M{\'e}nard}, {Micallef}, {Michaud}, {Montagnier}, {Morel}, {Moulin},
  {Perraut}, {Popovic}, {Rabou}, {Rochat}, {Rojas}, {Roussel}, {Roux},
  {Stadler}, {Stefl}, {Tatulli}, \& {Ventura}}]{LeBouquin11}
{Le Bouquin}, J.-B., {Berger}, J.-P., {Lazareff}, B., {et~al.} 2011, \aap, 535,
  A67

\bibitem[{{Lebreton} {et~al.}(2013){Lebreton}, {van Lieshout}, {Augereau},
  {Absil}, {Mennesson}, {Kama}, {Dominik}, \& A.}]{Lebreton13}
{Lebreton}, J., {van Lieshout}, R., {Augereau}, J.-C., {et~al.} 2013, \aap,
  submitted

\bibitem[{{Leconte} {et~al.}(2010){Leconte}, {Soummer}, {Hinkley},
  {Oppenheimer}, {Sivaramakrishnan}, {Brenner}, {Kuhn}, {Lloyd}, {Perrin},
  {Makidon}, {Roberts}, {Graham}, {Simon}, {Brown}, {Zimmerman}, {Chabrier}, \&
  {Baraffe}}]{Leconte10}
{Leconte}, J., {Soummer}, R., {Hinkley}, S., {et~al.} 2010, \apj, 716, 1551

\bibitem[{{Lisse} {et~al.}(2012){Lisse}, {Wyatt}, {Chen}, {Morlok}, {Watson},
  {Manoj}, {Sheehan}, {Currie}, {Thebault}, \& {Sitko}}]{Lisse12}
{Lisse}, C.~M., {Wyatt}, M.~C., {Chen}, C.~H., {et~al.} 2012, \apj, 747, 93

\bibitem[{{Maldonado} {et~al.}(2012){Maldonado}, {Eiroa}, {Villaver},
  {Montesinos}, \& {Mora}}]{Maldonado12}
{Maldonado}, J., {Eiroa}, C., {Villaver}, E., {Montesinos}, B., \& {Mora}, A.
  2012, \aap, 541, A40

\bibitem[{{Markwardt}(2009)}]{Markwardt09}
{Markwardt}, C.~B. 2009, in Astronomical Society of the Pacific Conference
  Series, Vol. 411, Astronomical Data Analysis Software and Systems XVIII, ed.
  D.~A. {Bohlender}, D.~{Durand}, \& P.~{Dowler}, 251

\bibitem[{{Mart{\'{\i}}nez-Arn{\'a}iz}
  {et~al.}(2010){Mart{\'{\i}}nez-Arn{\'a}iz}, {Maldonado}, {Montes}, {Eiroa},
  \& {Montesinos}}]{MartinezArnaiz10}
{Mart{\'{\i}}nez-Arn{\'a}iz}, R., {Maldonado}, J., {Montes}, D., {Eiroa}, C.,
  \& {Montesinos}, B. 2010, \aap, 520, A79

\bibitem[{{Matthews} {et~al.}(2010){Matthews}, {Sibthorpe}, {Kennedy},
  {Phillips}, {Churcher}, {Duch{\^e}ne}, {Greaves}, {Lestrade}, {Moro-Martin},
  {Wyatt}, {Bastien}, {Biggs}, {Bouvier}, {Butner}, {Dent}, {di Francesco},
  {Eisl{\"o}ffel}, {Graham}, {Harvey}, {Hauschildt}, {Holland}, {Horner},
  {Ibar}, {Ivison}, {Johnstone}, {Kalas}, {Kavelaars}, {Rodriguez}, {Udry},
  {van der Werf}, {Wilner}, \& {Zuckerman}}]{Matthews10}
{Matthews}, B.~C., {Sibthorpe}, B., {Kennedy}, G., {et~al.} 2010, \aap, 518,
  L135

\bibitem[{{Mawet} {et~al.}(2011){Mawet}, {Mennesson}, {Serabyn}, {Stapelfeldt},
  \& {Absil}}]{Mawet11}
{Mawet}, D., {Mennesson}, B., {Serabyn}, E., {Stapelfeldt}, K., \& {Absil}, O.
  2011, \apjl, 738, L12

\bibitem[{{McGregor}(1994)}]{McGregor94}
{McGregor}, P.~J. 1994, \pasp, 106, 508

\bibitem[{{Mennesson} {et~al.}(2013){Mennesson}, {Absil}, {Lebreton},
  {Augereau}, {Serabyn}, {Colavita}, {Millan-Gabet}, {Liu}, {Hinz}, \&
  {Th{\'e}bault}}]{Mennesson13}
{Mennesson}, B., {Absil}, O., {Lebreton}, J., {et~al.} 2013, \apj, 763, 119

\bibitem[{{M{\'e}rand} {et~al.}(2005){M{\'e}rand}, {Bord{\'e}}, \& {Coud{\'e}
  Du Foresto}}]{Merand05}
{M{\'e}rand}, A., {Bord{\'e}}, P., \& {Coud{\'e} Du Foresto}, V. 2005, \aap,
  433, 1155

\bibitem[{{M{\'e}rand} {et~al.}(2006){M{\'e}rand}, {Coud{\'e} du Foresto},
  {Kellerer}, {ten Brummelaar}, {Reess}, \& {Ziegler}}]{Merand06}
{M{\'e}rand}, A., {Coud{\'e} du Foresto}, V., {Kellerer}, A., {et~al.} 2006, in
  Proc. SPIE, Vol. 6268, {Advances in Stellar Interferometry}, 62681F

\bibitem[{{Millan-Gabet} {et~al.}(2011){Millan-Gabet}, {Serabyn}, {Mennesson},
  {Traub}, {Barry}, {Danchi}, {Kuchner}, {Stark}, {Ragland}, {Hrynevych},
  {Woillez}, {Stapelfeldt}, {Bryden}, {Colavita}, \& {Booth}}]{MillanGabet11}
{Millan-Gabet}, R., {Serabyn}, E., {Mennesson}, B., {et~al.} 2011, \apj, 734,
  67

\bibitem[{{Moerchen} {et~al.}(2010){Moerchen}, {Telesco}, \&
  {Packham}}]{Moerchen10}
{Moerchen}, M.~M., {Telesco}, C.~M., \& {Packham}, C. 2010, \apj, 723, 1418

\bibitem[{{Moerchen} {et~al.}(2007){Moerchen}, {Telesco}, {Packham}, \&
  {Kehoe}}]{Moerchen07}
{Moerchen}, M.~M., {Telesco}, C.~M., {Packham}, C., \& {Kehoe}, T.~J.~J. 2007,
  \apjl, 655, L109

\bibitem[{{Monnier} {et~al.}(2012){Monnier}, {Che}, {Zhao}, {Ekstr{\"o}m},
  {Maestro}, {Aufdenberg}, {Baron}, {Georgy}, {Kraus}, {McAlister}, {Pedretti},
  {Ridgway}, {Sturmann}, {Sturmann}, {ten Brummelaar}, {Thureau}, {Turner}, \&
  {Tuthill}}]{Monnier12}
{Monnier}, J.~D., {Che}, X., {Zhao}, M., {et~al.} 2012, \apjl, 761, L3

\bibitem[{{Monnier} {et~al.}(2007){Monnier}, {Zhao}, {Pedretti}, {Thureau},
  {Ireland}, {Muirhead}, {Berger}, {Millan-Gabet}, {Van Belle}, {ten
  Brummelaar}, {McAlister}, {Ridgway}, {Turner}, {Sturmann}, {Sturmann}, \&
  {Berger}}]{Monnier07}
{Monnier}, J.~D., {Zhao}, M., {Pedretti}, E., {et~al.} 2007, Science, 317, 342

\bibitem[{{Morales} {et~al.}(2012){Morales}, {Padgett}, {Bryden}, {Werner}, \&
  {Furlan}}]{Morales12}
{Morales}, F.~Y., {Padgett}, D.~L., {Bryden}, G., {Werner}, M.~W., \& {Furlan},
  E. 2012, \apj, 757, 7

\bibitem[{{Morales} {et~al.}(2011){Morales}, {Rieke}, {Werner}, {Bryden},
  {Stapelfeldt}, \& {Su}}]{Morales11}
{Morales}, F.~Y., {Rieke}, G.~H., {Werner}, M.~W., {et~al.} 2011, \apjl, 730,
  L29

\bibitem[{{Morales} {et~al.}(2009){Morales}, {Werner}, {Bryden}, {Plavchan},
  {Stapelfeldt}, {Rieke}, {Su}, {Beichman}, {Chen}, {Grogan}, {Kenyon},
  {Moro-Martin}, \& {Wolf}}]{Morales09}
{Morales}, F.~Y., {Werner}, M.~W., {Bryden}, G., {et~al.} 2009, \apj, 699, 1067

\bibitem[{{Mozurkewich} {et~al.}(2003){Mozurkewich}, {Armstrong}, {Hindsley},
  {Quirrenbach}, {Hummel}, {Hutter}, {Johnston}, {Hajian}, {Elias}, {Buscher},
  \& {Simon}}]{Mozurkewich03}
{Mozurkewich}, D., {Armstrong}, J.~T., {Hindsley}, R.~B., {et~al.} 2003, \aj,
  126, 2502

\bibitem[{{Neugebauer} \& {Leighton}(1969)}]{Neugebauer69}
{Neugebauer}, G. \& {Leighton}, R.~B. 1969, {Two-micron sky survey. A
  preliminary catalogue}

\bibitem[{{Nidever} {et~al.}(2002){Nidever}, {Marcy}, {Butler}, {Fischer}, \&
  {Vogt}}]{Nidever02}
{Nidever}, D.~L., {Marcy}, G.~W., {Butler}, R.~P., {Fischer}, D.~A., \& {Vogt},
  S.~S. 2002, \apjs, 141, 503

\bibitem[{{Owocki} {et~al.}(1994){Owocki}, {Cranmer}, \& {Blondin}}]{Owocki94}
{Owocki}, S.~P., {Cranmer}, S.~R., \& {Blondin}, J.~M. 1994, \apj, 424, 887

\bibitem[{{Padgett} {et~al.}(2013){Padgett}, {Liu}, {Stapelfeldt},
  {Fajardo-Acosta}, \& {Leisawitz}}]{Padgett13}
{Padgett}, D.~L., {Liu}, W., {Stapelfeldt}, K., {Fajardo-Acosta}, S., \&
  {Leisawitz}, D. 2013, \apj, in press

\bibitem[{{Peterson} {et~al.}(2006){Peterson}, {Hummel}, {Pauls}, {Armstrong},
  {Benson}, {Gilbreath}, {Hindsley}, {Hutter}, {Johnston}, {Mozurkewich}, \&
  {Schmitt}}]{Peterson06}
{Peterson}, D.~M., {Hummel}, C.~A., {Pauls}, T.~A., {et~al.} 2006, \apj, 636,
  1087

\bibitem[{{Plavchan} {et~al.}(2009){Plavchan}, {Werner}, {Chen}, {Stapelfeldt},
  {Su}, {Stauffer}, \& {Song}}]{Plavchan09}
{Plavchan}, P., {Werner}, M.~W., {Chen}, C.~H., {et~al.} 2009, \apj, 698, 1068

\bibitem[{{Reiners}(2006)}]{Reiners06}
{Reiners}, A. 2006, \aap, 446, 267

\bibitem[{{Rhee} {et~al.}(2007){Rhee}, {Song}, {Zuckerman}, \&
  {McElwain}}]{Rhee07}
{Rhee}, J.~H., {Song}, I., {Zuckerman}, B., \& {McElwain}, M. 2007, \apj, 660,
  1556

\bibitem[{{Richichi} {et~al.}(1996){Richichi}, {Calamai}, {Leinert},
  {Stecklum}, \& {Trunkovsky}}]{Richichi96}
{Richichi}, A., {Calamai}, G., {Leinert}, C., {Stecklum}, B., \& {Trunkovsky},
  E.~M. 1996, \aap, 309, 163

\bibitem[{{Rieke} {et~al.}(2005){Rieke}, {Su}, {Stansberry}, {Trilling},
  {Bryden}, {Muzerolle}, {White}, {Gorlova}, {Young}, {Beichman},
  {Stapelfeldt}, \& {Hines}}]{Rieke05}
{Rieke}, G.~H., {Su}, K.~Y.~L., {Stansberry}, J.~A., {et~al.} 2005, \apj, 620,
  1010

\bibitem[{{Roberge} {et~al.}(2012){Roberge}, {Chen}, {Millan-Gabet},
  {Weinberger}, {Hinz}, {Stapelfeldt}, {Absil}, {Kuchner}, \&
  {Bryden}}]{Roberge12}
{Roberge}, A., {Chen}, C.~H., {Millan-Gabet}, R., {et~al.} 2012, \pasp, 124,
  799

\bibitem[{{Royer} {et~al.}(2007){Royer}, {Zorec}, \& {G{\'o}mez}}]{Royer07}
{Royer}, F., {Zorec}, J., \& {G{\'o}mez}, A.~E. 2007, \aap, 463, 671

\bibitem[{{Santos} {et~al.}(2001){Santos}, {Israelian}, \& {Mayor}}]{Santos01}
{Santos}, N.~C., {Israelian}, G., \& {Mayor}, M. 2001, \aap, 373, 1019

\bibitem[{{Sch{\"u}tz} {et~al.}(2009){Sch{\"u}tz}, {Meeus}, {Sterzik}, \&
  {Peeters}}]{Schutz09}
{Sch{\"u}tz}, O., {Meeus}, G., {Sterzik}, M.~F., \& {Peeters}, E. 2009, \aap,
  507, 261

\bibitem[{{Selby} {et~al.}(1988){Selby}, {Hepburn}, {Blackwell}, {Booth},
  {Haddock}, {Arribas}, {Leggett}, \& {Mountain}}]{Selby88}
{Selby}, M.~J., {Hepburn}, I., {Blackwell}, D.~E., {et~al.} 1988, \aaps, 74,
  127

\bibitem[{{Smith} {et~al.}(2009){Smith}, {Wyatt}, \& {Haniff}}]{Smith09}
{Smith}, R., {Wyatt}, M.~C., \& {Haniff}, C.~A. 2009, \aap, 503, 265

\bibitem[{{Smith} {et~al.}(2012){Smith}, {Wyatt}, \& {Haniff}}]{Smith12}
{Smith}, R., {Wyatt}, M.~C., \& {Haniff}, C.~A. 2012, \mnras, 422, 2560

\bibitem[{{Soubiran} {et~al.}(2010){Soubiran}, {Le Campion}, {Cayrel de
  Strobel}, \& {Caillo}}]{Soubiran10}
{Soubiran}, C., {Le Campion}, J.-F., {Cayrel de Strobel}, G., \& {Caillo}, A.
  2010, \aap, 515, A111

\bibitem[{{Su} {et~al.}(2013){Su}, {Rieke}, {Malhotra}, {Stapelfeldt},
  {Hughes}, {Bonsor}, {Wilner}, {Balog}, {Watson}, {Werner}, \&
  {Misselt}}]{Su13}
{Su}, K.~Y.~L., {Rieke}, G.~H., {Malhotra}, R., {et~al.} 2013, \apj, 763, 118

\bibitem[{{Su} {et~al.}(2006){Su}, {Rieke}, {Stansberry}, {Bryden},
  {Stapelfeldt}, {Trilling}, {Muzerolle}, {Beichman}, {Moro-Martin}, {Hines},
  \& {Werner}}]{Su06}
{Su}, K.~Y.~L., {Rieke}, G.~H., {Stansberry}, J.~A., {et~al.} 2006, \apj, 653,
  675

\bibitem[{{Su} {et~al.}(2008){Su}, {Rieke}, {Stapelfeldt}, {Smith}, {Bryden},
  {Chen}, \& {Trilling}}]{Su08}
{Su}, K.~Y.~L., {Rieke}, G.~H., {Stapelfeldt}, K.~R., {et~al.} 2008, \apjl,
  679, L125

\bibitem[{{ten Brummelaar} {et~al.}(2005){ten Brummelaar}, {McAlister},
  {Ridgway}, {Bagnuolo}, {Turner}, {Sturmann}, {Sturmann}, {Berger}, {Ogden},
  {Cadman}, {Hartkopf}, {Hopper}, \& {Shure}}]{tenBrummelaar05}
{ten Brummelaar}, T.~A., {McAlister}, H.~A., {Ridgway}, S.~T., {et~al.} 2005,
  \apj, 628, 453

\bibitem[{{Tokovinin} {et~al.}(2010){Tokovinin}, {Mason}, \&
  {Hartkopf}}]{Tokovinin10}
{Tokovinin}, A., {Mason}, B.~D., \& {Hartkopf}, W.~I. 2010, \aj, 139, 743

\bibitem[{{Trilling} {et~al.}(2008){Trilling}, {Bryden}, {Beichman}, {Rieke},
  {Su}, {Stansberry}, {Blaylock}, {Stapelfeldt}, {Beeman}, \&
  {Haller}}]{Trilling08}
{Trilling}, D.~E., {Bryden}, G., {Beichman}, C.~A., {et~al.} 2008, \apj, 674,
  1086

\bibitem[{{Tuomi} {et~al.}(2013){Tuomi}, {Jones}, {Jenkins}, {Tinney},
  {Butler}, {Vogt}, {Barnes}, {Wittenmyer}, {O'Toole}, {Horner}, {Bailey},
  {Carter}, {Wright}, {Salter}, \& {Pinfield}}]{Tuomi13}
{Tuomi}, M., {Jones}, H.~R.~A., {Jenkins}, J.~S., {et~al.} 2013, \aap, 551, A79

\bibitem[{{Tuthill} {et~al.}(2006){Tuthill}, {Lloyd}, {Ireland}, {Martinache},
  {Monnier}, {Woodruff}, {ten Brummelaar}, {Turner}, \& {Townes}}]{Tuthill06}
{Tuthill}, P., {Lloyd}, J., {Ireland}, M., {et~al.} 2006, in Proc. SPIE, Vol.
  6272, {Advances in Adaptive Optics II}, 62723A

\bibitem[{{Valenti} \& {Fischer}(2005)}]{Valenti05}
{Valenti}, J.~A. \& {Fischer}, D.~A. 2005, \apjs, 159, 141

\bibitem[{{van Belle} \& {von Braun}(2009)}]{vanBelle09}
{van Belle}, G.~T. \& {von Braun}, K. 2009, \apj, 694, 1085

\bibitem[{{Watson} {et~al.}(2011){Watson}, {Littlefair}, {Diamond}, {Collier
  Cameron}, {Fitzsimmons}, {Simpson}, {Moulds}, \& {Pollacco}}]{Watson11}
{Watson}, C.~A., {Littlefair}, S.~P., {Diamond}, C., {et~al.} 2011, \mnras,
  413, L71

\bibitem[{{Wittenmyer} {et~al.}(2006){Wittenmyer}, {Endl}, {Cochran}, {Hatzes},
  {Walker}, {Yang}, \& {Paulson}}]{Wittenmyer06}
{Wittenmyer}, R.~A., {Endl}, M., {Cochran}, W.~D., {et~al.} 2006, \aj, 132, 177

\bibitem[{{Wright} {et~al.}(2004){Wright}, {Marcy}, {Butler}, \&
  {Vogt}}]{Wright04}
{Wright}, J.~T., {Marcy}, G.~W., {Butler}, R.~P., \& {Vogt}, S.~S. 2004, \apjs,
  152, 261

\bibitem[{{Wyatt} {et~al.}(2007){Wyatt}, {Smith}, {Greaves}, {Beichman},
  {Bryden}, \& {Lisse}}]{Wyatt07}
{Wyatt}, M.~C., {Smith}, R., {Greaves}, J.~S., {et~al.} 2007, \apj, 658, 569

\bibitem[{{Zhao} {et~al.}(2009){Zhao}, {Monnier}, {Pedretti}, {Thureau},
  {M{\'e}rand}, {ten Brummelaar}, {McAlister}, {Ridgway}, {Turner}, {Sturmann},
  {Sturmann}, {Goldfinger}, \& {Farrington}}]{Zhao09}
{Zhao}, M., {Monnier}, J.~D., {Pedretti}, E., {et~al.} 2009, \apj, 701, 209

\end{thebibliography}

\end{document}